# The Ultraviolet Spectrograph on ESA's Jupiter Icy Moons Explorer Mission (JUICE-UVS)

**Retherford, K. D.[1,2], P. M. Molyneux[1], T. K. Greathouse[1], M. W. Davis[1], G. R. Gladstone[1], S. C. Persyn[1], M. H. Versteeg[1], M. F. Araujo[1], F. Bagenal[3], T. M. Becker[1,2], A. Beth[4], R. K. Black[1], B. Bonfond[5], S. M. Brooks[7], E. J. Bunce[8], A. Carapelle[9], S. Cortinas[1], G. J. Dirks[1], B. Esquivel[1], P. D. Feldman[+], S. Ferrell[1], M. Ferris[1], L. N. Fletcher[8], M. A. Freeman[1], M. Galand[4], R. S. Giles[1], D. Grodent[5], V. Hue[10], E. Johnson[1], J. A. Kammer[1], C. Kempe[1], L. Lamy[10], A. Martin[11], M. A. McGrath[12], E. G. Nerney[3], C. Nuñez[1], Z. Olsen[1], N. Pelletier[1+], B. Perez[1], K. B. Persson[1], E. Quémerais[6], R. Raffanti[11], U. Raut[1,2], R. Rickerson[1], L. Roth[13], O. H. W. Siegmund[11], J. R. Spencer[14], A. J. Steffl[14], B. J. Trantham[1], J. V. Vallerga[11], T. J. Veach[1,2], M. A. Velez[1], B. C. Walther[1] and the JUICE-UVS Team**

## Abstract

The Jupiter Icy Moons Explorer (JUICE) mission, led by ESA, has an Ultraviolet Spectrograph (JUICE-UVS) contributed by NASA and built at Southwest Research Institute. JUICE-UVS is designed to provide a diversity of measurements to further our understanding of the potential habitability of icy ocean worlds at Jupiter and to study Jupiter and the Jovian system as an archetype for gas giants. JUICE-UVS observes photons in the 50-204 nm wavelength range at moderate spectral and spatial resolution along a 7.5° slit composed of 7.3°×0.1° and 0.2°×0.2° contiguous sections. JUICE-UVS performs a comprehensive study of icy satellite atmospheres, plumes, surfaces, and local space environments; Jupiter's atmosphere and aurora; Io and its Io Plasma Torus; and other Jupiter system targets (rings, small moons, etc.) as available. The variety of observational techniques employed include: nadir push-broom imaging, disk scans, limb stares, stellar and solar occultations, Jupiter transit observations, and neutral cloud/plasma torus stares and scans. This paper describes the UVS investigation's science plans, instrument details, concept of operations, and data formats in the context of the JUICE mission's habitability and Jupiter system goals.

### 1. Introduction

Ganymede's intrinsic magnetic field is unique among solar system satellites, and gives rise to fascinating auroral emissions and phenomena associated with its tenuous atmosphere. We describe here the Jupiter Icy Moons Explorer (JUICE) mission's Ultraviolet Spectrograph (JUICE-UVS) instrument and its role in completing the overall goals and objectives for the project, which emphasize Ganymede aurora studies once in orbit around this moon of Jupiter, in addition to many UV spectral imaging observations of the myriad targets within the Jovian system. As described in Witasse et al. (2026), Grasset et al. (2013) and Witasse et al. (2020), two of the top science themes for JUICE are A) Emergence of habitable worlds around gas giants; and B) Jupiter system as an archetype for gas giants. The JUICE-UVS instrument design is tailored to address these top goals with objectives to: 1) Explore the atmospheres, plasma interactions, and surfaces of the Galilean satellites; 2) Determine the dynamics, chemistry, and vertical structure of Jupiter's upper atmosphere, from equator to pole, as a template for giant planets everywhere; and 3) Investigate the Jupiter-Io connection by quantifying energy and mass flow in the Io atmosphere, neutral cloud, and torus. In short, these objectives are summarized as Icy satellites, Jupiter, and the Io system, respectively. Other Jupiter system targets (rings, small moons, etc.) will be investigated as available.

The JUICE-UVS imaging spectrograph obtains both spatial and spectral information simultaneously on a detector capturing an image of its long slit's field-of-view (FOV) at each wavelength. Its optical grating disperses extreme-UV (EUV) to far-UV (FUV) wavelengths instantaneously onto the image plane, hence the name "spectrograph" rather than "spectrometer" (which in the original historical sense scans frequency). Designed, built, assembled, aligned, and calibrated at the Southwest Research Institute (SwRI), the instrument was delivered to the European Space Agency (ESA) as shown in Fig. 1. The versatility of the UVS instrument and its observational techniques (Section 3) enable several key science measurements (Section 2) as guided by the JUICE project's approach for planning copious measurements of the Jupiter system prior to entering Ganymede orbit for intensive study.

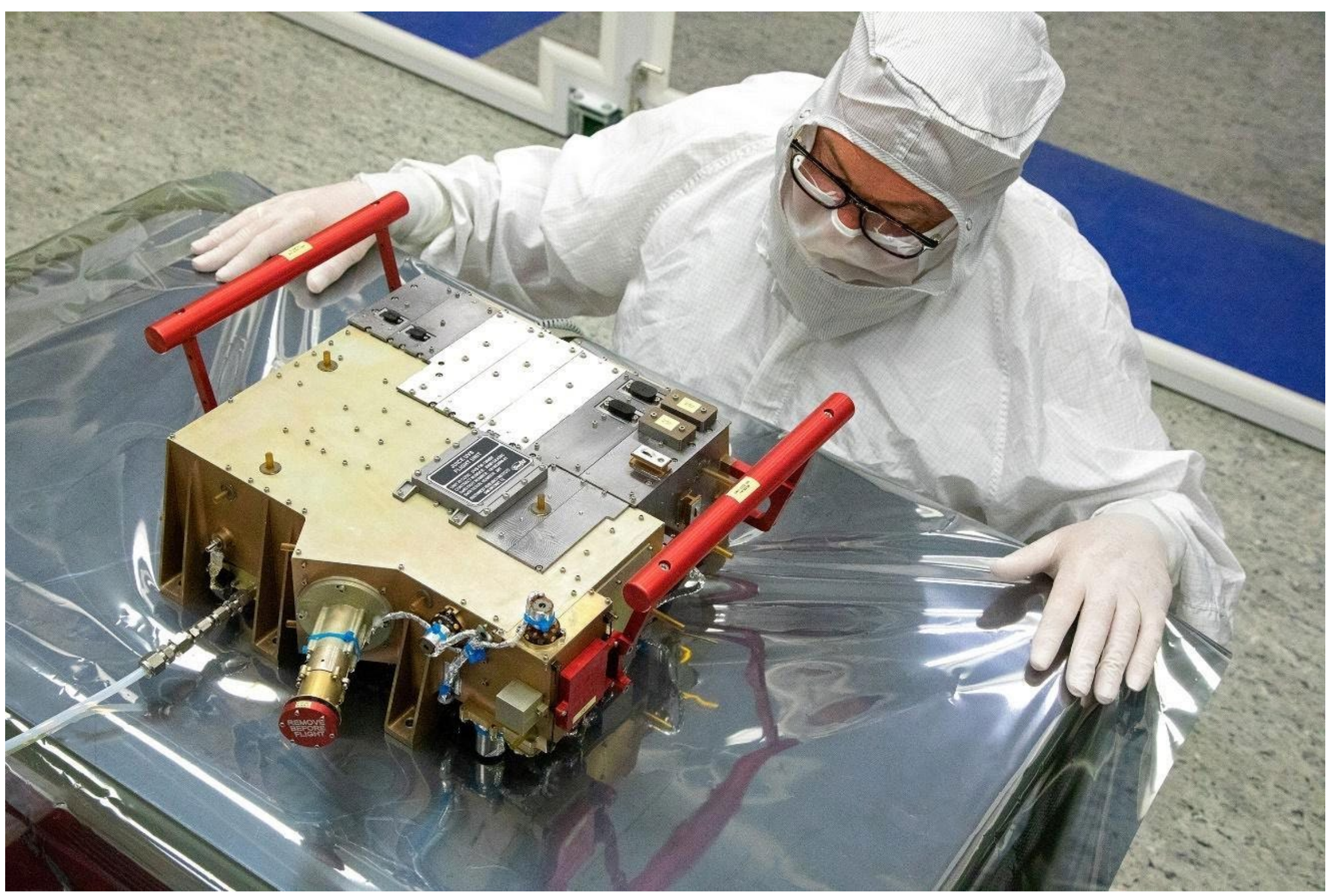

**Fig. 1** JUICE-UVS photo at time of delivery in February 2020. With a volume of 35.4 cm × 41.1 cm × 15.8 cm its size is approximately that of a briefcase

## 2. Scientific Objectives

To achieve the baseline Level-1 JUICE mission requirements with JUICE-UVS, a few dozen measurement requirements were defined and supplemented with additional guidelines. In the following sections, we summarize the motivation and rationales for these planned measurements and discuss opportunities for additional investigations such as those enabled by joint measurements with the Europa Clipper mission (Pappalardo et al. 2024) and its “sister” instrument Europa-UVS (Retherford et al. 2024). The JUICE-UVS science and observation techniques are best discussed in terms of studying icy moons (atmospheres, surfaces, and local space environment), Jupiter, and the Io system. Several questions guided the development of the proposed science case for JUICE-UVS, as repeated in Fig. 2. Outstanding issues raised in the review by McGrath et al. (2004) still remain, *viz.*:

• What is the relative abundance of O vs. $O_2$ in the Europa and Ganymede atmospheres and how are they distributed spatially?

• Why is Europa’s neutral atomic oxygen emission (indicative of its $O_2$ atmosphere) apparently non-uniform?

• What is the average $O_2$ atmospheric density at Ganymede? Do the UV and visible oxygen emissions detected at Ganymede have a common (auroral) origin?

• What is the source of Callisto's atmosphere, endogenic or exogenic? Is the major species inferred from the ionospheric measurements $O_2$, as is presumed?

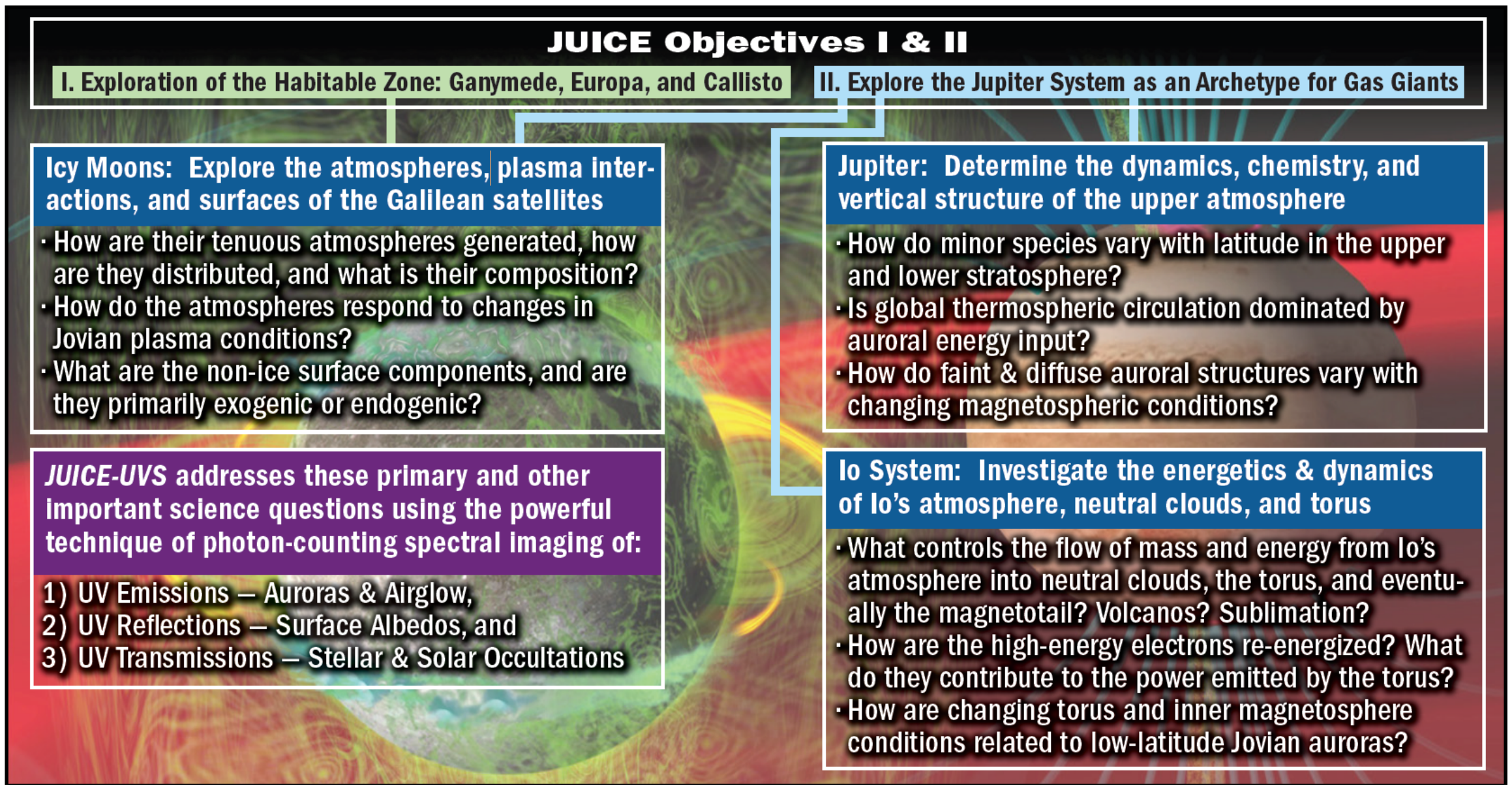

**Fig. 2** JUICE-UVS science objectives include studies of icy moons, Jupiter, and the Io System

## 2.1. Icy Moons

### 2.1.1. Tenuous atmospheres

The icy moons of Jupiter (Europa, Ganymede, and Callisto) have tenuous atmospheres, but little is known about them except that they are generally surface-bounded and nearly collisionless (McGrath et al. 2004, 2009, Roth et al. 2024). Few details are known about their composition, except that their surfaces are dominated by water species. UV emissions of atomic oxygen (130.4 nm, 135.6 nm) and hydrogen (121.6 nm) have been observed at all three moons (Hall et al. 1995, 1998; Feldman et al. 2000; Cunningham et al. 2015; Roth et al. 2017a, 2017b), and the ratio of the observed UV oxygen emissions is diagnostic of the excitation mechanism that produces them. Dissociative electron impact excitation of $O_2$ produces a 135.6 nm / 130.4 nm intensity ratio of ~2.3 (Kanik et al. 2003), while electron impact on O or $H_2O$ leads to emission ratios <1. Roth et al. (2021a, 2021b) and Leblanc et al. (2023) used maps of the UV oxygen emission ratio to demonstrate that the atmospheres of Europa and Ganymede are globally dominated by $O_2$, with water vapor dominating on Europa's sunlit trailing hemisphere, and on the sunlit hemisphere of Ganymede. Ganymede's $H_2O$ component is clearly related to sublimation, which is the dominant atmospheric source near the subsolar point. At Europa, the colder surface temperatures and higher surface sputtering rates mean that the production mechanism(s) responsible for the $H_2O$ component are more ambiguous (Roth et al. 2021b). $H_2O$ and $O_2$ are also present in the atmosphere of Callisto, likely having the densest, most collisional atmosphere of the three (Cunningham et al. 2015, Carberry Mogan et al. 2020, 2021, 2022).

The presence of $H_2O$ and $O_2$ at all three icy moons was confirmed by optical observations (de Kleer et al. 2023).

Atomic hydrogen, H, has also been detected—both from Galileo and by the Hubble Space Telescope (HST). Modeling shows that other species must also be present, particularly O, $H_2$ and OH. $CO_2$ has been detected on Callisto (Carlson 1999a, Cartwright et al. 2024) and at Ganymede (Bockelée Morvan et al. 2024). However, the global distribution of these species is unknown, as is the possible correlation with surface features that give rise to the initial sputtered or sublimated water molecules.

JUICE-UVS will produce multiple full-disk maps of the UV oxygen and hydrogen emissions at Europa, Ganymede, and Callisto during each moon flyby period (Section 3.7.2), including nightside maps not achievable by Earth-based observatories. These observations will allow comprehensive studies of the composition, distribution, and temporal variability of the satellite atmospheres, providing new insight into atmospheric production and losses. The UVS bandpass also encompasses many emission lines of potential minor atmospheric species, including S, C, CO, Cl, etc. Limb stare observations will be performed to build up high signal-to-noise ratio (SNR) spectra of the satellite near-surface atmospheres, constraining the abundances of these and other trace species.

Further information about the tenuous icy moon atmospheres will be obtained via absorption spectroscopy. JUICE-UVS will perform stellar occultation measurements (Section 3.7.5), observing the attenuation of starlight as a function of wavelength as the line-of-sight between the instrument and the star passes through a satellite atmosphere. This technique will be used to monitor UV-absorbing species including $H_2O$ and $O_2$. Similarly, the absorption of Jupiter airglow by satellite atmospheres as they transit Jupiter (Roth et al. 2017a, 2017c, 2023a, Sparks et al. 2016, 2017, Giono et al. 2020) will be mapped (Section 3.7.7). These transit observations are particularly useful for identifying localized regions of denser atmosphere on the satellite limb, potentially associated with plume activity.

### 2.1.2. Response to Jovian plasma interaction

Images of far-UV neutral atomic emissions from the icy satellites provide information on the interaction between the Jovian plasma environment, in addition to the exospheric gases themselves. The emission brightness is a function of neutral gas density, electron density, and electron temperature, with the electron properties closely related to the overall bulk fields and particles environment. Auroral emissions are the primary features in Europa and Ganymede's far-UV imagery, while airglow processes dominate Callisto's emission as a result of its relatively more conductive ionosphere deflecting the impingement of incoming plasma flux tubes. Sections 2.1.4 to 2.1.6 provide more details for each icy moon. Saur et al. (2015) demonstrated that the locations of Ganymede's auroral ovals provide a diagnostic method for constraining the conductance of a subsurface ocean. Roth et al. (2017b) use Io's auroral morphology to similarly constrain the conductance of a possible subsurface magma layer. The use of far-UV auroral imagery to constrain Europa's subsurface ocean remains to be demonstrated (e.g., Schlegel et al. submitted 2025), while Callisto's airglow features are presumably unsuitable for investigating its interior properties. UVS provides a global view of the interaction region, complementing the Particle Environment Package (PEP, Barabash et al. *this collection*), Radio and Plasma

Wave Investigation (RPWI, Wahlund et al. 2024), and JUICE Magnetometer (J-MAG, Dougherty et al. *this collection*) in situ investigations.

### 2.1.3. Non-ice surface components

The surfaces of the icy moons of Jupiter are not just pure water ice; they contain minor species such as $CO_2$ (Villaneuva et al. 2023, Trumbo & Brown 2023, Cartwright et al. 2024, 2025, Hibbitts et al. 2003), $SO_2$ (Hendrix et al. 2011, Becker et al. 2022), and $H_2O_2$ (Carlson et al. 1999b, Trumbo et al. 2023) volatiles, NaCl (Trumbo et al. 2022), and salts, acids and their hydrated forms (Trumbo et al. 2019, Dalton et al. 2013, McCord et al. 2010, Brown and Hand 2012). The occurrence and distribution of these minor species provide clues to the complex history of the endogenic (e.g., tectonics and icy volcanism) and exogenic (e.g., energetic particle and micro-meteor bombardment, solar UV) processes that modify them over the age of the solar system. Compositional studies of Galilean moons at near-IR wavelengths maturing with recent James Webb Space Telescope (JWST) observations (e.g. Villanueva et al. 2023, Cartwright et al. 2025, Raut et al. 2026) are usefully complemented by observations at FUV wavelengths (e.g., Becker et al. 2024). In particular, with respect to the icy moons of Jupiter, surface composition is an area where mid-UV (200-300 nm) observations have been useful, detecting $O_3$ in the surface ices of Ganymede (Noll et al. 1996) and $SO_2$ at Callisto (Noll et al. 1997) and Europa (Noll et al. 1995, Carlson et al. 2009). Follow-up observations indicated that the ozone features on Ganymede are uncorrelated with major geological surface units (e.g., Spencer et al. 1999), although there remains an intriguing dependence on solar zenith angle (e.g., Hendrix et al. 1999). More recently, FUV surface reflectance observations of the icy moons have advanced our understanding of the composition on the topmost surface veneer, which is strongly modulated by radiolytic and photolytic processes (e.g., Hendrix et al. 2016, Becker et al. 2018, 2022, Molyneux et al. 2020, Mamo et al. 2025). During Ganymede-orbiting phases especially, JUICE-UVS will be mostly nadir pointed, building up brightness and albedo maps with reflected sunlight and Interplanetary Medium (IPM) Lyman-α (121.6 nm), much like LRO-LAMP at the Earth's Moon since 2009 (Gladstone et al. 2012).

### 2.1.4. Ganymede

Ultraviolet oxygen emissions from Ganymede were first detected by the Galileo mission (Hall et al. 1998) and later found to include auroras in HST Space Telescope Imaging Spectrograph (STIS) images (Fig. 3), highlighting the particle-focusing power of Ganymede's intrinsic magnetic field (Feldman et al. 2000). As Ganymede generates its own magnetic field (Kivelson et al. 1996), the auroral emissions form two ovals around its poles. As with all aurora and airglow phenomena, the morphology and spectroscopy of the emissions yield information about the ambient atmosphere and, for the case of the aurora, the nature of the precipitating particles. More recent observations made by Juno UVS, the predecessor to JUICE-UVS, during a close, 1046 km altitude (Hansen et al. 2022), flyby of Ganymede allowed for the highest spatial resolution observations of Ganymede's aurora to date; they showed the two distinct northern/southern bands of emissions most probably mapping the open/closed field line boundary, large variations in auroral brightness with longitude, and patchy structure down to the smallest resolved scale of 4 km in the observations (Greathouse et al. 2022). These observations, along with the rich in situ data set gathered during the same Juno flyby, have set off a burst of modeling studies to further understand either what precipitating particle distributions are required to form the emissions assuming a given neutral background (Benmahi et al. 2025) or to constrain the exospheric structure using non-coincident observed electron fluxes (Waite et al. 2024). The details of how these emissions vary with time and location is currently unknown, and understanding them is an important JUICE objective. The JUICE mission's plan to enter into a circular polar orbit about Ganymede at altitudes of 500 km and then 200 km will allow for unprecedented spatial and temporal resolution mapping of the aurora, while measuring in situ auroral electrons and atmospheric neutrals. Combining both coincident datasets, complemented by detailed neutral and plasma modeling (e.g., Beth et al. 2025, Vorburger et al. 2024), will allow retrieval of critical information on the currently poorly-constrained neutral atmosphere.

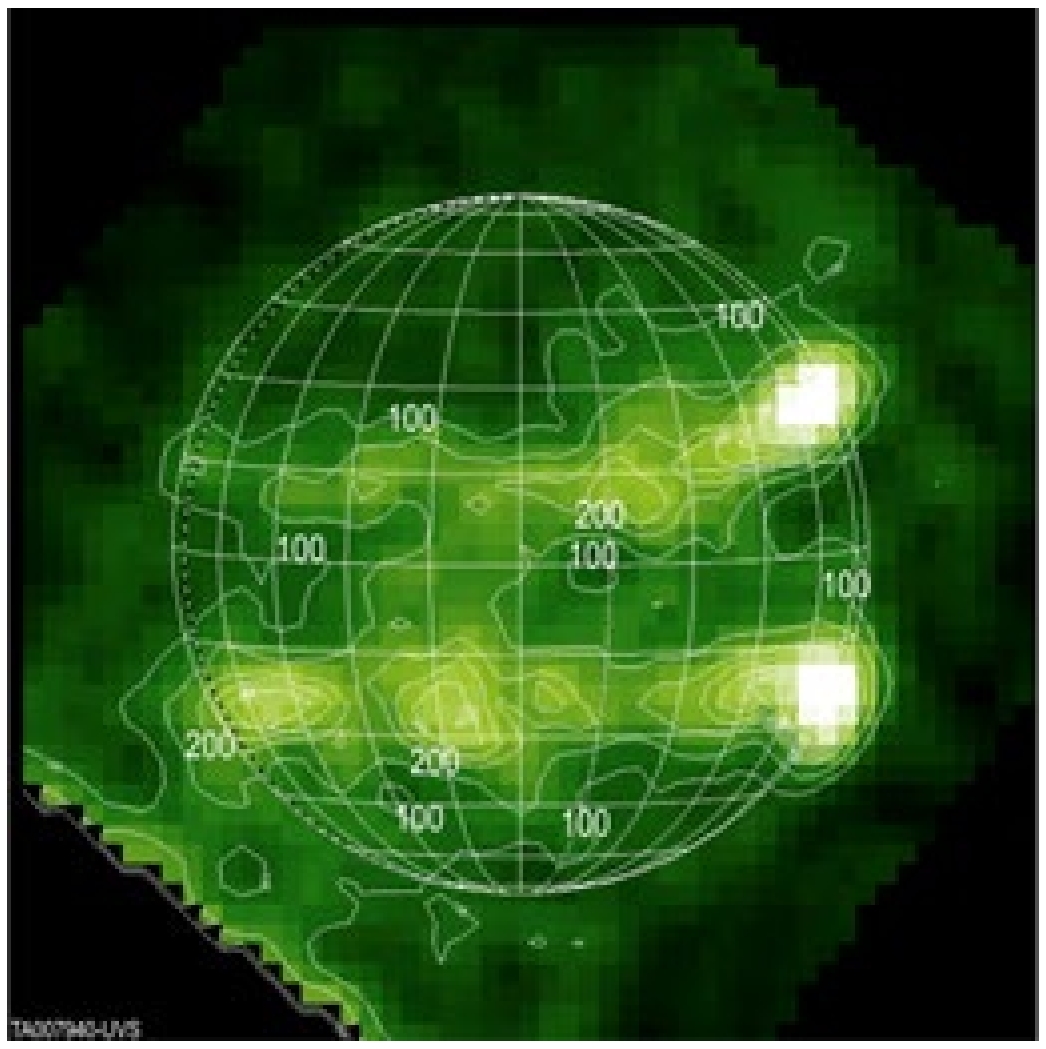

**Fig. 3** Ganymede auroral emissions seen by the HST STIS instrument (brightness contours in Rayleighs; Feldman et al. 2000). UVS observations from JUICE's ultimate 200-km altitude in Ganymede orbit will resolve such structures at ~1-km resolution to help us understand how the magnetospheres of Ganymede and Jupiter interact

JUICE-UVS is also well suited to characterizing Ganymede's patchy $CO_2$ atmosphere, which was first detected by JWST (Bockelée-Morvan et al. 2024). The $CO_2$ abundance was observed to peak over the northern polar region on the leading hemisphere, but the reason for this localized enhancement is not yet understood, and the relative importance of sublimation and sputtering to the generation of the $CO_2$ component of the exosphere is similarly poorly constrained. UVS will address these open questions by mapping and monitoring carbon emissions at 156.1 nm and 165.7 nm, and by measuring absorption of starlight at wavelengths <120 nm by $CO_2$ during stellar and solar occultations (see Sections 3.7.5 and 3.7.6). Observations covering a wide range of longitudes, latitudes and local times will be used to determine the sources and sinks of Ganymede's $CO_2$ and potentially other minor atmospheric species.

The UV reflectance of Ganymede's surface will be mapped and monitored to determine the composition and distribution of volatiles including $O_2$, $O_3$, $H_2O_2$, $CO_2$, $NH_3$, etc. that, if present, will be important sources of atmospheric species, as well as water ice. Previous studies have noted that Ganymede is darker at UV wavelengths >165 nm than expected based on the $H_2O$ abundance derived from visible and near-IR observations (Molyneux et al. 2020, 2022), likely due to modification of the ice by the impinging Jovian plasma. However, it is not clear if this surface darkening is a chemical process, e.g., via radiolytic production of UV-absorbing contaminants, or the result of physical damage to the ice matrix. JUICE-UVS will address this question by mapping the surface reflectance near 165 nm, longward of which water ice becomes highly reflective, to determine where the ice is most modified and whether this is correlated to the charged particle surface access, surface temperature, the presence of additional surface materials, etc. UVS will also identify and map two absorption features detected during the Juno UVS flyby of Ganymede: a broad absorption centered around 180 nm seen in Tros crater and potentially related to $NH_3$, and another at >190 nm that may be the start of an ozone absorption band (Molyneux et al. 2022).

JUICE-UVS will observe the icy moon atmospheres over a wavelength range between 55 to 180 nm, which enables the separation and identification of atomic oxygen emission lines at 130.4 and 135.6 nm and atomic hydrogen features at 121.6 nm (Fig. 4). To ensure that scans of the atmosphere yield observations with sufficient signal to make compositional and spatial measurements, JUICE-UVS is able to pre-configure the acquired datasets with variable spectral and spatial bins (see Section 3.6) to achieve the required signal quality in each programmable acquisition.

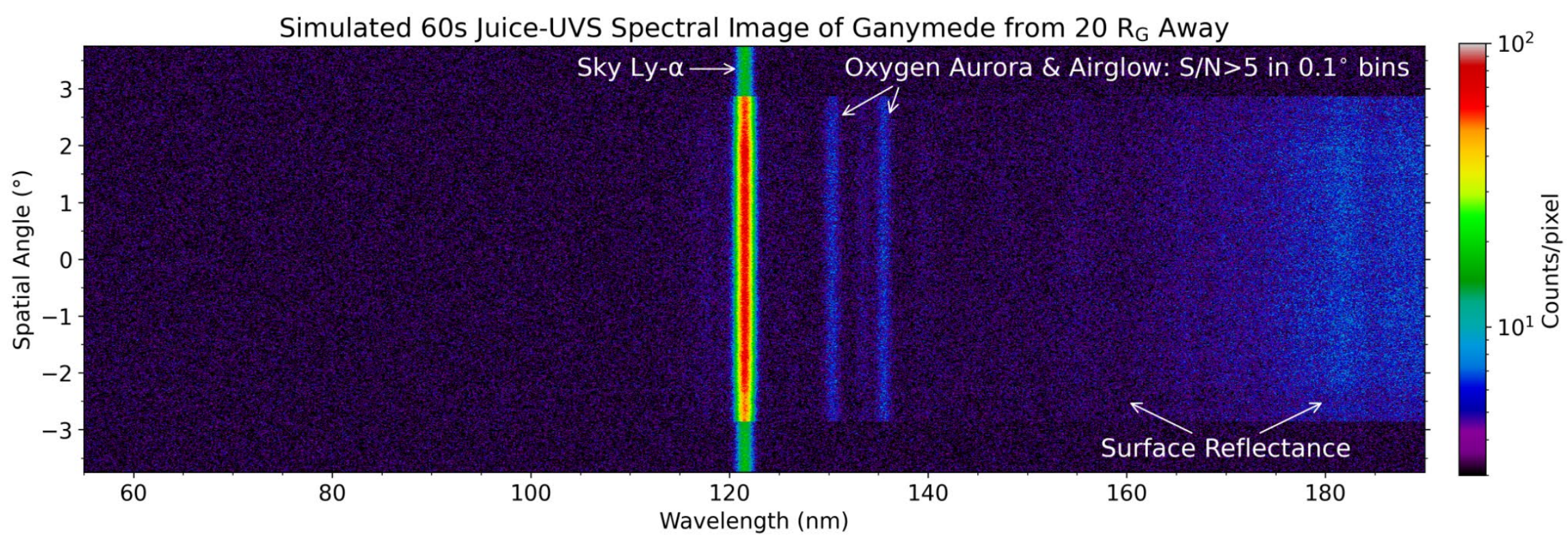

**Fig. 4** Proof-of-concept spectral image demonstrating JUICE-UVS expected counts per pixel for a nearly filled slit of Ganymede's sunlit disk from 20 $R_{Ganymede}$ away. Reflected sunlight at far-UV wavelengths increases in brightness at longer wavelengths. Simulated atomic oxygen emissions from dissociated $O_2$ are shown averaged across the disk at expected brightnesses. Lyman-α skyglow routinely fills the slit as a background source and is also shown at 121.6 nm. Simulated radiation noise counts are included across the entire detector image

### 2.1.5. Europa

JUICE-UVS studies of Europa's tenuous atmosphere, plasma interactions, and the non-ice surface distribution will support a key overarching science objective for the moon: to search for ongoing geophysical activity. The first tentative evidence for such activity was in the form of outgassing water vapor plumes, detected as coincident FUV emissions of H and O from HST (Roth et al. 2014a, 2014b, 2026), which was then followed by detections of absorption from a potential plume off the limb of Europa as the moon transited in front of Jupiter (Sparks et al. 2016, 2017, Giono et al. 2020). Additional evidence (Blöcker et al. 2016, Paganini et al. 2020), including a reanalysis of magnetometer data from the Galileo mission (Jia et al. 2018), supports the prospect of active plume activity, though a series of non-detections by HST (e.g., Roth et al. 2014b) and JWST (Villanueva et al. 2023) indicate that such plumes must be stochastic if present at all. Roth et al. (2026) reanalyzed the original Roth et al. (2014a) data set as informed by numerous subsequently obtained data sets and concluded that the evidence for plumes in that data set is no longer valid. A summary of previous detections and a layout for the types of needed observations at Europa to confirm the presence of activity are described by Roth et al. (2025b) and Tosi et al. (2024).

JUICE-UVS is particularly capable of detecting and characterizing potential actively-released water vapor in both emission and absorption if such plumes reach a few tens of km above Europa's surface. During the two Europa flybys, JUICE-UVS will make surface, aurora, and airglow observations that can identify any short term changes or variability. The majority of these observations will be performed as a series of scans over the full disk to a distance of ~0.5 $R_E$ above the surface, collecting data simultaneously with JUICE's other remote-sensing instruments. Nominally, a total of eight scans will be conducted during the inbound and outbound portions of the flybys, with limb stares conducted in between the scans. The scans will map the distribution of emissions from the atmosphere over the course of the flyby and will be used to look for coincident emissions produced from water dissociation products within plumes silhouetted against space, as shown in HST data by Roth et al. (2014a, 2014b, 2026).

The improved spatial resolution of JUICE-UVS FUV surface reflectance observations obtained from the scans will be used to map $H_2O$ ice concentrations across the surface. Variations in both non-ice detections and the purity of fresh, less-irradiated ice which can be observed as a stronger reflectance enhancement at ~165 nm, could be used to identify freshly-deposited ice on the surface (Mamo et al. 2025, Becker et al. 2024, Raut et al. 2023, Velez et al. 2026).

Water vapor plumes can also be identified through FUV absorption with JUICE-UVS. Europa transits across Jupiter's disk are a powerful means of detecting telltale absorptions by water vapor (and potentially other constituents) against the backdrop of the planet all along the limb of Europa. This same observation technique was implemented by Sparks et al. (2016, 2017) and Roth et al. (2017c) for investigating Europa's atmosphere and searching for plumes with HST/STIS. Active plumes can also be discovered and characterized through stellar occultations by Europa. If present at sufficient levels, plume constituents such as $H_2O$, $CO_2$, $O_2$, $CH_4$, $NH_3$, $H_2$ and $CO_2$ will be measured.

In addition to plume searches, JUICE-UVS will address outstanding questions about Europa's atmosphere including: 1) likely hemispherical asymmetries in its global atmospheric density, 2) the scale heights of its main constituents, 3) the location of its exobase, and 4) the contribution of material to Jupiter's magnetosphere (Bagenal and Dols 2020). The suite of JUICE-UVS observational techniques (Section 3.7) will constrain atmospheric density and vertical structure for each constituent species in Europa's atmosphere. The combined global and detailed perspective that the UVS remote-sensing method provides will be highly complementary to other atmospheric measurements acquired by the JUICE mission and will provide bounding conditions for computational models aiming to explain the various time-dependent variations of Europa's atmosphere.

### 2.1.6. Callisto

Callisto is thought to possess the densest atmosphere of the three icy Galilean moons (Carberry Mogan et al. 2020, 2021), but it is also the most well shielded from the Jovian plasma, which is relatively tenuous and cold at the location of Callisto's orbit. Remote detection of the atmosphere is therefore more challenging, since Callisto lacks the brighter electron-impact driven auroral emissions seen on Europa and Ganymede. However, faint oxygen emissions at 130.4 nm and 135.6 nm have been detected by HST (Cunningham et al. 2015), with disk-averaged brightnesses ranging from <1 R to 15-25 R. The only instrument sensitive enough to detect these UV emissions from Earth orbit is HST's Cosmic Origins Spectrograph (COS), which does not have imaging capabilities at FUV wavelengths. As a result, little is known about the morphology and variability of these airglow features, and the spatial and temporal coverage obtained by UVS during JUICE's 23 Callisto flybys is expected to provide a significantly improved understanding of the distribution, composition, and sources and sinks of oxygen-bearing species in the atmosphere. UVS will also search for UV emissions of atomic carbon, since $CO_2$ is known to be an important atmospheric species (Carlson 1999a, Cartwright et al. 2024).

As solar and stellar occultations provide a key means for determining atmospheric composition and structure (see Sections 3.7.5 and 3.7.6), Fig. 5 shows example circumstances and the SNR expected for a solar occultation by Callisto. The results show that the major species, e.g., $O_2$, $H_2O$, $CO_2$ (Liang et al. 2005, Carberry Mogan et al. 2022) will be easily detected, with useful upper limits placed on $H_2$, CO, and $SO_2$, if not new detections. UVS measurements of its structure and diurnal properties may confirm that Callisto's atmosphere is collisional, rather than being a surface-bounded exosphere, and helping to constrain the relatively high-conductivity of its ionosphere that efficiently diverts much of the incoming plasma flow from the magnetosphere and affects magnetic induction signals (Carberry Mogan et al. 2020, 2021).

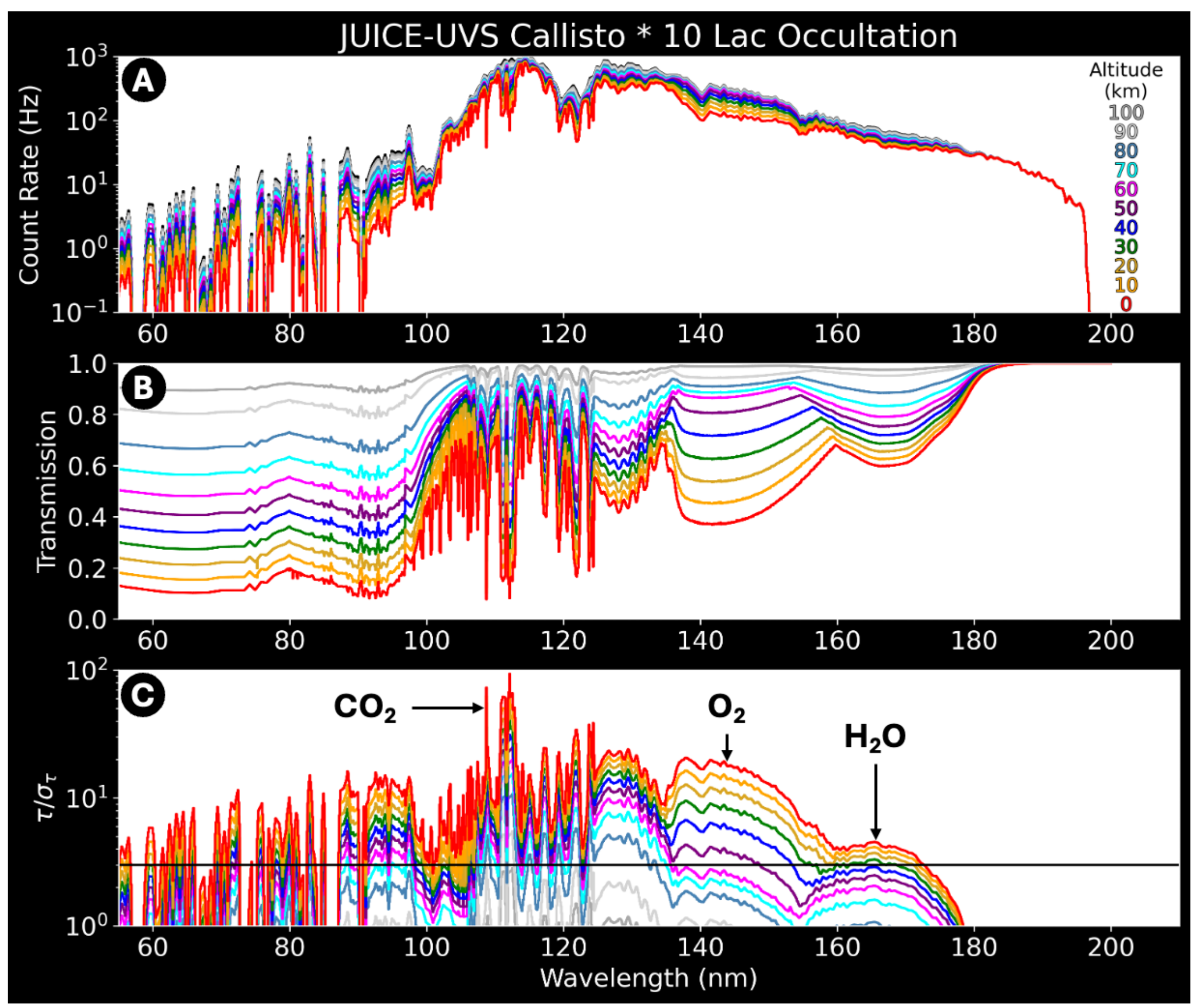

**Fig. 5** Simulated UVS A) stellar occultation spectra for example UV-bright star 10 Lacertae occulted by Callisto, B) gas transmission spectra, and C) SNR per resolution element ($\tau/\sigma_\tau$) for several Callisto tangent altitudes assuming the compositional abundances in Carberry Mogan et al. (2021). The star used in this example has a JUICE-UVS signal rate of ~26,000 counts/s and we use a 1-second time resolution at each altitude level. As indicated, the primary absorber is $O_2$, with contributions from $CO_2$ at λ<120 nm. Occultations allow very high vertical resolution of atmospheric species' densities, impossible to achieve from current and planned facilities at Earth

As at Ganymede and Europa, UV surface observations are also important for understanding the volatile inventory that contributes key species to Callisto's atmosphere. $CO_2$ and $SO_2$ ices, for example, have previously been observed on Callisto's surface (e.g., Hibbits et al. 2000, Cartwright et al. 2024) and both have spectral features within the JUICE-UVS bandpass. Comparisons between the UV reflectance of $H_2O$-rich regions on Callisto and those on Europa and Ganymede will also allow an assessment of the importance of radiolytic surface processing to the UV-darkening of the satellite surfaces, since Callisto's surface is the oldest yet located in the most benign radiation environment.

## 2.2. Jupiter

FUV spectroscopy of Jupiter's atmosphere is an ideal technique to study the interface between the external charged particle environment and the churning atmospheric weather layer. Jupiter's upper troposphere and extensive stratosphere is a stable region whose composition is shaped by rich photochemical pathways (a 'carbon cycle' initiated by photodissociation of $CH_4$), meridional circulation, and hazes of uncertain chemical makeup (e.g., Moses et al. 2005, Hue et al. 2024). Above the methane, the thermosphere is a heterogeneous region of $H_2$, He, and H, where solar EUV is the dominant ionizing source at low latitudes, while auroral electron precipitation increasingly dominates towards higher latitudes. A variety of well-defined spectral absorption features in the FUV provide access to the stratospheric hydrocarbons, e.g.: absorption bands in reflected sunlight of $CH_4$, $C_2H_2$, $C_2H_6$ and higher-order products at $\lambda<180$ nm (e.g., Gladstone & Yung, 1983, Melin et al. 2020, Giles et al. 2021, 2023)(Fig. 6); vertical temperature structure via $H_2$ and H opacity (Gladstone et al. 2004); UV-absorbent hazes and dust (via their scattering cross-sections and phase functions); and tropospheric $NH_3$ (>160 nm) and $PH_3$ (160-180 nm) in their photochemical depletion regions (e.g., Edgington et al. 1998). The blue-absorbing nature of localized hazes, such as those above prominent anticyclones like the Great Red Spot, may provide the key diagnostics of the chromophores responsible for Jupiter's vivid colors.

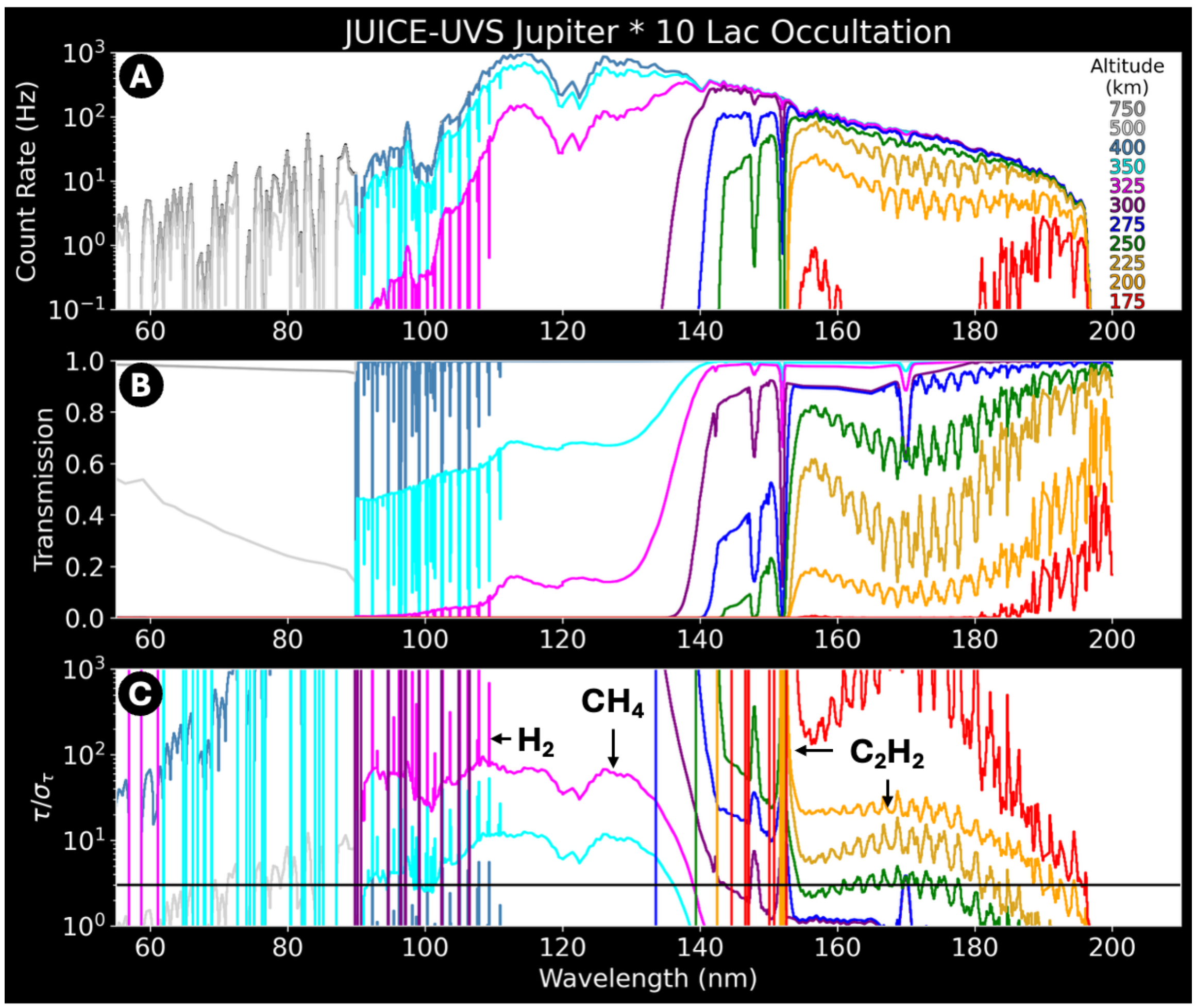

**Fig. 6** Simulated UVS A) stellar occultation spectra for example UV-bright star 10 Lacertae occulted by Jupiter, B) gas transmission spectra, and C) SNR per resolution element ($\tau/\sigma_\tau$) for several Jupiter tangent altitudes. The star used in this example has a JUICE-UVS signal rate of ~26,000 counts/s and we use a 1-second time resolution at each altitude level. At large tangent altitudes $H_2$ band absorption occurs at $\lambda<111$ nm, and continuum $CH_4$ absorption at $\lambda<140$ nm. At medium tangent altitudes, absorption by higher hydrocarbons are detectable at $\lambda>140$ nm (e.g., $C_2H_2$, $C_2H_4$)

The UVS instrument on NASA's Juno spacecraft has been exploring the atmosphere and auroras of Jupiter since 2016 (Gladstone et al. 2017). A major advantage of JUICE over the spinning Juno spacecraft, however, is that JUICE allows UVS to point and stare, enabling the study of much fainter auroral and atmospheric features than Juno-UVS, as well as enabling stellar/solar occultations (see Sections 3.7.5 and 3.7.6). Thus, the focus for JUICE-UVS auroral science is on targets such as the diffuse aurora and secondary-auroral arcs (e.g., Radioti et al. 2011, Gray et al. 2017, Grodent 2015), injection signatures (Dumont et al. 2018) and dawn storms (Bonfond et al. 2021), as well as synoptic observations to correlate with other variables (e.g., solar wind strength, satellite activity). UVS will observe Jupiter's aurora from a considerable distance (typically 10-20 $R_J$), but always with sufficient spatial resolution to provide synoptic data on the northern and southern emitted power for each of the three main aurora groups (i.e., main emissions, polar region, and equatorward emissions including diffuse and satellite footprint auroras together with injection signatures (Head et al. 2026), for comparison with in situ particles and fields data (Fig. 7). UVS includes a new mode (Section 3.2), customized to provide HST-quality imaging of Lyα aurora and airglow at higher spatial resolution but requiring longer exposures at lower cadence.

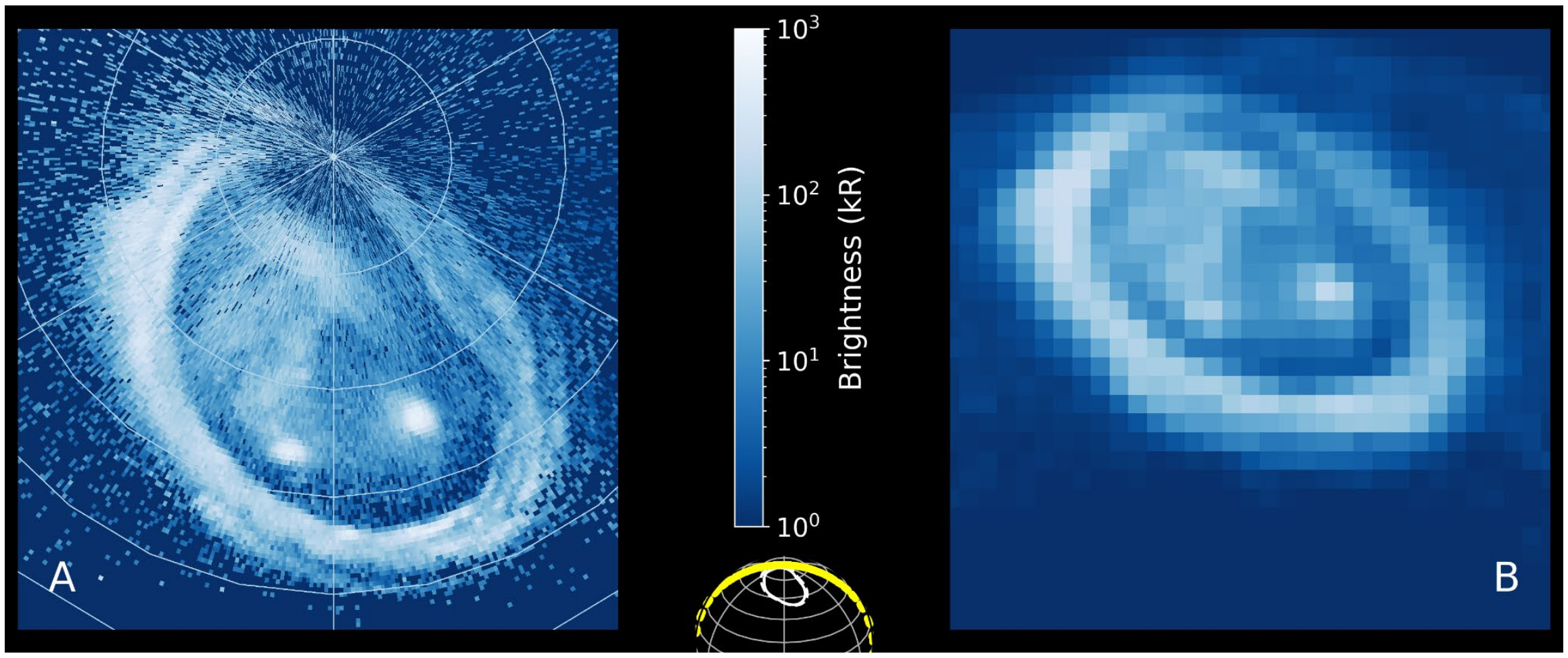

**Fig. 7** Comparison of high-spatial resolution Juno-UVS images (A) with a simulated JUICE-UVS image of Jupiter's aurora (B) to help set expectations of lower-resolution imagery. A UVS scan across Jupiter's auroral region from a distance of 20 $R_J$ would produce results comparable to this re-sampled Airglow Port (AP) mode image; at this range, UVS images with its new High-spatial-resolution Port (HP) mode have comparable spatial resolution to HST-STIS images

### 2.2.1 Dynamics of the upper troposphere and stratosphere

Jupiter's upper troposphere and stratosphere interface between the 'weather layer' (circulations on the scale of Jupiter's prominent bands and zonal jets) and the radiatively-controlled middle atmosphere (dominated by waves and large-scale atmospheric circulations, e.g., Fletcher et al. 2020). JUICE-UVS will map the temperature and composition of Jupiter's middle atmosphere via a combination of dayside FUV observations (to determine the spatial distribution of minor species and aerosols through their effect on scattered sunlight), and solar and stellar occultations (to determine vertical density, temperature, and abundance profiles (e.g., Greathouse et al. 2010) at a range of latitudes and local times to assess variability. By deriving temperature profiles from stellar occultations at many latitudes, and thus deriving horizontal temperature gradients, JUICE-UVS will help to reveal the vertical windshear. This windshear knowledge will connect the zonal and meridional motions observed by tracking clouds in the troposphere to the motions shaping the stratosphere (supporting similar observations with Gravity and Geophysics of Jupiter and the Galilean Moons (3GM, Iess et al. 2026, this collection) radio occultations and Sub-millimeter Wave Instrument (SWI; Hartogh et al. 2026, this collection) sub-millimetre sensing, e.g., Fletcher et al. 2023).

The general circulation of the middle atmosphere is studied using long-lived hydrocarbons as tracers (e.g., Melin et al. 2020, Giles et al. 2021), determining the strength of vertical mixing from place to place and the global variation of the $CH_4$ homopause (mapped over the dayside using He 58.4 nm and Lyα resonance emissions). Meteorology in the upper troposphere can be addressed by mapping the spatial variation of condensible volatiles like $NH_3$-- (and potentially disequilibrium species like $PH_3$) in the photochemical depletion region of the upper troposphere and lower stratosphere, particularly in association with belt/zone contrasts, polar vortices and discrete phenomena (e.g., the Great Red Spot). Both of these regions exhibit extreme temporal variability in the form of horizontally-propagating planetary (e.g., Rossby) waves (e.g., Orton et al. 1991, Fletcher et al. 2017), and vertically-propagating patterns like the quasi-quadrennial oscillation (also known as Jupiter's Equatorial Stratospheric Oscillation, Antunano et al. 2021). Horizontal waves can cause temperature and aerosol contrasts, such that their origins, lifetimes, and propagation can all be diagnosed via long-term UVS monitoring. Vertically-propagating waves will cause oscillations in the high-resolution stellar and solar occultations, helping to reveal where waves break and deposit their energy in the stratosphere.

Finally, Jupiter's ever-changing meteorology is tied to the availability of water that drives moist convection in the weather layer, evidenced by frequent lightning activity (e.g., Brown et al. 2018). Detecting and mapping potential transient luminous events in the UV (Giles et al. 2020) could enable investigations of the origins and distribution of thunderstorm activity in Jupiter's deeper troposphere.

Taken together, these observations will allow UVS to compare the redistribution of energy and material (chemicals, aerosols) from Jupiter's tropics (from powerful tropospheric upwelling and the quasi-quadrennial oscillation of the stratosphere), to the mid-latitudes (dominated by tropospheric belt/zone circulation and stratospheric waves), to the cold polar vortices (radiative cooling from unique aerosols, and energy injection via auroras), providing a complete view of this archetypal giant planet's dynamics.

### 2.2.2. Chemistry, aerosols, and airglow in the middle/upper atmosphere

Wide phase-angle coverage is planned to map UV-absorbing hazes, determine the optical properties of stratospheric aerosols (e.g., composition, size, number density, phase function and scattering properties), provide contrast of polar and low-latitude hazes, and investigate the haze asymmetry between the northern and southern poles (and its implications for upper atmospheric transport). Solar and stellar occultations reveal the vertical distributions of several key species such as $H_2$, $CH_4$, $C_2H_2$, $C_2H_4$, $C_2H_6$, $NH_3$, etc. Chemistry of higher-order hydrocarbons is initiated by methane photolysis by solar EUV at lower latitudes (Gladstone et al. 1996, Moses et al. 2005), with a likely secondary production source triggered by precipitating electrons at high latitudes, as suggested by numerous observing campaigns (Sinclair et al. 2017, 2018, 2019, 2023; Giles et al. 2023; Rodriguez-Ovalle et al. 2023, 2024). Coupling with horizontal transport processes controls the meridional distributions of the longer-lived species (Nixon et al. 2007, Zhang et al. 2013, Fletcher et al. 2016, Hue et al. 2018, Giles et al. 2021). UVS occultations will be crucial in determining the chemical cycles, and their interaction with aerosols and atmospheric circulation, across the planet (see the review of Hue et al. 2024).

Jupiter's upper atmosphere can also be investigated using FUV airglow emissions. The FUV dayglow is comprised of resonantly scattered solar emissions of Lyα (~10 kR at 121.6 nm) and HeI (~5 R at 58.4 nm), $H_2$ bands (~2 kR at $80<\lambda<160$ nm; excited by photoelectrons and fluoresced sunlight, in roughly comparable amounts), and Rayleigh-scattered sunlight ($\lambda>160$ nm, around P~1-10 mbar), which have the absorption signatures of the major species. Jupiter's nightglow has not been well studied; there might be $H_2$ band emissions, but it is likely scattered Lyα from the IPM.

UVS observations of Jupiter's airglow will be important for resolving several outstanding issues. In addition to the energetics of the dayglow mentioned above, the existing nightglow observations from Voyager, Galileo, Cassini, and New Horizons are also contradictory, indicating either substantial particle-excited $H_2$ and H emissions on the nightside, or none at all (Gladstone et al. 2007). Likewise, the Lyα bulge discovered by Voyager was observed with IUE to broaden considerably in longitude (McGrath et al. 1990), but has been observed with HST only sporadically.

### 2.2.3 Jupiter's Aurora

Jupiter's auroras are extremely bright, occasionally reaching several tens of megaRayleighs (i.e., ~1000x brighter than Earth's auroras) and a total power of several TeraWatts in the FUV. Typically, the UV Jovian auroras are comprised of Lyα (~15%), $H_2$ Lyman bands (~40% at 80 nm < λ < 170 nm), $H_2$ Werner bands (~40% at $80<\lambda<130$ nm), and $H_2$ Rydberg bands (~5% at 80 nm < λ < 90 nm). The auroral emissions fall into three categories: 1) main (fairly stable on hourly timescales); 2) polar (highly variable on minute timescales and extremely bright); and 3) equatorward auroras that include diffuse emissions, injection signatures, and satellite footprints connected to each of the Galilean satellites (quite variable, depending strongly on local plasma conditions near each satellite). Jupiter's auroras have been studied extensively (e.g., Bhardwaj & Gladstone 2000, Clarke et al. 2004).

FUV imaging with HST has provided the vast sum of our understanding of auroral morphology (Nichols et al. 2009, Clarke et al. 2009, Grodent et al. 2018). In the past decade, the UVS spectrograph on Juno has provided views over the Jovian auroras, otherwise inaccessible from an Earth-based perspective. For instance, Juno-UVS revealed local-time dependence of the polar auroral emission (Greathouse et al. 2021), witnessed the formation and evolution of auroral dawn storms (Bonfond et al. 2021), and allowed characterizing new types of auroral features (Hue et al 2021a, Haewsantati et al. 2021, Head et al. 2025), among others. While fewer studies have made use of FUV spectral data, the spectral signatures of hydrocarbon absorption, as observed by Juno-UVS, provide valuable information on the penetration depth and the mean energy of the precipitating electrons, ionospheric conductances, and hydrocarbon distributions (e.g., Yung et al. 1984, Gustin et al. 2006, Benmahi et al. 2024a, 2024b, Hue et al. 2024). This information can then be used to infer the conductance of the ionosphere, which is a key feature needed to understand the coupling between the atmosphere and the magnetosphere (Gérard et al. 2021, Sicorello et al. 2025). JUICE-UVS will build on the previous successes of Juno-UVS, a spinning spacecraft, by offering the ability to point and stare at the auroral region for long enough durations to greatly improve measurement signal quality. This will dramatically expand our understanding of the Jovian auroras as it targets and observes specific auroral features as Jupiter rotates.

Another powerful means to investigate the Jovian magnetosphere is simultaneously observing both the magnetospheric plasma with in situ instruments and their auroral emission counterparts imaged at the other end of the magnetic field lines (e.g. Mauk et al. 2002, Guo et al. 2021, Yao et al. 2022, Nichols et al. 2023, Giles et al 2025). The majority of these studies have used complementary observations from HST, but more recently, Juno-UVS has obtained synoptic observations of Jupiter's aurora during the apojove period of the spacecraft's elliptical orbit, which have been directly compared to in situ observations from the same spacecraft (Giles et al. 2025). The orbital distance of JUICE means that the mission will likewise be able to perform distant in situ observations and remote sensing imaging of the aurora from the same platform; these contemporaneous observations are key to understanding the mechanisms that drive auroral activity.

## 2.3 Io System

### 2.3.1 Io's atmosphere

Io, a volcanic moon, has a diversity of both large and small erupting plumes that continuously supply $SO_2$ and other volatiles to Io's surface. Sublimation of $SO_2$ frost on Io's dayside surface is the primary supplier of its atmosphere, while volcanoes emit less gas at any given moment; a growing consensus points to a largely sublimated dayside atmosphere with volcanos directly providing 15-50% (Tsang et al. 2015, de Pater et al. 2020, Roth et al. 2025a). However, Io's nightside atmospheric densities remain unconstrained by observations. The question of the extent to which its atmosphere partially collapses during eclipse or by inference at night requires confirmation, and several contradictions and exceptions in absolute densities and latitudinal gradients remain problematic even on the dayside (de Pater et al. 2021). Gas escape is primarily caused by the interaction with the surrounding plasma, and most of the material outgassed at volcanoes does not directly escape (e.g., Tsang et al. 2016). Despite numerous observations of various types in the decades following Voyager's detection of volcanic plumes, progress in understanding the overall Io system and the connections between Io's volcanoes, surface volatiles, atmosphere, magnetospheric plasma-interaction, and beyond to Io's extended neutral clouds, the Io Plasma Torus, and Jupiter's ionosphere all remain difficult to quantify and understand. Uncertainties in Io's atmospheric asymmetries, composition and behavior entering night-time or eclipse remain key to the remaining puzzles.

Retherford et al. (2007) report New Horizons Jupiter encounter Alice observations of Io's far-UV emissions from atmospheric oxygen and sulfur atoms, which varied with time before and after eclipse. Despite being obtained at large distances these data constrained plasma-atmosphere model predictions for the auroral brightnesses and found similar constraints to the partial collapse of Io's atmosphere in shadow as the current consensus (based on more precise and direct techniques), and provide a good example for JUICE-UVS observational monitoring of the time variability of these auroral emissions for the purpose of better understanding Io's atmosphere and plasma interaction. JUICE will be close enough to Io on a few occasions to resolve hemispherical scale auroral features during disk-scan measurements, such as its equatorial spots (Retherford et al. 2000) and/or polar limb glow (Retherford et al. 2003), depending on the scan direction, and more extended emissions from Io's escaping atmosphere out to several tens of Io radii, e.g., as in Wolven et al. (2001).

JUICE-UVS plans several stellar occultations of Io to measure vertical profiles of species such as $SO_2$, $O_2$ and possibly $H_2S$, the latter two of which are expected from photochemical modeling but have yet to be detected (Denk et al. 2026). JUICE's unique vantage point, compared with Earth-based observations, enables stellar occultation profiles of Io's nightside, where volcanic gas components are expected to be relatively more abundant than the previously sublimated species that remain uncondensed above Io's cold nightside surface. While several stellar occultations of Io were planned with New Horizons Alice for these same motivations, safing events caused by radiation environment noise signal spikes prohibited any data collection at the planned event times. Io stellar appulses, i.e., a sort of "near-miss" occultation event, will also be valuable for constraining the vertical extent to which $SO_2$ remains the dominant species. Io transit observations are also useful for investigating the asymmetries in its equatorial versus polar gas distributions (Retherford et al. 2019). How much $SO_2$ remains intact compared to its dissociative products (e.g., SO, O and S) as these gases escape radially and enter the neutral cloud remains an important question for understanding variability in Io plasma torus mass loading rates.

### 2.3.2 Io's neutral clouds

The interaction between Jupiter's magnetic field and Io results in a large mass loss from Io's atmosphere. The canonical value for this mass loss rate is ~1000 kg/s, although it varies with time (e.g., Delamere & Bagenal 2003, Delamere et al. 2004, Steffl et al. 2008) ranging 700-2400 kg/s (Thomas et al. 2004, Dols et al. 2008, Smyth and Marconi 2003). Most of this material leaves Io as neutrals, and only ~20% is thought to be ionized near Io (Bagenal 1997). Neutrals escape Io's atmosphere at speeds of a few km/s. They are produced primarily by electron-impact dissociation of $SO_2$ and dissociative recombination of molecular ions, with smaller contributions from charge exchange and atmospheric sputtering. They form large clouds along Io's orbit, extending both in front of and behind Io (Roth et al. 2025a, Bagenal & Dols 2020). These neutral clouds are made of O and S atoms formed by dissociations of the primary molecules, with traces of neutral atomic Na,

K, Cl, etc. and are the main plasma source for the torus. The distance from Io where the majority of photodissociations and electron impact dissociations occur remains largely unconstrained by observations, since the neutral $SO_2$ molecule remains undetected beyond Io's near-surface collisional atmosphere.

Models of Io plasma torus dynamics and energetics (Nerney et al. 2017, Nerney et al 2025) unfortunately rely strongly on the poorly constrained neutral input parameter. Through electron impact ionization and charge exchange reactions (roughly equally important overall, with charge exchange the dominant process for O and electron impact the dominant process for S; Nerney et al. 2020), the neutral clouds become ionized. Through $\mathbf{E} = -\mathbf{v} \times \mathbf{B}$ forces, an electric field accelerates the fresh ions to the local plasma flow velocity of ~58 km/s and energizes S and O ions by the pickup process to form the Io Plasma Torus. Coulomb collisions between ions and electrons redistribute this energy between species, such that typical torus ion temperatures are ~100 eV and thermal electron temperatures are ~5 eV.

Relatedly, the interaction of magnetospheric plasma with Europa's atmosphere also produces a neutral torus. This torus is composed primarily of $H_2$ molecules and O atoms. Europa's O torus is very dim, with peak brightnesses predicted to be of order 0.5 R. The spatial distribution is strongly peaked near Europa. JUICE-UVS observations with long duration stares at Europa are needed to constrain the radial distribution of resonantly scattered oxygen 130.4 nm emissions to investigate the source of the Europa torus.

### 2.3.2 Io Plasma Torus

The power supplied to the torus from pickup ions is not enough to sustain the ~$10^{12}$ W of emitted FUV radiation while also maintaining an electron temperature of ~5 eV and a plasma ionization state of ~1.5 (e.g., Shemansky 1988, Smith et al. 1988). This shortfall, termed the "energy crisis," implies there is a significant additional source of energy for the torus. One possible source is from hot electrons. Both the Voyager and Galileo plasma instruments detected a high-energy tail in the electron energy distribution function (e.g., Scudder et al. 1981, Sittler and Strobel 1987), while analysis of Ulysses plasma wave observations showed the electron distribution function in the torus is not Maxwellian but more of a κ-distribution (Meyer-Vernet et al. 1995). Such a high-energy tail equilibrates very rapidly (tens of minutes) with the thermal electron population, and thus must be continually re-energized. This hot electron source must provide between 20-60% of the total energy input to the torus, but is still poorly understood.

Mass is lost from the Io torus in two ways: escaping fast neutrals and outward radial transport. Energy is lost from the Io plasma torus primarily through EUV and FUV emissions observable within the JUICE-UVS bandpass. For example, the 68.0 nm emission from $S^{++}$ shown in Fig. 8 is the most efficient at emitting energy. Roughly 1.5 TW of power was emitted by the torus, as observed during the Cassini flyby (Steffl et al. 2004). This accounts for 50-90% of the energy budget, with the remainder split nearly equally between escaping fast neutrals and outward transport (Delamere & Bagenal 2003, Bagenal & Delamere 2011).

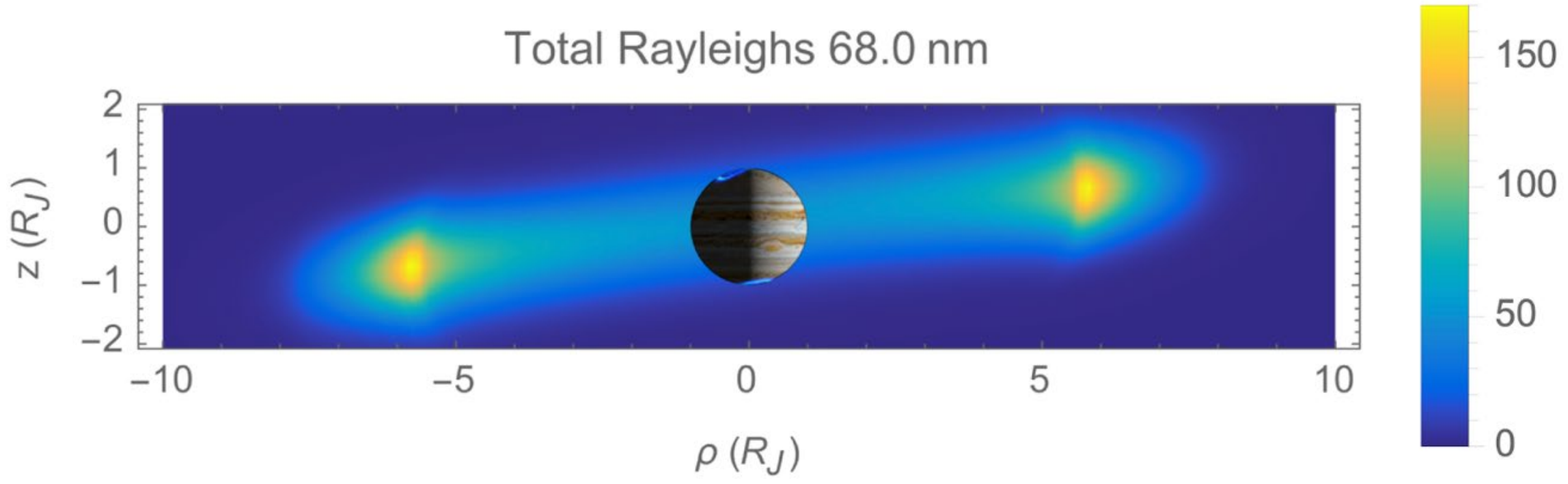

**Fig. 8** Nominal Io plasma torus $S^{++}$ 68.0 nm emission in Rayleighs viewed from an arbitrary system III longitude. The image of Jupiter's aurora is from Juno UVS superimposed on an image of Jupiter (Bonfond 2020/ULiège/SwRI/NASA)

The UV spectrum of the Io torus consists of a few dozen spectral features at 64 nm < λ < 173 nm. At UVS resolution each feature will comprise several lines, often from different ion species. Modeling of these features yields the densities and ionization states of the O & S ions in the torus plasma (Fig. 8, and cf Figs. 18-19). Plasma flowing down Jupiter's magnetotail produces a dusk-to-dawn electric field. This field adiabatically shifts the Io torus a fraction of an $R_J$ toward the dawn side, compressing the dusk ansa plasma relative to that of the dawn ansa, so that the dusk ansa appears ~25% brighter.

Juno-UVS observations away from perijove altitudes capture parts of the torus at wavelengths >70 nm but not the key 68.0 nm feature and have very limited quality owing to the short dwell time for each spacecraft spin. Stares and much slower rate scans of the Io plasma torus planned for JUICE-UVS, in additional to copious serendipitous opportunities for lines of sight intercepting various key areas of the torus from a variety of distances, will provide an incomparably valuable dataset for addressing these key questions regarding the energy balance of the torus and will provide global context for magnetospheric variability as also investigated by JUICE's fields and particles instruments.

## 2.4. Cross-cutting science objectives

### 2.4.1. Related Measurement Requirements for JUICE

NASA manages its contributions to the JUICE payload through a Planetary Missions Program Plan, with an appendix for JUICE-UVS called the Program Level Requirements Agreement (PLRA) document. The requirements are consistent with the science objectives of the JUICE mission as described in the Red Book (issue 1, 2014) and SciRD (JUI-EST-SGS-RS-001, issue 2, rev 5, 2014). The UVS team listed baseline requirements, ordered Group 1 as most important (threshold) and Group 2 important, with additional desirable Group 3 science objectives. The focus of the Group 1 measurement requirements relates to Ganymede and Europa spectral mapping of the surface; Ganymede, Callisto and Europa limb scans; Europa transits; Ganymede, Callisto, Europa, Io and Jupiter stellar occultations; and Ganymede and Europa aurora and airglow atmospheric mapping. Group 2 baseline measurement requirements relate to Callisto spectral mapping; Ganymede and Callisto transits; Jupiter aurora and airglow; and Io plasma torus and Europa tori spectral imaging. Additional Group 3 objectives list desired Io and Jupiter spectral mapping (hemispherical coverage); Io transits; Callisto and Io aurora and airglow; Jupiter spectral mapping and limb scans; Ganymede neutral cloud (search); inner small moons and rings; and assorted measurements jointly with JUICE instruments and also Europa Clipper coordination. In this section we elaborate further on the latter items.

### 2.4.2. Additional topics: rings, shape models, small moons, etc.

In addition to performing detailed studies of the icy moons, JUICE also aims to characterize Jupiter's rings and small moons (e.g., Amalthea, Thebe, Kalichore; Denk et al. 2026). These targets are small and dark at UV wavelengths, so UVS will not drive the planned remote sensing observations, but will ride-along with the other instruments and attempt to at least place upper limits on the UV reflectance. If any suitable stellar occultations of the small moons can be identified, UVS will use these to refine shape models and ephemerides. Given the low optical depth of Jupiter's main ring, low-incidence stellar occultation geometries would be needed to permit the characterization of the band of parent bodies at the outer edge of Jupiter's main ring.

### 2.4.3. Relations to Working Groups (WGs)

Working Group 1 (WG1) has the responsibility for the geophysical characterization of moon interiors (Van Hoolst et al. 2024). While UVS cannot directly observe the subsurface or interiors of the moons, key properties can be inferred from remote UVS observations. In particular, temporal variations in the location of auroral emissions can be used to constrain the depth and salinity of conducting sub-surface oceans, as demonstrated by Saur et al. (2015) using HST images of Ganymede. JUICE-UVS spectra of any recent plume deposits observed on the moon surfaces may also be used to constrain the composition of the subsurface liquid layer.

Stellar occultations are planned to constrain the shape of all three icy moons using astrometry, which as demonstrated for Europa-UVS by Abrahams et al. (2021) helps improve knowledge in coordination with other measurement types for WG1, e.g., as planned for Radar for Icy Moon Exploration (RIME, Bruzzone et al. 2026), Ganymede Laser Altimeter (GALA, Hussman et al. 2025), and 3GM (Iess et al. 2026). The extent to which these UVS astrometric stellar occultations can be used to help improve the ephemerides of the Jupiter system, including Jupiter's tidal-Q parameter, is in discussion with JUICE's Planetary Radio Interferometer & Doppler Experiment (PRIDE, Gurvits et al. 2023) team.

Working Group 2 (WG2) covers topics related to the geology, surface composition, and exospheres of Jupiter's moons (Tosi et al. 2024). JUICE-UVS will comprehensively address WG2 science requirements using a range of techniques, including global and regional spectral imaging of both the moon surfaces and exospheres, limb scans and stares, and stellar and solar occultation measurements, as described in the previous sections. Flyby segments within ±12 hours of closest approach provide the highest resolution datasets, but regular distant monitoring observations will also be performed and are particularly important for characterizing any ongoing plume activity on Europa and volcanism on Io (Denk et al. 2026).

For WG2 icy moon studies UVS will observe a few specific lines of interest: HI Lyman-α, OI 130.4 nm, OI 135.6 nm. At icy moons, scrutinising Lyman-α emission will support characterisation of the hydrogen corona and indirectly $H_2$ (species with the most extended exosphere), as hydrogen is mainly produced through $H_2$ dissociation (Naesenius 2025). In contrast, probing OI 130.4 nm and OI 135.6 nm will support the characterisation of the $H_2O$ and $O_2$ exosphere components. The ratio between both emissions will help us to infer the ratio between their neutral number densities, with prior knowledge of the electron energy distribution relying on in situ PEP Jovian Electron and Ion analyzer (JEI) observations, as the emission cross sections are different. These characterizations will be performed mainly over the auroral regions where the emissions are the strongest. Note that exospheric densities will be obtained from stellar and solar occultations (see Sections 3.7.5 and 3.7.6).

The Magnetosphere and Plasma Working Group (WG3) is in charge of coordinating science activities in connection with Jupiter's magnetosphere and the plasma environment that surrounds the icy moons (Masters et al. 2025). While JUICE has a series of plasma instruments (e.g. PEP, RPWI, J-MAG) to probe in situ plasma properties (e.g. ion number density, ion speed, ion and electron energy distribution), UVS will perform remote-sensing observations of the emissions and auroras from the moons. These emissions stem from the interaction between the energetic charged particles from the Jovian magnetosphere (mainly electrons, responsible of dissociation and excitation of neutral molecules) and the different constituents of the exosphere (e.g. $O_2$) embedding Ganymede, Europa, and Callisto. However, from the emission solely, it is impossible to derive the local electron energy distribution and neutral number densities independently. Coupled with in situ measurements (neutral composition from PEP Neutral and Ion Mass Spectrometer (NIM) and electron energy distribution from PEP/JEI), UVS will link observed emissions with those modeled, driven by the former. Similar multi-instrument comparisons are planned to investigate the auroral footprints of the moons for flux tubes connecting to Jupiter's ionosphere through complex plasma processes, as discussed.

The Jupiter Working Group (WG4) is designed to coordinate multi-instrument campaigns for Jupiter's atmosphere and auroras within the broader context of the JUICE Jupiter science goals (Fletcher et al. 2023). While UVS provides diagnostics of circulation, chemistry, auroral emissions, airglow, and upper-atmospheric energy deposition, its measurements are complementary to observations made by other JUICE instruments, working in synergy to constrain atmospheric structure, dynamics, and composition across multiple wavelengths.

Thermal and compositional information recorded by SWI (Hartogh et al. 2026, *this collection*), the Moons and Jupiter Imaging Spectrometer (MAJIS; Poulet et al. 2024), and the radio-sounding experiment 3GM (Iess et al. 2026, *this collection*) are crucial for linking UVS measurements to the underlying atmospheric state. Infrared and sub-millimeter observations constrain temperatures, clouds, and trace species in the upper troposphere and stratosphere and therefore provide boundary (or overlapping) conditions for the interpretation of UVS observations. 3GM measurements through Earth radio occultations allow reconstruction of the atmospheric density profile from ~1 mbar. Vertical temperature profiles from UVS stellar/solar occultations, complemented by 3GM radio occultations and SWI spectral measurements, will reveal the role of winds, waves, and vertical transport in shaping the atmospheric structure and composition.

Visible and near-infrared observations, such as those provided by the JANUS instrument (Palumbo et al. 2024), further complement UVS measurements for WG4 by characterizing haze and cloud layers, and by characterising discrete dynamical features in the weather layer. These combined measurements enable correlations between variability observed at ultraviolet wavelengths in the upper aerosol distribution and meteorological features at deeper levels, such as convective activity, wave patterns, and compositional or temporal changes in the aerosol distributions.

Combined sub-millimeter observations provide a link between Jupiter's temperatures, winds, and trace chemical species from millibar to microbar pressure levels, complementing UVS measurements of hydrocarbons and precipitating particles through mapping of auroral emissions. This synergy is particularly important in the polar regions, where ion–neutral chemistry is likely to play a role in shaping the observed chemical distributions (e.g., Hue et al. 2024, and references therein).

Finally, we note that the Jupiter WG4 is also responsible for coordination of supporting remote sensing observations from other facilities, including ground-based observations and space telescopes (like Hubble and JWST). Complementary observations, particularly mid-infrared studies of stratospheric temperatures and composition (e.g., Fletcher et al. 2016), and high-resolution auroral observations from HST and JWST (e.g., Nichols et al. 2025) will set the close-in observations of JUICE-UVS in their wider temporal, spatial, and spectral context.

### 2.4.4. Joint JUICE- Europa Clipper working group

It is fortunate that the JUICE mission science phase overlaps with NASA's Europa Clipper mission (Pappalardo et al. 2024), a mission to study Europa in great detail. The concurrence of both JUICE and Europa Clipper with nearly identical UVS instrument capabilities on these two missions (Retherford et al. 2024) is expected to provide a unique opportunity to understand the Jovian system at a much deeper level than what is obtainable through the individual missions. Recently, the Europa Clipper team has started discussions of Jupiter System Science (JSS) activities, planning observations of targets in addition to Europa as available, e.g., prioritizing other Jupiter system targets during orbits without Europa flybys. Numerous opportunities for synergistic measurements are present.

In addition to providing the JUICE mission with a powerful remote-sensing tool for atmospheric, aurora, plume and surface studies, JUICE-UVS can be used to address a multitude of other cross-cutting science objectives important to understanding the Jovian System in its entirety. Some of these topics include constraining the global shape of the icy satellites using stellar occultations (Abrahams et al. 2021), and high-energy (>7 MeV) particle flux monitoring (as on Juno, Kammer et al. 2018; Meitzler et al. 2023).

Understanding the interior of the icy satellites is one of JUICE's fundamental goals, as it is for Europa Clipper at Europa. Although JUICE-UVS and Europa-UVS are primarily intended to study satellite atmospheres and surfaces, they have the ability to function as a high-precision altimeter (Abrahams et al. 2021). Measuring moon global shape precisely with 1-ms resolution astrometric stellar occultations allows JUICE to test predicted long-wavelength variations in ice shell thickness (Ojakangas and Stevenson 1989) and better understand interiors. In particular, Europa's global shape can reveal Europa's elastic thickness (Araki et al. 2009, Nimmo et al. 2011), interior rheology (Fu et al. 2017), and surface tectonics (Ermakov et al. 2019). JUICE-UVS's contributions to improving constraints on the global shape of icy moons will help us understand their global tectonics and their link to its interior.

To interpret UV emission from Ganymede's atmosphere and emissions from Europa and Callisto as well, it is important to understand its plasma environment and plasma interaction. The UV aurora emission is primarily generated by electron impact excitation of the $O_2$ atmospheres, and the semiforbidden OI 135.6 nm emissions are particularly useful for plasma-related studies since its crosssection for resonant scattering is low, unlike for OI 130.4 nm. Therefore, observing and modeling the UV emissions from ions in the plasma environment in conjunction with the UV aurora observations will help us understand these tenuous atmospheres. The morphology of and asymmetries in the UV emission from the atmosphere are strongly controlled by the plasma. For example, the north-south asymmetries and the total UV brightness at different magnetospheric latitudes for Europa reported in Roth et al. (2016) are limited by the amount of electron energy being able to get into the atmosphere (Saur et al. 1998). Constraining the plasma interaction is very important, as the plasma magnetic field can be of the same order as the induction magnetic fields that the PIMS investigation likewise constrains (Kivelson et al. 2004, Schilling et al. 2007) for Europa Clipper and PEP and RPWI constrain for JUICE (Barabash et al. 2026, this collection, Wahlund et al. 2024). Remote observations by UVS during encounters by the alternate spacecraft, when their in situ measurements are characterizing the plasma environment, add value even at large distances where UVS only provides disk-averaged trends in brightness variability.

Regardless of which spacecraft arrives first at the Jupiter system, the second spacecraft can serve en route as a solar wind monitoring system to investigate connections with Jupiter's magnetospheric phenomenon. There is much speculation about how Jupiter's magnetosphere reacts to solar wind variations, but little to no data taken simultaneously from inside and outside Jupiter's magnetosphere exist that could provide decisive conclusions (most correlations are captured using models of solar wind propagation from Earth out to Jupiter and those are highly uncertain). Understanding this missing feature of plasma sheet and magnetodisk variability is needed to improve models of the plasma interaction that creates the auroral emissions at all four moons targeted by both UVSs. JUICE-UVS observations of the Io plasma torus during the Jupiter approach phase benefit from Europa Clipper's position within the Jupiter magnetosphere.

The UV reflectance from a surface is controlled by material optical constants, space weathering due to solar UV irradiation and the high-energy electron and ion bombardment, and microphysical properties such as porosity. Thus, measurement of the variations of the surface albedo can inform studies on the chemical composition and age, important for numerous JUICE (see Section 2.1.3) and Europa Clipper objectives. These attributes to be investigated with both UVSs, coupled with detailed surface mapping performed by the Jovis, Amorum ac Natorum Undique Scrutator (JANUS) (Palumbo et al. 2025) and MAJIS (Poulet et al. 2024) instruments and especially when coordinated with the Europa Imaging System (EIS; Turtle et al. 2024) and Mapping Imaging Spectrometer for Europa (MISE; Blaney et al. 2024) instruments, can lead to a more complete understanding of the formation and evolution of geologic structures on Europa's surface (Tosi et al. 2024, Dauber et al. 2024).

## 3. Instrument Description

### 3.1. UVS Overview

JUICE-UVS is preceded by four similar UV spectrograph investigations: 1) Rosetta-Alice visited comet 67P/C-G (Stern et al. 1998, Slater et al. 2001); 2) Pluto-Alice on New Horizons explored the Jupiter system, Pluto system, Arrokoth, and continues with Interplanetary Medium (IPM) studies (Stern et al. 2005, 2008; Slater et al. 2005); 3) Lunar Reconnaissance Orbiter (LRO) Lyman Alpha Mapping Project (LAMP), launched in 2009, studies lunar volatile, exosphere, and mineral composition with nearly continuous data taking (Gladstone et al. 2010); and 4) Juno-UVS at Jupiter (Gladstone et al. 2017) observed Ganymede, Europa, and Io's UV aurora and airglow in 2021-2024 during its extended mission (Fig. 9). Europa-UVS is the sixth in the series of investigations (Retherford et al. 2024, Davis et al. 2022), which launched in October 2024 and will arrive at Jupiter shortly before JUICE.

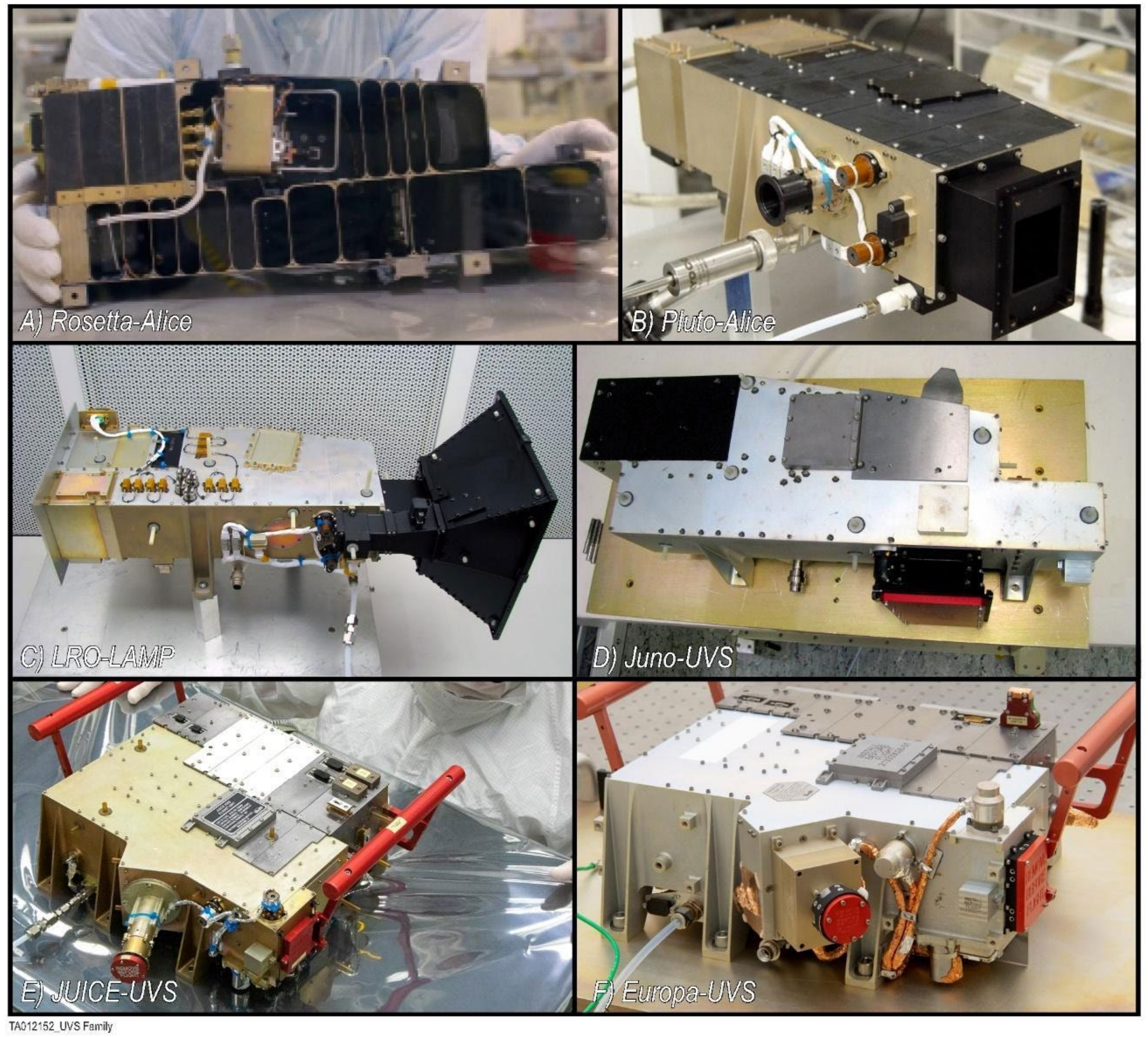

**Fig. 9** "Family photo" of JUICE-UVS with its five other closely related instrument predecessors and successor UVS/Alice type UV spectrographs. A) Rosetta-Alice is shown with the top cover removed; B) New Horizons Pluto-Alice instrument prior to shipment to JHU/APL; C) LRO LAMP instrument prior to shipment to NASA/GSFC; D) Juno UVS instrument prior to shipment to Lockheed Martin; E) JUICE-UVS (Fig. 1); and F) Europa-UVS prior to shipment to JPL. JUICE-UVS and Europa-UVS have the most commonality, being developed together (Europa-UVS being typically ~1.5 years behind) for similar Jupiter radiation environments and with 2 years (within ~2 weeks) between delivery dates

The UVS instrument is a single assembly comprising: i) a telescope section, ii) a spectrograph & detector section, and iii) an electronics section. Figure 10 shows these sections from left to right in a view of its design with the cover off.

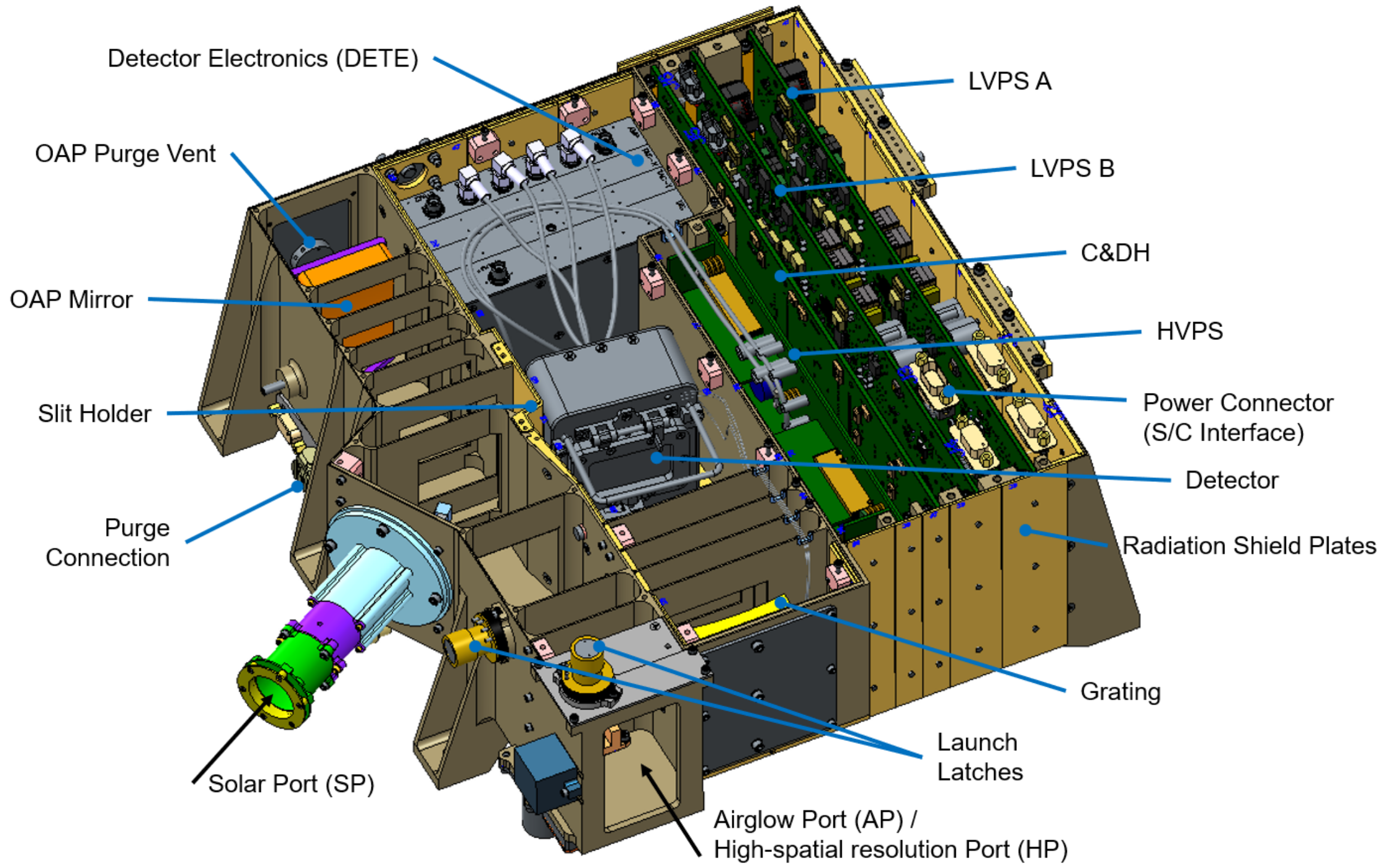

**Fig. 10** JUICE-UVS optomechanical schematic showing major subsystems, mechanisms, and key components. Fig. 13 shows a similar top view of how rays enter the telescope from bottom through the AP, HP or SP apertures (black arrows). A mix of internal and external radiation shielding plates are used, with those on the instrument electronics plus spectrograph sections shown here in gold color. Ultra-pure filtered dry $N_2$ gas enters the instrument through the purge connection to maintain instrument cleanliness through the time of launch

The UVS telescope feeds a 15-cm Rowland circle spectrograph with a spectral bandpass of 50-204 nm. The telescope has an input aperture 4 × 4 cm$^2$ and uses an off-axis parabolic (OAP) primary mirror with 120-mm focal length. The light is then focused onto the spectrograph entrance slit (Fig. 11), which has two contiguous segments with field of view (FOV) shapes of 0.1° × 7.3° and 0.2° × 0.2° projected onto the sky. The purpose of the wider square box at the end of the slit is to accommodate the Sun's 0.11°-diameter viewed near 5.2 AU during solar occultations.

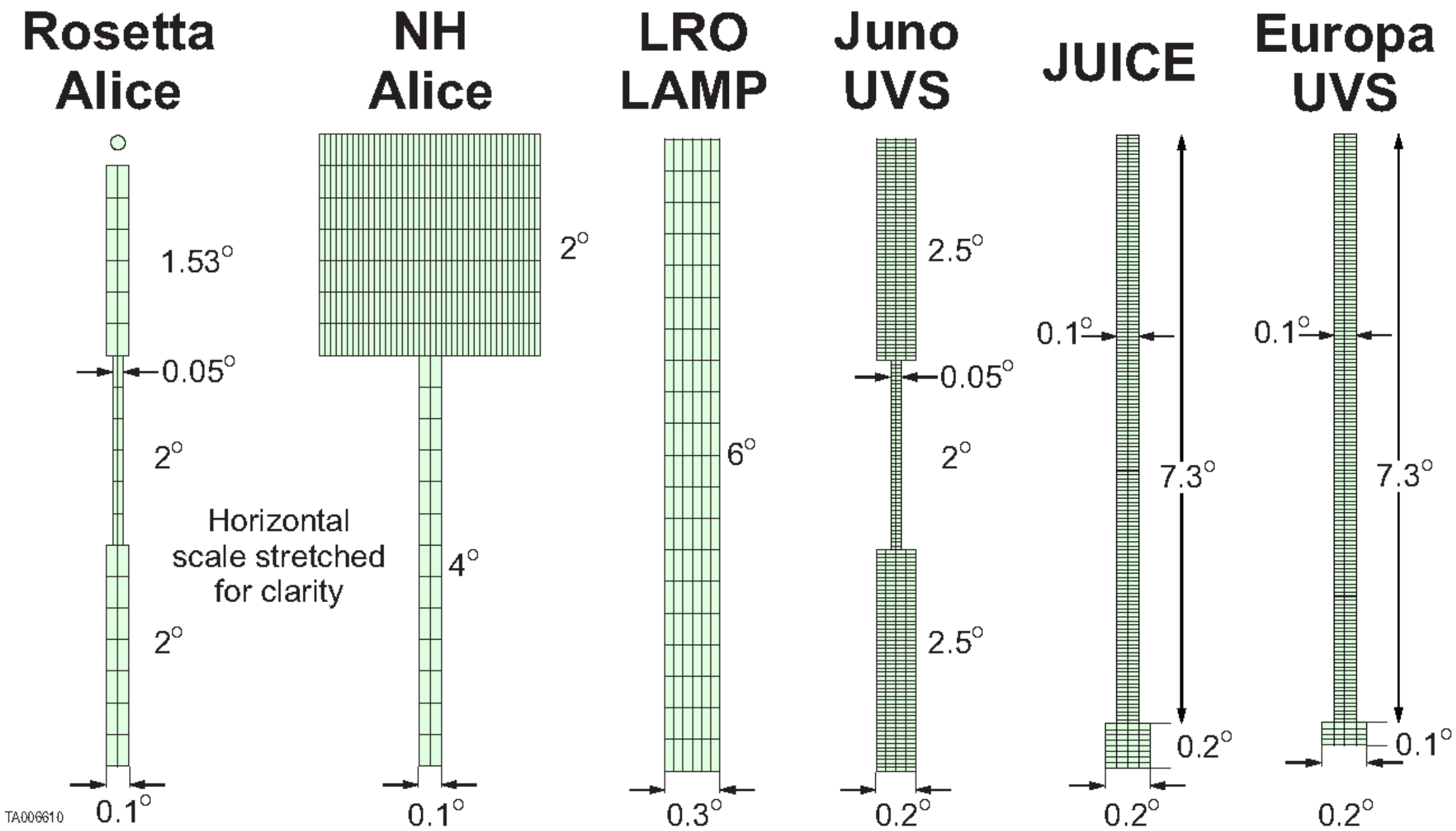

**Fig. 11** Comparisons of JUICE-UVS's slit field of view with other heritage unobstructed FOVs. For Europa-UVS it was determined post-launch that its bottom ~0.1° portion of the intended 0.2° × 0.2° box region is imaged just below the active area of the detector collecting surface, effectively reducing its slit length at least for certain wavelengths and still being calibrated at this time

Light entering the slit is dispersed by a toroidal diffraction grating that focuses the UV bandpass onto a microchannel plate (MCP) detector with a solar-blind, UV-sensitive cesium iodide (CsI) photocathode read out by cross-delay line (XDL) electronics. Tantalum/Tungsten (TaW) plates contiguously surround the detector and electronics assemblies, shielding the detector and sensitive parts from general particle radiation and high-energy electrons. The detector electronics (DETE) are located behind the detector (Fig. 10).

In a chamber beside the spectrograph are located the Low Voltage Power Supply (LVPS), High Voltage Power Supply (HVPS), and Command and Data Handling (C&DH) electronics boards, connected through a common backplane. The C&DH controls the heater/actuator activation electronics, and event-processing electronics of the DETE.

**Heritage & changes from previous UVS/Alice builds.** JUICE-UVS builds heavily upon the heritage of the previous SwRI UV spectrographs, especially Juno-UVS, while incorporating a few modifications for the JUICE mission that improve spatial resolution, maximum instantaneous count rates, and radiation background rejection when compared to previous spectrographs (Stern et al. 2007, 2008, Gladstone et al. 2010, 2017) and similar radiation tolerance as for Europa-UVS (Retherford et al. 2024). First, the detector has been upgraded to feature Atomic Layer Deposition (ALD) coated lead glass microchannel plates, making JUICE-UVS resilient to gain degradation by fluence induced charge depletion effects. The detector anode mask is expanded to widen the spectral bandpass with more illuminated physical area on the MCP. This change allows greater spectral coverage of atomic helium lines at 58.4 nm while obtaining solar spectral signals at the longest wavelengths possible with the CsI photocathode.

JUICE-UVS's grating was contributed by France's Centre National d'Études Spatiales (CNES) and led by coinvestigators at the Laboratoire Atmosphères et Observations Spatiales (LATMOS). The grating has an epoxy layer on top of an aluminum baseplate in which a holographically ruled pattern is impressed into it using a metallic master grating having its inverse toroidal pattern. Optical coatings are deposited onto this ruled epoxy surface for far-UV optimization. A new master grating was remade by vendor Horiba/Jobin-Yvon using improved facilities relative to the previous four heritage spectrographs, improving peak to trough shape and slightly better surface roughness properties for better overall performance. This grating improvement benefitted Europa-UVS as well.

A High-spatial-resolution Port (HP) is designed to improve the spatial resolution of the base UVS design by a factor of up to 4 at a cost of up to 16 times worse throughput. In practice, there are variations in optical aberrations that do not always scale with aperture size, so the HP spatial resolution varies as a function of wavelength with improvements being on average ~x2 improved spatial resolution and ~x10 decreased throughput (at the boresight).

The Solar Port (SP) is derived from the Solar Occultation Channel ("SOCC") that flew on New Horizons's Pluto-Alice. The solar port was modified from Pluto-Alice so that it is pointed 60° from the main boresight (rather than 90°) for improved accommodation on the JUICE spacecraft (i.e., avoiding the solar arrays), as highlighted in Fig. 12. The pinhole has also been resized from 1-mm to 0.25 mm to reduce solar flux. Lastly, the pickoff mirror is coated with gold to further reduce incoming solar flux to a level within the detector's dynamic range.

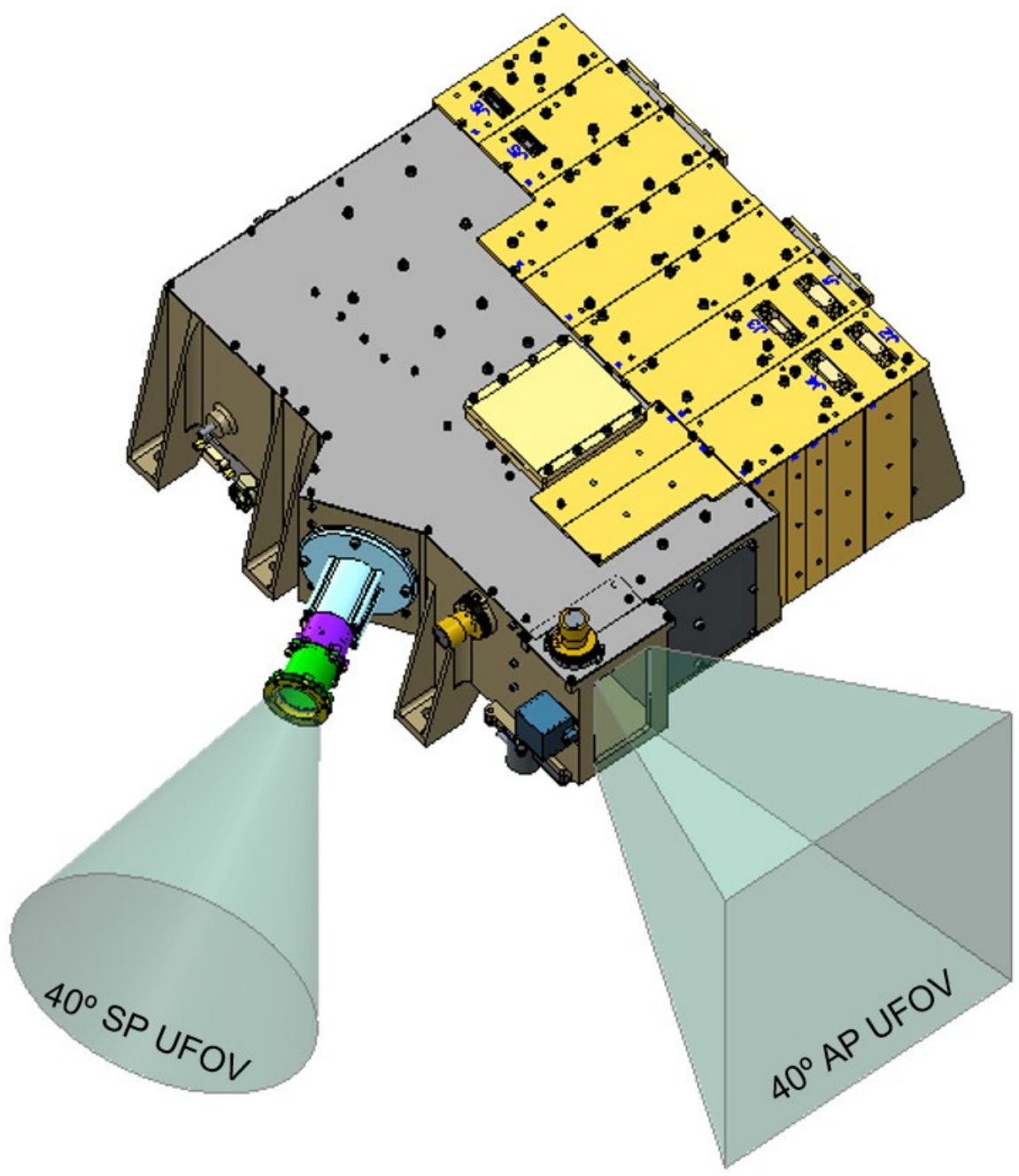

**Fig. 12** JUICE-UVS fields of regard (also described as unobstructed fields of view, UFOV). The SP is pointed 60° from the main boresight (AP/HP) to avoid the solar arrays from causing glints or obscuration

Table 1 lists the salient specifications for all six UVS/Alice family instruments as flown (e.g., Retherford et al. 2009, Davis et al. 2011, Hue et al. 2021b, Retherford et al. 2024). Note that the telescope and spectrograph sections have almost identical optical specifications in terms of focal length 120 mm, gratings 1600 grooves/mm and 55×55 $mm^2$ substrate, and excluding differences in optical coatings. The OAP mirror substrate of ~44×66 $mm^2$ for the first four builds was updated to ~44×72 $mm^2$ for JUICE-UVS and Europa-UVS to allow for the field-of-view increase from 6° to 7.5°.

**Table 1** Characteristics & Spacecraft Resources of Heritage Instruments as Flown

| **Characteristic** | **Rosetta/ Alice** | **New Horizons/ Alice** | **LRO-LAMP** | **Juno-UVS** | **JUICE-UVS** | **Europa-UVS** |
|---|---|---|---|---|---|---|
| **Science Target(s)** | Comet C-G /67P | Pluto, Kuiper-belt | The Moon | Jupiter Poles, Satellites | **Jupiter system & Ganymede** | Europa |
| **External Dimensions** | 41 × 20 × 12 $cm^3$ | 46 × 16 × 12 $cm^3$ | 55 × 16 × 12 $cm^3$ | 44 × 24 × 22 $cm^3$ + 13 × 6 × 9 $cm^3$ (Ebox) | **35.4 cm x 41.1 cm x 15.8 cm** | 36.2 cm × 35.3 cm × 15.8 cm |
| **Mass** | 3.0 kg | 4.4 kg | 6.1 kg | 21.6 kg | **19.40 kg** | 18.8 kg |
| **Power** | 4.0 W | 4.4 W | 4.5 W | 9.8 W | **7.5 W** | 7.9 W |
| **Principal Investigator (PI) or Inst. Lead; successor leads with start year** | Stern; 2016: Parker | Stern | Stern; 2007: Gladstone; 2011: Stern; 2012: Retherford | Gladstone; 2023: Greathouse | **Gladstone; 2025: Retherford** | Retherford |
| **Bandpass** | 70–205 nm | 52–187 nm | 57–196 nm | 68–210 nm | **50–204 nm** | 55–206 nm |
| **Input Channels** | Airglow | Airglow & Solar Occultation Channel | Airglow & Pinhole; Fail-safe Door Open after Oct. 2016 | Airglow | **Airglow Port, High-spatial-resolution Port & Solar Port** | Airglow Port, High-spatial-resolution Port & Solar Port |
| **Mirror Substrate** | 44×66 $mm^2$ | 44×66 $mm^2$ | 44×66 $mm^2$ | 44×66 $mm^2$ | **44×72 $mm^2$** | 44×72 $mm^2$ |
| **Optical Coating** | SiC | SiC + Au pick-off | Al/$MgF_2$ | Al/$MgF_2$ | **Al/$MgF_2$ + Au pick-off** | Al/$MgF_2$ + Au/$MgF_2$ pick-off |

| **Entrance Slit FOV shape (l×w)** | 2.0°×0.1° + 2.0°×0.05° + 2.0°×0.1° | 4.0°×0.1° + 2.0°×2.0° | 6.0°×0.3° | 2.0°×0.2° + 2.0°×0.05° + 2.0°×0.2° | **7.3°×0.1° + 0.2°×0.2°** | 7.3°×0.1° + 0.2°×0.2° |
|---|---|---|---|---|---|---|
| **Detector Photocathodes** | KBr (70-120 nm) CsI (123-205 nm) | KBr (52-118.1 nm) CsI (125.1-187 nm) | CsI | CsI | **CsI** | CsI |
| **MCP Readout Electronics** | Double delay-line | Double delay-line | Double delay-line | Cross delay-line | **Cross delay-line** | Cross delay-line |
| **MCP Pulse Height Distribution Bins** | 16 | 64 | 64 | 32 (17 per event) | **256** | 256 |
| **Detector Dead-Time Metrics** | 15.7 µsec, 4.075 kHz at 10% DQE loss | 6.172 kHz at 10% DQE loss | 17.3 µsec (6.172 kHz at 10% DQE loss) | 1.2 µsec | **825 ns** | 825 ns |
| **Detector Global Background Rate (total array in flight)** | 18 cnts sec$^{-1}$ | 98.5 cnts sec$^{-1}$ (includes RTG noise) | 27 cnts sec$^{-1}$ | 50 cnts sec$^{-1}$ | **<31 cnts sec$^{-1}$** | <16 cnts sec$^{-1}$ |
| **HVPS** | 1 (non-redundant) | 2 (redundant) | 2 (redundant) | 2 (redundant) | **2 (redundant)** | 1 (non-redundant) |
| **Acquisition Modes** | Histogram and Pixel List | Histogram and Pixel List | Histogram and (mainly) Pixel List | Pixel List only | **Programmable Histogram and Pixel List** | Programmable Histogram and Pixel List |

| **Unique Component** | None | Solar Occultation Port, RTG Radiation, Jupiter Radiation and Outer Solar System Environments | Terminator Sensor, Large Baffle, Aperture Door Pinhole | Scan Mirror w/ Baffle, Vaulted Electronics, Jupiter Radiation Environ. | **Solar Occultation Port, Jupiter Radiation Environments, Lead Glass MCP with ALD coating** | Solar Occultation Port, Jupiter Radiation Environments, Borosilicate Glass MCP with ALD coating |
|---|---|---|---|---|---|---|

JUICE-UVS was designed to achieve the measured performance capabilities in Table 2. The final pre-launch ground-calibrated performance (Section 4) all met or exceeded the requirements needed to realize its science objectives (Davis et al. 2020). Results from commissioning and early payload checkout activities are reported in Davis et al. (2025). Additional calibration details were provided by measurements obtained during JUICE's Lunar-Earth Gravity Assist (LEGA) maneuver (Molyneux et al. 2026), which offered UVS's first filled-slit measurements at wavelengths other than for Lyα and He I 58.4 nm skyglow (see Section 4.2.2.2 and Fig. 34).

**Table 2** JUICE-UVS Measured Performance Summary

| **Attribute** | **Measured Performance** |
|---|---|
| **Bandpass:** | 50 – 204 nm |
| **Spectral Resolution:** | ~0.55-0.8 nm (Point source at slit center; wavelength dependent) |
| | ~1.5 nm (Point source near edge of slit) |
| | ~3.1 nm AP, ~1.0 nm HP (Filled slit average FWHM at slit center, extended source) |
| **Spatial Resolution:** | 0.07°-0.15° (AP, near 130 nm at center of slit) |
| | 0.29° (AP, near edge of slit and <60 nm) |
| **Pixel Plate Scale** | 0.0092º ± 0.0001º per pix (spatial); ~0.106 nm/pix (average spectral) |
| **Peak Effective Area:** | 0.7 $cm^2$ (115 nm) |

### 3.2. Optomechanical Design

The JUICE-UVS optomechanical design is the latest evolution of the heritage UVS/Alice design (Davis et al. 2020) as detailed in Table 3 with respect to spectral imaging performance, which is still being calibrated and assessed at the time of this writing for both JUICE-UVS and Europa-UVS. The OAP and grating are coated with aluminum and a magnesium fluoride overcoat ($Al/MgF_2$) to maximize reflection at FUV wavelengths >115 nm while still providing measurable throughput at EUV wavelengths 50–115 nm. The toroidal holographic grating has 1600 grooves/mm and a radius-of-curvature of 150 mm in the dispersion plane. The grating blaze wavelength is 90 nm, optimizing diffraction between 50-200 nm, with a corresponding radius-of-curvature of 143.3 mm for astigmatic focus at 90 nm. The light dispersed by the grating is focused onto a MCP detector with a XDL anode. The 2D MCP detector is coated with a solar-blind CsI photocathode with responsivity to EUV/FUV photons up to 204 nm. The detector and its accompanying electronics are inside local contiguously-arranged tantalum shields to reduce the background noise on the detector while also mitigating the total MeV electron dose on the electronics.

**Table 3** In-Flight Performance of JUICE-UVS Spectral Imaging Compared with Other UVS/Alice's

| Attribute | Rosetta/ALICE | New Horizons/ ALICE | LRO/LAMP | Juno/UVS | JUICE-UVS | Europa-UVS |
|---|---|---|---|---|---|---|
| **Spectral Pixel Plate Scale** | 0.1739±0.0017 nm per spectral pixel | 0.1832±0.0003 nm per spectral pixel | 0.182±0.001 nm per spectral pixel | 0.085 nm to 0.105 nm per spectral pixel | **~0.106 nm per spectral pixel** | ~0.075 nm per spectral pixel |
| **Spectral Resolution Element Filled Slit** | 0.8-1.2 nm FWHM | 0.90±0.14 nm FWHM | 3.64±0.18 nm FWHM (2.7 nm | Top & Bottom: 2-3 nm FWHM | **AP: ~1.4 nm HP:** | ~1.4 nm |

|  |  |  | boxcar width) | Middle: 1.3 nm |  |  |
|---|---|---|---|---|---|---|
| **Spectral Resolution Element Point Source (at slit center)** | 0.4-0.9 nm FWHM | 0.30-0.35 nm FWHM (SOCC), 0.30-0.45 nm FWHM (airglow) | 0.39-0.57 nm FWHM | 0.4-0.6 nm FWHM | **~0.55-0.8 nm** | 0.6-0.8 nm |
| **Spatial Plate Scale** | 0.019° per spectral pixel | 0.4-1.5 spatial pixels | 0.30±0.18° FWHM | 0.041° ± 0.001° | **0.0092º ± 0.0001º** | 0.0075º ± 0.0002º |
| **Spatial Resolution Element** | 0.3° per spatial pixel | 0.27±0.01° per spatial pixel | 0.29±0.03° per spatial pixel | Top & Bottom: 0.17°; Middle: 0.07° | **AP: 0.07°-0.25° HP:0.06°-0.1°** | AP: 0.1°-0.3° HP: 0.12°-0.16° |
| **Stray Light Rejection at $\theta_{off}$** | $<10^{-4}$ at >4° off boresight | $<8.3 \times 10^{-7}$ at >7° off boresight (spatial axis) | $<8 \times 10^{-6}$ at >7° off boresight | $<7 \times 10^{-6}$ at >7° off boresight | **$<5 \times 10^{-6}$ at >7° off boresight** | $<7 \times 10^{-6}$ at >7° off boresight (spatial axis) |

Light enters UVS via one of three apertures, depending on the desired observation type: a 40×40-mm square Airglow Port (AP, the main aperture described previously), a 10×10-mm square High-spatial-resolution Port (HP), or a 0.25-mm diameter pinhole in the Solar Port (SP). The HP is located in an additional door in front of the AP along the optical path that changes the depth of focus for the instrument to sharpen the spatial resolution while reducing throughput (Davis et al. 2019). The SP is canted ~60º from the main telescope boresight to avoid other instrument FOVs when pointing at the Sun, and to avoid interference with the spacecraft solar arrays. The SP pickoff mirror is coated with gold, which is ~35% reflective in the UV, reducing the solar flux to better match the detector dynamic range. The red-tag handles and covers shown in Fig. 1 (i.e., circular for SP and square for AP) are removed before flight. The JUICE-UVS spectrograph section consists of an aberration-corrected toroidal grating that focuses spectrally dispersed light onto the cylindrically curved MCP detector (Fig. 10).

UVS was assembled in an ISO7 clean room and kept on dry-GN2 purge pre-launch, as UV optics are especially susceptible to darkening by contaminants such as hydrocarbons deposited on the surfaces and subsequently polymerized. The mounts for the UVS optical elements (OAP, grating, and SP pickoff mirror) include heater elements to desorb contaminants from their surfaces in flight. Devaud et al. (2018) reported a detailed study informing the duration and cadence of optical heater decontamination operations for Europa-UVS, which is equally relevant for JUICE-UVS given the commonality of designs.

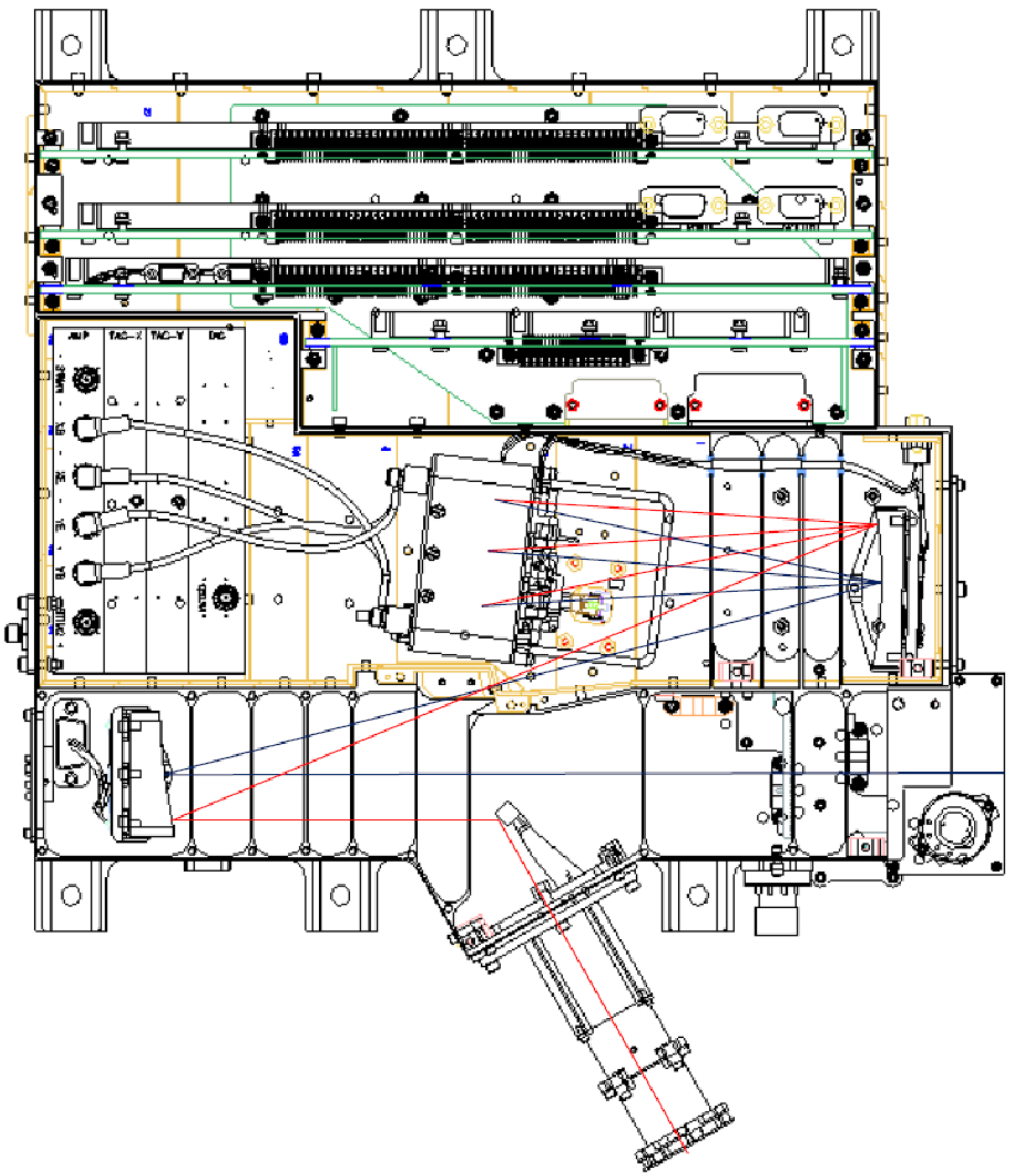

**Fig 13.** JUICE-UVS Rowland circle spectrograph top-view optomechanical schematic with ray trace for AP (blue) and SP (red) pathways. See Fig. 10 for labeled subcomponents, mechanisms, and apertures. Light enters the AP from the right, is reflected by the OAP mirror (at left), focused at the slit (where the rays cross near the middle; see Fig. 11 with plate aligned into the page), and reimaged on the grating at middle right. The grating disperses the image of the slit onto the detector (three red and three blue rays at middle) for each wavelength of light in first-order. Light entering the HP follows a similar path, but with its smaller aperture reducing the throughput of rays. Rays of light entering the 0.25 mm SP pinhole aperture (red) from bottom right, is reflected by a small pick-off mirror (small square), and follows the same optical path from the OAP, through the slit, to the grating, and then imaged on the detector. The boresight is offset (into the page in this view)

to align the SP optical path with the wider 0.2° × 0.2° box part of the physical slit ("bottom" of slit in Fig. 10), but otherwise follows the same path through the telescope and spectrograph sections

The UVS housing is made from a single billet of aluminum which provides a reliable and consistent thermal conduction across the instrument. The instrument can be sensitive to thermal gradients that would drive optical component movement, so keeping a uniform temperature throughout the environmental conditions is important for the instrument's operation. The tantalum shielding provides significant thermal mass helping prevent quick swings in temperature. The non-operating thermal limits are -35°C to +55°C, close to the instrument maximum allowable temperature of +60°C (Table 4). The operational limits are set to -15°C to +40°C.

**Table 4** Thermal operating and non-operating limits with 5 K margins.

| **Limit Type** | **Non-Operating (Survival)** | | | **Operating** | | |
|---|---|---|---|---|---|---|
| | **Qualification** | **Acceptance** | **Design** | **Qualification** | **Acceptance** | **Design/Performance** |
| **Temperature Range** | -35°C, +55°C | -30°C, +50°C | -25°C, +45°C | -15°C, +40°C | -10°C, +35°C | -5°C, +30°C |

The telescope section contains several baffles designed to minimize direct viewing of the housing walls by the OAP, thereby greatly reducing stray and scattered light effects. However, JUICE-UVS incidentally has one baffle that is ~0.050 in misplaced on one side, reducing throughput between the slit and the grating by a few percent.

A combination of rad-hard parts selection and high-Z metal shielding from energetic particles are a key element of the JUICE-UVS design to ensure that the minimum mission lifetime requirement and science requirements are met. Needed analyses included: i) analysis of the mission environment to provide clarity on the radiation dose, the Single Event Effects, and the radiation background flux requirements; ii) calculation of a rough order estimate of the expected background count rate levels for science, including what the background rate tolerance would be for the MCP detector (next section); iii) definition of the radiation shielding requirements and determination of any alterations needed to meet the radiation environment survivability and operation requirements; and iv) assessments of dielectric discharge levels, and any other top-level parts analyses necessary to find cost savings opportunities.

This analysis compared the effectiveness of the nominal shielding of the Juno-UVS and Europa-UVS shielding configurations. The housings for both JUICE-UVS and Europa-UVS are 50 mil Al enclosures. Additional TaW (2.5% W) plates ranging from 50 to 310 mil (nominally 100 mil) add additional shielding around the housing. Lastly, 0.2 mm (7.87 mil) MLI blankets cover the instrument, providing stopping power equivalent to another 6.6 mil Al. The vast majority of TID comes from electrons rather than ions for this level of shielding, and both particles create gamma rays in the stopping process. Hence additional shielding mass has diminishing returns in terms of blocking the gamma ray photon TID and detector noise sources in a relative sense.

Carapelle et al. (2019) summarize a test of a heritage UVS MCP detector using a new low electron flux facility in the 0–3.5 MeV range developed in Liège, Belgium for the study of induced signal in both UVS and MAJIS JUICE instruments in Jupiter's radiation environment. The results provided information ensuring the planned shielding strategies would reduce science measurement noise backgrounds as needed.

Planned radiation dark measurements for both JUICE-UVS and Europa-UVS (section 3.7.9) enables the verification of this overall radiation model simulation and analysis approach. A TID sensing dosimeter within the Europa-UVS instrument further helps verify the shielding approach for the Europa Clipper mission, but is not included in JUICE-UVS.

The electronics systems were assessed with a sector analysis of TID, targeting a <100 kRad level within the electronics section of the instrument overall and in some sectors achieving much lower TID values. The rad-hard parts selection process accounted for these variations in internal radiation dose. An analysis is ongoing for a certain subset of MOSFET parts that were found to be more susceptible to radiation than designed, given subsequently known manufacturing quality issues; routine heating of the LVPS and HVPS electronics boards to anneal these parts shows promise as a mitigation to this issue.

## 3.3. Detector and Detector Electronics

The JUICE-UVS detector configuration (Figs. 4 & 7) consists of three MCPs with ~12-µm pores arranged in a Z-stack. The CsI photocathode is on the top surface of the Z-stack. When a UV photon hits the CsI photocathode, the resulting photoelectron is accelerated down a microchannel, where collisions with the walls produce an avalanche of electrons that emerges from the other end of the Z-stack. The amplified charge cloud leaves the back end of the MCP and is accelerated across the MCP-anode gap, landing on the anode and initiating pulses that propagate in both the +X and –X directions and +Y and –Y directions along integral delay lines to the detector electronics. The detector electronics then output the X and Y pixel locations to the C&DH based on the time delay between the two opposing pulses in each axis, defining an event.

The MCP material is lead glass with atomic-layer-deposited (ALD) magnesium oxide (MgO) on the bottom plate (Siegmund et al. 2020). The ALD coating results in comparatively reduced changes in gain due to charge extraction relative to uncoated lead-glass plates, greatly improving operational lifetime (McPhate et al. 2019). Each of the three nested MCPs within the Z-stack has a cylindrical 7.5-cm radius of curvature, designed to match the 150-mm Rowland circle diameter for spectral focus and best spatial focus across the JUICE-UVS bandpass. The MCP format is 4.6-cm wide in the spectral axis by 3.0-cm height in the spatial axis with a channel length-to-diameter (L/D) ratio of 80:1 per plate. The XDL anode size of 4.4 cm × 3.0 cm, with an active array area of 3.85 cm × 1.9 cm, covers the entire 50-204-nm instrument bandpass in the long dimension (1442 spectral pixels) and 7.5º in the short dimension (~784 spatial pixels). The approximate resistance per MCP is ~50 MΩ. A 20 lines per inch repeller grid is situated above the MCP photocathode to enhance quantum efficiency by turning escaping photoelectrons back toward the MCP pores.

The XDL MCP is housed in a vacuum enclosure with a door containing a UV-grade, fused-silica window for long-wavelength FUV throughput testing pre-flight (and for low fluorescence in-flight). The detector door is spring-loaded for a one-time opening (once in flight) via a wax-pellet actuator. The vacuum enclosure includes a vacuum pump port and a small, polished region that reflects zero-order light from the grating into a light trap on the instrument housing. The vacuum enclosure also has four female connectors for the anode signals, and high-voltage (HV) connectors for the MCP and anode gap voltages.

The DETE, located next to the detector housing, is composed of five boards: 1) the amplifier board, which comprises two fast amps for the X direction (spectral dimension), two fast amps for the Y direction (spatial dimension), and two charge amps for total event charge; 2 & 3) time amplitude converter (TAC) boards for the X and Y axes, which encode 4096 pixels in the X-axis and 4096 pixels in the Y-axis by event arrival time differences; 4) the digital board (DIG), that provides control signals and interface logic, and 5) a delay-line board to delay the End signals. The x and y detector addresses of each event are sent to the C&DH electronics for further processing into either pixel list data or histogram data at programmable spectral and spatial resolution.

The detector electronics also produce 8-bit analog sum signals for each event that are used for generating pulse-height distributions (PHDs). The MCP pulse-height information is output as 8 bits, which, together with the 12 bits of spectral and 12 bits of spatial information, results in a 4-byte output for each event.

The PHDs of detected UV photons differ from noise counts created by both electrons and gamma rays. The JUICE-UVS DETE features two improvements to the pulse height recording to enhance rejection of background radiation. First, the resolution of recorded pulse heights increased from 32 bins on Juno-UVS to 256. This improvement allows finer rejection of background events with large pulse height, potentially reducing background noise rates by 30%. Second, an additional analog-to-digital converter was added to report pulse heights accurately even at high count rates.

A pulse generator within the detector electronics provides two stimulation (STIM) pixels at two locations in the array. These STIM pixels are useful for both checking data throughput and acquisition modes without the HVPS activated to power the detector, and for correcting the wavelength scale for temperature effects.

The detector electronics require input DC voltages of ±6.12V and + 3.3V. The detector MCP high voltage is raised to a room temperature operational voltage of about –4.45 kV. The gap between the MCP output and the anode array requires a voltage drop of approximately –400 V. Both the MCP and the anode gap voltages are supplied by the high-voltage power supply (HVPS). The overall detector gain is ~$6\times10^6$ (~1 pC). At an expected average count rate of 2000 count/s, the amount of charge pulled from the MCP as a function of time is ~0.06 Coulomb/year.

### 3.4. Instrument Electrical Design

The JUICE-UVS instrument support electronics include redundant features for certain important functions. These electronics include a dual-string LVPS, redundant actuator electronics, single-string C&DH electronics, a redundant optics decontamination heater system, and a redundant HVPS to feed the detector (with a single DETE). These subsystems are all controlled by an 8051 core processor implemented in a radiation-hardened field-programmable gate array (FPGA) with 64 kbit of programmable read-only memory (PROM), 16 Mbit of magnetoresistive random-access memory (MRAM), 128 kbit of static random-access memory (SRAM), and two 16-Mbit acquisition memory banks. Command and data flow is through a SpaceWire connection to the JUICE spacecraft. Fig. 14 shows a detailed electronics block diagram.

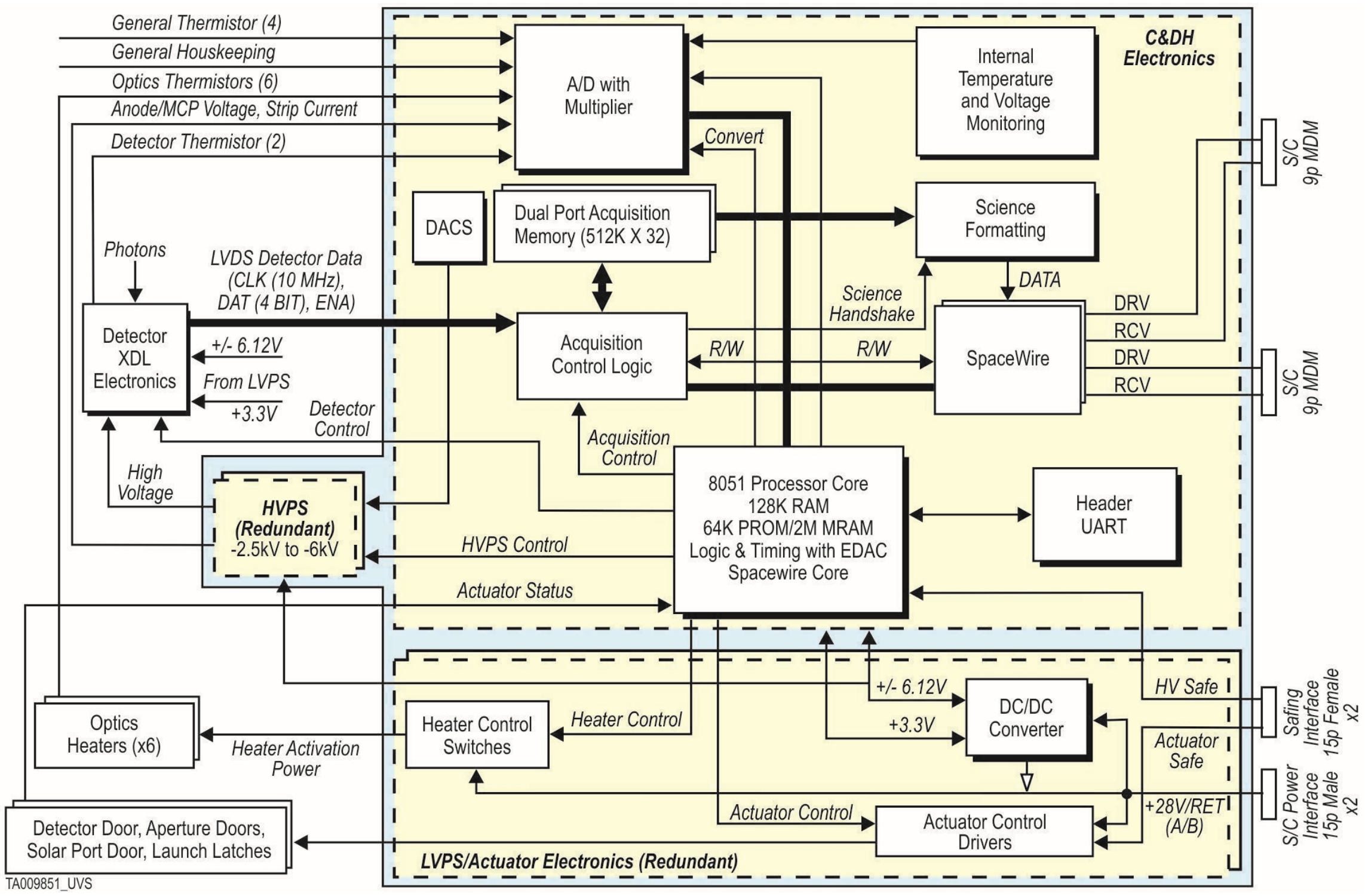

**Fig. 14** JUICE-UVS electronics block diagram. Individual circuit boards are shown in yellow boxes with dashed-line borders. Thick arrows show data flow. Thin arrows show control logic, power flow and voltage control, e.g., of mechanisms. Connections with the spacecraft interface are depicted on the right side of the block

### 3.5. Operations and Data Collection Modes

At any given point in time the UVS instrument can be in any of five states (including off). Four of these states are operational ones. The different instrument states and associated state transitions are shown in Fig. 15.

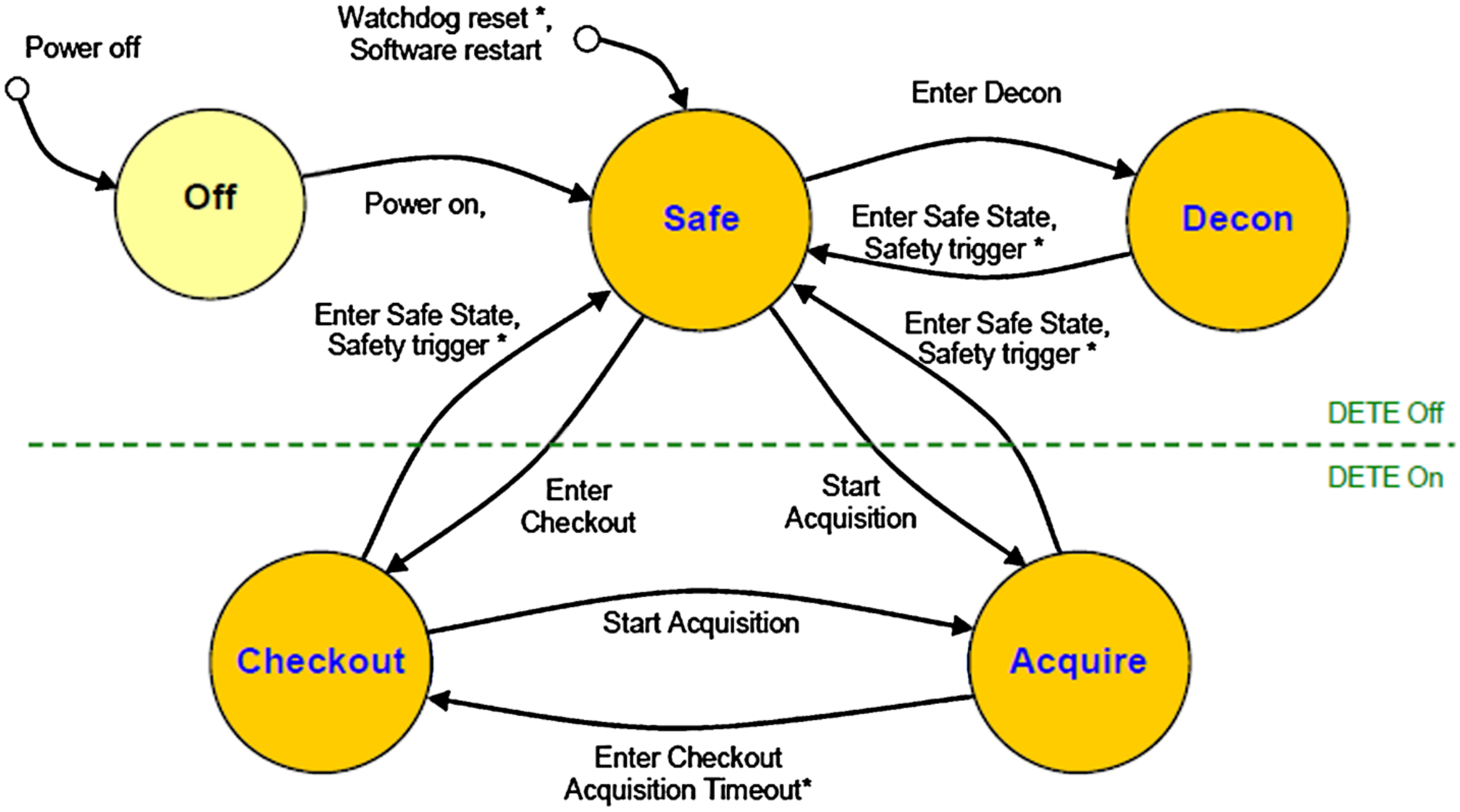

**Fig. 15** JUICE-UVS instrument states, state diagram. Three states with the detector off (top) and two states with detector electronics on (bottom) are shown. The "safe" state is both the one entered during a instrument safety-event triggered by a parameter limit exceedance and a nominal interim state for UVS following a power-on commanded by the spacecraft or a state transition commanded by UVS. Decontamination heaters ("decon" mode) are used periodically for optics cleanliness throughout the mission. Science data is obtained in "Acquire" mode

JUICE-UVS implements two main science-acquisition modes: histogram and pixel list. These modes both capture the information from the detected photon events over a specified acquisition period but differ in the way this information is captured and stored for transmission to the ground. The pixel list-acquisition mode captures each detected photon event with a spectral, spatial and pulse-height value and stores these in a list. To provide timing information, special entries called time-hacks are added to this event list at a regular specified rate. The histogram mode counts the number of detector events for a specified exposure time and duration based on their spectral and spatial coordinates in a 2-D array and can also count the number of detector events based on their pulse-height coordinate in a separate 1-D array. The histogram mode in most cases results in smaller and deterministic data volumes and is expected to be used for the majority of JUICE-UVS science observations. The pixel list mode can be considered a raw mode that captures all the information from the detector.

In histogram mode the maximum time resolution (exposure duration) is 10 ms, selectable in 10-ms steps up to 2.55 s and then followed by 1-second steps from 3 s to 18.1 h. In pixel list mode the maximum time-resolution capability is 1 ms, selectable in 1-ms steps up to 255 ms.

The JUICE-UVS detector electronics assign detected photon events in a 4096 × 4096 × 256 array space, where the three dimensions correspond to the spectral, spatial and pulse-height 'location' of the detected event. Therefore, each raw detector event is represented in 12 + 12 + 8 = 32 bits. Within the 4096 × 4096 addressable spectral and spatial ranges, the active detector occupies an area subset of about 1442 × 784 pixel elements, located around the spectral and spatial positions ("[x,y]") of top left [780,1961] and bottom right [2222,1177] measured pre-flight at room temperature. Note however that image distortions near the left and right edges of the detector (i.e., low and high x value columns) result in an active area space on the array that isn't a perfect rectangle and with reduced spectral resolution near these edges; compared to heritage UVS/Alice MCP detectors the benefit of accepting these regions of degraded linearity and resolution is ~10 nm of expanded bandpass. The ongoing in-flight calibrations could slightly adjust these numbers, e.g., being temperature dependent, and possibly the associated bandpass of 50-204 nm listed in Table 1 as the mapping of wavelength to the distorted regions near the edges improves. As mentioned, the detector electronics also include a STIM (stimulator) fiducial within the detector space that generates artificial events at two locations, allowing for thermal correction of the data. These STIM pixels are outside the detector's active area and reside near the spectral and spatial positions of [598,2448] and [2393,694], respectively (at room temperature, pre-launch).

After detector events are received from the electronics two simple configurable filtering functions can be applied to reduce the number of invalid events: a discriminator function limiting the accepted pulse-height values, and ten separate defined masks. The discriminators provide an upper and lower limit to the accepted pulse-height value for processing events, allowing us to filter out low pulse-height events marginally detected due to a low detector-gain value, and very high pulse-height events most likely caused by radiation instead of photons. Masking allows us to exclude events in a defined spectral/spatial range (i.e., a rectangle in the detector space) from any further processing, reducing the number of digital events. This may be used to mask out hot spots in the detector or to reduce the data volume, e.g., by excluding any improperly classified events from the edge of the detector. Events are only passed through the masking phase when they fall outside all of the (up to ten) defined rectangular masks, which allows us to apply complex masking patterns if necessary.

In addition to the two main acquisition modes, the instrument can capture either parallel or standalone count-rate data. This allows reporting of the total number of counts over a configurable period from 100 ms up to 25 seconds in steps of 100 ms. In addition, any number of the following nine of different count rate product types can be selected for reporting: i) raw (fast) analog detector, ii) fully captured digital, iii) discriminated, iv) masked, and v) five different spectral-band subsets. Even at the highest rates and most inclusive collection this count-rate data provides only a very modest-sized data volume. It can be used for various functions within the system. For example, it could be used as a selector to determine or update the priority for which data should be downlinked. It can track captured analog and digital rates to estimate detector dead-time for calibration (Section 4). With the possibility of defining up to five separate spectral bands, this last type of count rate can be used as a low-fidelity science capture mode, capturing just photon rates in a few spectral bands suitable for quick-look assessments (Section 6.4). Likewise, one or more spectral bands could be defined to provide a low-rate, high-energy radiation monitor using relevant detector columns expected to be absent of photon-based signals.

### 3.6. Programmable Acquisition Modes

The histogram and pixel list modes can be configured to tailor the digital selection of spectral and spatial resolution and range of detector area to downlink the desired science information using a minimum of needed data volume per observation. Supported by hardware, each acquisition can be commanded with a set of Lookup Table (LUT) parameters that are custom designed by the JUICE-UVS team for each activity type (Section 3.7) to return the information subset of interest.

After the initial discrimination and masking steps (Fig. 16), all the remaining events are converted in resolution and range for each of the spectral, spatial and pulse-height values of the event. The conversions of these three value types are implemented as LUTs that can be filled with any conversion function. The photon events in the detector space are converted to a required science space (excluding inactive detector space) for the specific science observation, tailoring the subset of spatial and spectral ranges to remove unnecessary information. Furthermore, the events are also encoded using the minimum number of bits needed for the specific science observation. For pixel list acquisitions this programming simply reduces the number of bits from the raw 32-bit detector event to a number just sufficient to obtain the measurable quantity of interest (i.e., based on spectral and spatial ranges and resolutions).

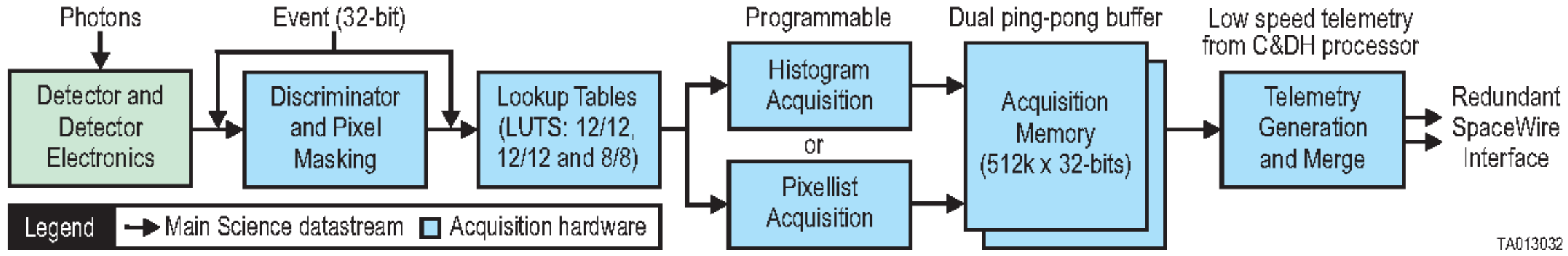

**Fig. 16** UVS instrument acquisition hardware flow chart with masking and Lookup Table (LUT) steps

Note that this lookup-based mechanism maximizes operational flexibility by maximizing science return while minimizing data volume. For example, it allows us to select the atomic oxygen emission line wavelength locations at full resolution but sample the longer-wavelength surface reflected sunlight continuum (Fig. 4) at a lower spectral resolution. Histogram exposures can be commanded in units of 10 ms. Such a fast rate for the full array size would result in prohibitively large data volumes. This selectable reduction in spectral and spatial resolutions enables more flexible plans for data acquisitions.

For the histogram acquisitions the system provides an additional data-reduction method in the form of setting a digital image saturation level. As mentioned above, the histogram acquisitions count the number of events for each spectral and spatial coordinate in a 2-D array, where each value is represented as a 16-bit counter, allowing 0 to 65,535 counts to be captured. The counter will saturate at the maximum count in the case when more events are captured during a single exposure in a specific mapped spectral and spatial coordinate. Based on the observation planning, including expected brightness and resolutions, a required dynamic range may be defined that is smaller than the provided 16 bits; for example, a reduced bin depth may be defined at 10 bits, reducing the downlinked data volume by an additional 37%. For instance, in cases where the bright Lyman-α sky feature is of no importance to the specific science goal (e.g., perhaps for relatively dim stellar occultations), then the bins capturing those events can simply be left to saturate.

## 3.7 Science Operations Activity Types

### 3.7.1 JUICE Operations Summary

The JUICE-UVS investigation will achieve its scientific requirements through a series of activity types (Table 5). Each activity is designed to address specific science goals; however, the activities are not limited to a single detector setup where all detector regions are recorded equally. Instead, they can be programmed to achieve any combination of spatial and spectral resolutions, exposure times, etc. as described in Section 3.6. This flexibility within each activity allows for the science objectives to be met while optimizing the required data volume resource.

**Table 5** UVS Observation Types & Library (subject to evolve during the mission)

| Target | Observation name | Purpose |
|---|---|---|
| Jupiter | UVS_JUP_AP_LIMB_SCAN | Monitor auroral and airglow emissions in limb scans, requiring a continuous S/C motion to point to limb and scan over planetary limb. Uses AP port. |
| | UVS_JUP_HP_LIMB_SCAN | Same as UVS_JUP_AP_LIMB_SCAN but through the HP port |
| | UVS_JUP_AP_SCAN_MAP | Using the Airglow (AP) port, scan at a constant rate across Jupiter to produce a map. |
| | UVS_JUP_HP_SCAN_MAP | Same as UVS_JUP_AP_SCAN_MAP but for High spatial resolution Port (HP). |
| | UVS_JUP_AP_STELL_OCC | Inertial pointing at a moderately bright star for ~10 - 20 minutes to observe the change in stellar flux as the star is occulted by Jupiter's atmosphere |
| | UVS_JUP_HP_STELL_OCC | For bright stars, use the High spatial resolution port (HP) for higher contrast of star signal to Jupiter background signal. Used also as calibration reference standards. |
| | UVS_JUP_HP_FEATURE_SCAN | To assess the evolution of discrete phenomena (e.g., H Ly-alpha bulge, plumes, auroral features, etc.) using the HP port and pixellist mode. |
| | UVS_JUP_DEFAULT | Default pointing to be inserted at the start and end of the timeline |
| | UVS_JUP_AP_AIRGLOW_STARE | Monitoring auroral and airglow emissions in stare mode using the Airglow Port (AP). Used if/when it is possible to hold the slit along Jupiter's North/South |
| | | axis and on the central meridian, while Jupiter rotates below S/C creating a map. Histogram mode. |
| | UVS_JUP_HP_AIRGLOW_STARE | Same as UVS_JUP_AP_AIRGLOW_STARE but using the HP aperture. |
| | UVS_JUP_MONITORING_AP | Used for Jupiter monitoring observations when a N/S slit alignment is not possible. UVS performs a slow scan across the disk instead of a stare. |
| | UVS_JUP_MONITORING_HP | As above, focusing on auroral regions where brighter emissions are better observed using HP mode. |
| | UVS_JUP_ROLL_SCAN | Alternative method of scanning Jupiter's disk. Point to nadir, then rotate about nadir so that UVS scans a circle (or a fraction of a circle - e.g. covering the auroral regions) over Jupiter's disk. Rotation rate ~0.1 degree per second |
| | UVS_JUP_SP_SOL_OCC | Inertial pointing at the Sun through the SP aperture. The large solar disk and the substantial distance from Jupiter mean that solar occultations will not provide the same vertical resolution as stellar occultations, but they are useful for measurements of minor/trace constituents due to high SNR. |
| Icy satellites | UVS_SAT_DISK_SCAN_AP | Scan the UVS aperture across the satellite disk to construct spectral image cubes of multiple atmospheric emission line features (up to 1024 selectable spectral bins with a minimum of 3 key emissions: H Lyman-α, OI 130.4 nm, OI 135.6 nm), with repeated scans to investigate highly time-variable auroral dynamics. |
| | UVS_SAT_DISK_SCAN_HP | As above, but using the HP aperture |
| | UVS_SAT_LIMB_SCAN_AP | Similar to disk scan observations, but with the pointing defined relative to the limb during flyby sequences. |
| | UVS_SAT_LIMB_SCAN_HP | As above, but using the HP aperture |
| | UVS_SAT_LIMB_STARE_AP | Search for faint atmospheric emissions by building signal to noise through long integrations. |
| | UVS_SAT_LIMB_STARE_HP | As above, but using the HP aperture |
| | UVS_SAT_SOL_OCC | UVS solar port stares at Sun as the satellite occults it. |
| | UVS_SAT_STELL_OCC | UVS airglow port stares at a fixed RA and DEC as the satellite occults the star. Bright stars will require use of HP mode |
| | UVS_SAT_SURF_AP | Pushbroom observations near flyby closest approach to investigate surface composition |
| | UVS_SAT_SURF_HP | As UVS_SAT_SURF_AP but using the high resolution port for improved spatial resolution in key surface regions |
| | UVS_SAT_TRANSIT | Measure absorption of Jupiter airglow by satellite atmospheres as they transit Jupiter's disk, to constrain satellite atmospheric composition and variability. Either point to nadir and wait for moon to transit, or scan across the satellite as it transits, depending on available resources (including time) |
| Io + small moons | UVS_IO_SCAN | Similar to UVS_DISK_SCAN, but including extra emission lines e.g. from S and Cl. Also requires different spatial binning since Io is more distant |
| | UVS_IO_TORUS_SCAN | Map emissions from the Io torus. Slit aligned parallel with Jupiter's equator, scanned N-S across one ansa of the torus, then move in four steps to the other ansa, repeating the N-S motion each time |
| | UVS_IO_TORUS_STARE | Monitor emissions from the Io torus. Slit aligned parallel with Jupiter's equator. |
| | UVS_NC_STARE | Characterize the Io/Europa neutral clouds in the immediate vicinity of the satellite. Center satellite in slit. Align the slit with the satellite orbital plane |
| | UVS_IRR_SAT | Obtain reflectance spectra of irregular satellites |
| Ganymede (orbital phase) | UVS_GCO_HISTOGRAM_001 | Monitor auroral emissions and surface reflectance during GCO. Limited spectral resolution. |
| | UVS_GCO_HISTOGRAM_002 | Similar to observation 001 but with Increased time sampling to capture auroral morphology and variability |
| | UVS_GCO_HISTOGRAM_003 | Similar to observation 001 but with increased spectral resolution to achieve < 2 nm resolution between 100 and 200 nm as specified in SciRD |
| | UVS_GCO_HP | High spatial resolution observations of Ganymede's aurora to look for small scale features |

| Operational | UVS_DECONTAMINATION | Not a true observation, but included in observation plans to allow proper consideration of instrument resource requirements |
|---|---|---|
| | UVS_CALIBRATION | Generic calibration observation currently used as placeholder - may include star stare, flip ridealong, or dark/radiation observations. |

### 3.7.2 JUICE-UVS Scans

Many of the planned JUICE-UVS observations of both Jupiter and the icy moons are scans, during which spacecraft slews are used to move the UVS slit at a constant rate in the across-slit direction over the target. During satellite flybys, each ~30-minute full-disk scan begins and ends approximately ±1.5 $R_{satellite}$ away from the satellite's center to capture any extended auroral and airglow emissions above the limb. The scans enable the long, yet narrow, JUICE-UVS slit to build up a complete composite image over an entire hemisphere of the moon, resulting in a relatively high-resolution snapshot of any auroral or airglow emissions, along with a spectral/spatial map of the surface. During most flybys, four scans are nominally scheduled to occur during the -12 to -2 hour period before closest approach and four scans are scheduled during departure (+2 to +12 hours after closest approach). The scans are nominally planned at regular intervals to sample a range of local magnetic conditions, which vary over Jupiter's approximately 10-hour rotation period. The changing interaction between Jupiter and its satellites leads to variations in the satellite auroral morphology, which UVS will use to obtain global information regarding the plasma environment and its interaction with satellites as modified by the induced magnetic fields and subsurface oceans.

Similar full-disk scans are performed during Jupiter perijove segments, both in AP and HP mode. AP mode scans are used to map reflected sunlight to probe hydrocarbon species distribution, stratospheric hazes, and to search for faint airglow emissions on Jupiter's nightside. A range of sub-spacecraft longitudes and local times are targeted, allowing studies of atmospheric variability over a range of timescales. HP mode scans provide the highest resolution view of the aurora and are performed at sub-spacecraft longitudes close to 180° W to target the northern aurora, or 45° W to target the southern aurora (due to the offset between Jupiter's rotation and magnetic axes). In addition to full-disk scans, each perijove segment typically includes ~2 "feature scans". These longer (~1 - 2.5 hour) observations comprise multiple scans back and forth over key regions - for example the aurora or the terminator - to investigate short-timescale variability. The large variation of FUV opacity with wavelength allows UVS to probe an enormous range in altitude, addressing JUICE's goal of studying horizontal and vertical coupling processes which redistribute energy and mass throughout the upper atmosphere. In particular, the high inclination mission phase will allow UVS to study latitudinal contrasts expected in Jupiter's polar vortices and their unique aurora-related ion-neutral chemistry. Finally, regular monitoring scans are performed outside of Jupiter perijove periods. These more distant scans are particularly useful for monitoring the total auroral power, and provide important context for in situ measurements.

### 3.7.3 Limb Pointing

The full-disk scans during satellite flybys will capture bright aurora/airglow emissions above the surface out to ±1.5 $R_{satellite}$, but longer integration times may be required to detect any fainter emissions of potential trace species, including C, S, CO, Cl, etc. For this reason, ~30-minute limb stares are performed in between the disk scans: three before closest approach, and three on departure. During these limb stares, the slit may be oriented in one of two ways: 1) just above the surface at a tangent to the limb, or 2) perpendicular to the limb with the boresight at the surface. The first orientation provides better coverage of the densest atmosphere at the lowest altitudes, while the second provides information on vertical structure.

Limb observations are also important for understanding Jupiter's vertical structure. Jupiter limb observations are scans, similar to the AP disk scans described above but with the UVS boresight – where the spatial resolution is best – placed at the limb rather than the center of Jupiter's disk.

### 3.7.4 Satellite Surface, Aurora and Airglow Nadir Stares

During the pushbroom phase of each satellite flyby (typically within ±1 hour of closest approach), JUICE-UVS will perform a series of short nadir-pointed stare observations to obtain high resolution surface observations while also capturing any auroral or airglow emissions falling within the FOV. Outside of flyby segments, distant satellite stares are planned for monitoring purposes, providing information about atmospheric emission variability over a range of timescales throughout the Jupiter tour. The long duration series of these observations is needed to constrain how the gas distribution and variable auroral emissions are coupled to the inhomogeneous plasma environment, providing the required baseline to interpret more sporadic events such as potential plume emissions. Distant stare observations are most important for Europa and Io, since JUICE will only perform two Europa flybys and rarely gets within 500,000 km of Io.

Ganymede's FUV albedo is about 5% (Feldman et al. 2000, Molyneux et al. 2022) and Benmahi et al. (2026) reports a variation with up to 8% in the 140-205 nm range. For a solar Lyα flux of ~$5 \times 10^{11}$ photons/$cm^2$/s (at 1AU), the expected UVS count rate in a filled 0.1° pixel is 580 counts/s, assuming an UVS effective area at Lyα of 0.64 $cm^2$ (Fig. 4). During the GCO-200 phase (orbital period ~9500 s, Boutenet et al. 2024), the time for a point on the surface to move through the 0.1°-wide UVS slit is 0.2 s. Thus, each 0.35-km region of Ganymede in a 35-km wide swath forms a mapping element with reasonably good quality at >100 counts. As with LRO-LAMP at the Moon, UVS's dynamic range allows it to take data on the night-side of Ganymede as well the dayside, using reflected Lyα light from the all-sky IPM and Jupitershine. Such observations are useful near the terminator, where shadows cause problems for photometric analyses using sunlight. The Lyα count rate on the nightside is much lower, however, dropping from the 580 counts/s/0.1°pixel dayside value down to ~80 counts/s/0.1°pixel.

### 3.7.5 Stellar Occultations

JUICE-UVS performs regular (~1 - 3 per week) observations of stellar occultations by Jupiter's Galilean satellites to characterize their atmospheres and search for spatial and temporal variability, for example plume activity at Europa. This activity involves centering the slit on a UV-bright star when that star passes within 150 km (~0.1 $R_{satellite}$) of the satellite's limb from the spacecraft's perspective (Fig. 17). The primary goal of the stellar occultation activities is to identify and map the vertical distribution of atomic and molecular species in the satellite atmospheres. The data set will constrain column density, spatial distribution, and temporal variations of these atmospheric constituents (e.g., $O_2$, $H_2O$, $CO_2$, $H_2$, at the icy satellites; $O_2$, $SO_2$, $H_2S$ at Io). The observations are planned to be distributed in latitude, longitude and local time to provide a comprehensive view of atmospheric composition and distribution. Stellar occultation observations are designed to achieve a 50-km scale height resolution in order to adequately probe the vertical structure of the atmosphere, with this resolution improving to at least 30-km for characterizing the composition of any plumes, if detected.

Stellar occultations will also be used to probe the composition, vertical structure, and variability of Jupiter's upper atmosphere (Yelle and Miller, 2004). As the star is a point source, they have the advantage to provide high spatial resolution which is key for detecting waves (e.g., Yelle et al. 1996, Brown et al. 2022) as well as deriving the homopause location and the vertical eddy diffusion coefficient (e.g., Vervack and Moses, 2015; Brown et al. 2024). Serendipitous multiple occultations, with more than one star in the slit, allow UVS to correlate wave phenomena in the upper atmosphere, studying the distribution and frequency of vertically propagating waves from tropospheric sources (e.g., moist convective events and plume disturbances) to assess the contribution of wave forcing to the upper atmospheric energy budget, which is still not understood four decades after Pioneer found exospheric temperatures wildly in excess of predictions (e.g., Yelle et al. 2004). A minimum of 50 Jupiter stellar occultations are planned to characterize key atmospheric species including $H_2$ (and hence thermospheric temperature), hydrocarbons ($CH_4$, $C_2H_2$, $C_2H_4$, $C_2H_6$, $C_6H_6$) in the FUV, $NH_3$, etc. Again, a range of longitudes, latitudes, and local times are targeted, with occultations of the auroral regions a particularly high priority.

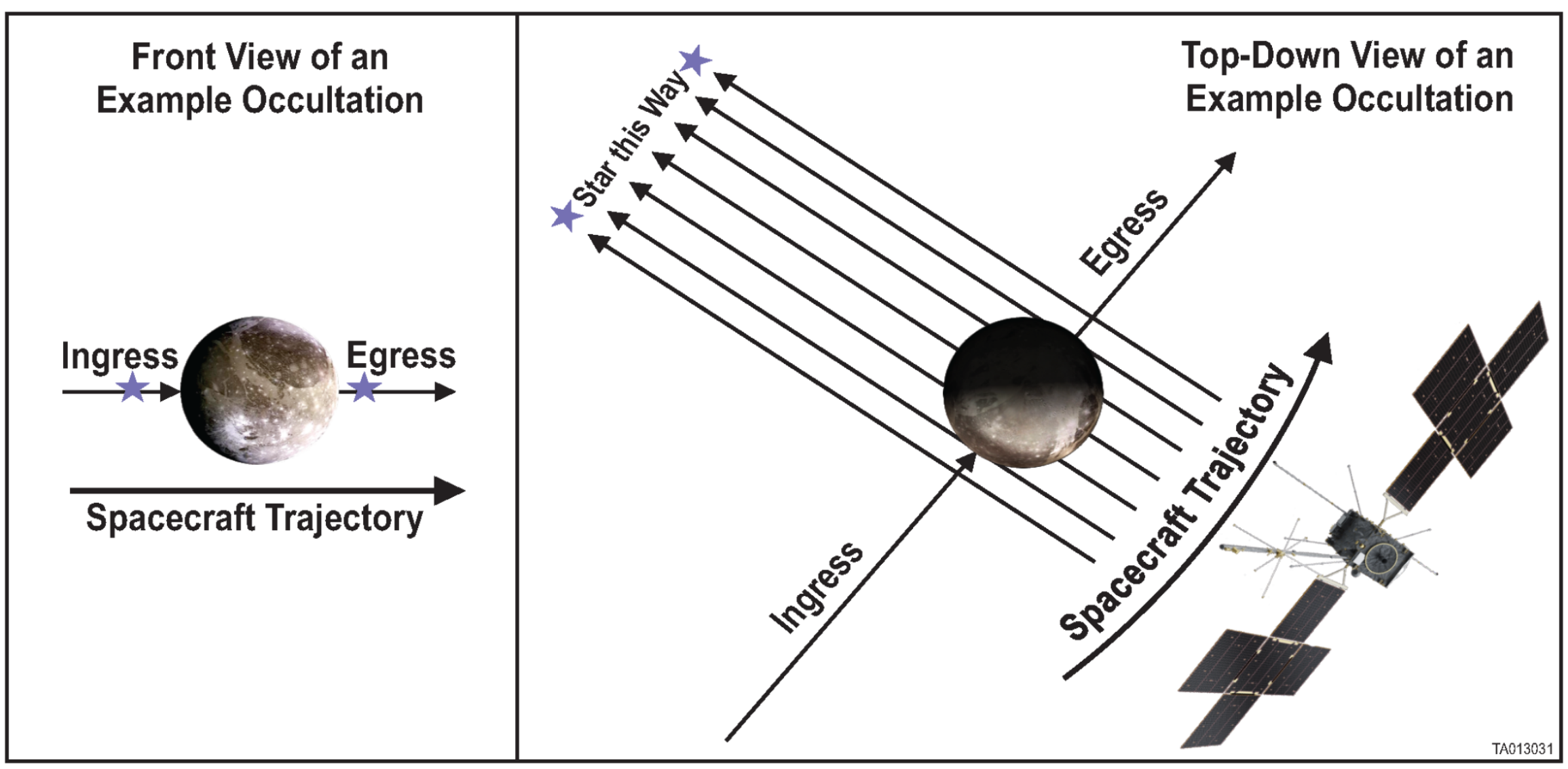

**Fig. 17** Occultation geometry sketch. With inertial spacecraft pointing towards a star, Ganymede flies into (ingress) and out of (egress) the UVS field of view based on the spacecraft velocity along its trajectory as shown. The angular velocity of Ganymede's tangent point along these lines of sight determines the available exposure time per atmospheric scale length of interest. Gases around Ganymede (and the other moons) attenuate the starlight as a function of wavelength, leaving a potentially discernible fingerprint in the stellar photons observed with JUICE-UVS

Stellar occultations can be observed during ingress (the star is observed prior to being occulted by the target) or egress (the star is observed after emerging from behind the target). JUICE-UVS also observes grazing stellar occultations (appulses), when the star crosses behind a satellite's atmosphere but is never blocked by the satellite itself. In all three cases, the occultation observations are designed to acquire a baseline of the unocculted stellar signal for direct comparison with the occulted stellar spectrum as a function of distance to the satellite surface. Stellar occultations can also probe the exact shape and size of the moons, constraining the topography with a precision on the order of meters when measuring the surface intercept point at 1-ms time resolution (Abrahams et al. 2021).

The Catalog of UV Bright Stars (CUBS) was created (Velez et al. 2024) to select the set of stars for the planned occultations. Consisting of ~90,000 stars, CUBS provides spectra for each star by using IUE (International Ultraviolet Explorer) observational data where available (1746 stars) and Kurucz models elsewise. Every potential occultation is run through a prioritization scheme to rank its quality. The scheme starts with a numerical analysis of the estimated SNR for each star contingent on its brightness in the UV bandpass of $O_2$ absorption (130-175 nm) and the speed of the star relative to the target from the spacecraft's point of view (Fig. 17). This bandpass was chosen for Europa occultations as an example, and for other targets such as Jupiter a more appropriate bandpass (Table 2 in Velez et al. 2024) can be selected, e.g., for methane, while performing the same calculations. Fig. 5 and Fig. 6 show simulated stellar and solar spectra at different tangent point altitudes, and SNR per resolution element derived for the normalized transmission spectra in expected average cases. After assigning each occultation a quality factor of "Very Good", "Good", "Ok", and "Bad", each candidate event is further analyzed using a list of subjective criteria to increase or decrease its rank accordingly (Velez et al. 2024). For example, "Good" events that might be coordinated with Europa-UVS observations would be increased to "Very Good" in the priority scheme. For another example, any occultation opportunity viewing a line of sight that passes near the spacecraft trajectory's footprint on the surface would be of value for direct comparisons with in situ measurements by the PEP-NIM instrument (Barabash et al. 2026, *this collection*). Events with SNR<3 are "Bad" and not considered further. See Section 5 for a more complete list of criteria considered during instrument operations planning. The molecular oxygen ($O_2$), water ($H_2O$), hydrocarbons ($C_xH_y$), and other species of interest are all detectable by JUICE-UVS using these occultations during the entirety of the mission.

### 3.7.6 Solar Occultations

Solar-occultation observations are similar to stellar occultations, except the occulted light source is the Sun rather than a distant stellar point-source. The benefits of solar occultations are their increased signal-to-noise and ability to observe atmospheric absorptions of light at wavelengths <91.2 nm, which are not covered by the vast majority of stellar occultations because those wavelengths are typically absorbed by the interstellar medium.

For the Jovian atmosphere, solar occultations offer the possibility of deriving the H density profile in altitude and are more robust than stellar occultations to derive the $H_2$ density profile from the homopause all the way to the exosphere (e.g., Yelle and Miller 2004, Vervack and Moses 2015, Stephenson et al. 2025). H density can be derived from the analysis of the attenuated solar flux from the ionisation continuum of H, especially from 80.4 nm up to 91.2 nm. $H_2$ density can be derived from the $H_2$ ionisation continuum (<80.4 nm), instead of only relying on the electronic band system above 91.2 nm (in the case of stellar occultations). The band system not only depends on temperature and vibrational/rotational level population but its analysis is sensitive to contribution from the lower thermosphere. $H_2$ density profiles can then be used to derive the atmospheric temperature profile. $CH_4$ can be derived beyond 110 nm (where the contribution from $H_2$ drops), similarly to stellar occultations.

For the icy moons, complementing the hundreds of potential stellar occultations, there are several solar occultations by each Galilean satellite during the JUICE mission, enabling targeting of specific surface features (e.g., recently active sites on Europa to search for possible plumes). An example of a solar occultation by Callisto atmosphere is discussed in Section 2.1.6.

Solar occultations better constrain the column densities of several trace species (see Retherford et al. 2024 Table 2 for detectability of several example species). To enable solar occultations, the Sun is observed through the solar port (Section 3.2), whose boresight is designed to place the Sun in the 0.2° × 0.2° box at one end of the instrument's slit-defined FOV (Fig. 11). These occultation observations are designed to typically last ~10 minutes for moon and ~30 minutes for Jupiter during either the ingress or egress portion of the occultation, depending on the availability to schedule a spacecraft pointing maneuver in the planning sequence. With several solar occultation events expected for Jupiter amongst the several dozen perijoves in the trajectory, a comprehensive analysis of the dataset and forward modelling of the transmission of the solar flux through the planetary atmosphere makes it possible to extract $H_2$, H and $CH_4$ profiles with a suitable altitude resolution, even with the projected solar disk size exceeding the atmospheric scale height (e.g., Stephenson et al. 2025). We expect JUICE's initial high-altitude Ganymede orbit phases to enable a few dozen opportunities for high-quality Ganymede solar occultations.

### 3.7.7 Jupiter Transits

Jupiter transits provide detailed information about the atmospheric composition and structure around the entire limb of a satellite. These activities use Jupiter as the background illumination source to constrain the column density, global spatial distribution, and temporal variations of the primary atmospheric constituents of the target (i.e., $O_2$ and $H_2O$) and for hunting multiple plumes locations around the limb, similar to the technique used to make putative plume detections by Sparks et al. (2016, 2017). The activity occurs in two primary ways: (1) the JUICE-UVS FOV is centered on Jupiter while the satellite passes through the slit; and (2) one or more scanning activities are conducted while the satellite is passing in front of Jupiter. This activity occurs on the daylit side of Jupiter (e.g. solar phase angle less than 120°) a minimum of 5, 2, 10 and 5 times each for Io, Europa, Ganymede and Callisto, respectively, throughout the mission. Transits by Ganymede and Callisto are less frequent but will be observed when practical, with a goal of achieving at least 3 - 5 observations per satellite. When possible, the observations target Europa at times when viewing the brighter subsolar or Lyman-α bulge regions of jovian dayglow emissions. For Io, observations with solar phase angles closer to 90° are also important for comparisons of the dayside and nightside atmospheres to constrain the relative importance of sublimation and volcanism as atmospheric production mechanisms.

### 3.7.8 Neutral Cloud and Torus Stares

Io's volcanic source of predominantly neutral $SO_2$ is the ultimate origin of the Io plasma torus and majority of Jupiter's magnetospheric plasma. Material from Io's atmosphere is both ionized, picked-up and then swept away by Jupiter's magnetic field, and also ejected as neutrals to form an extended neutral cloud that is shaped via gravity and electromagnetic interactions (Bagenal and Dols 2020). This extended source of neutral atomic sulfur and oxygen is ionized into sulfur and oxygen ions ($S^+$, $S^{++}$, $S^{+++}$, $O^+$, and $O^{++}$), again producing magnetospheric plasma that is subsequently transported to Europa. This corotating plasma is excited to higher-energy states via electron impact excitation; then, through spontaneous emission of forbidden transitions, it decays back to a lower-energy state, emitting UV light that indicates plasma conditions (Steffl et al. 2004 & Nerney et al. 2017). By measuring these UV emissions routinely during the monitoring segments of each orbit, JUICE-UVS is able to investigate Io as an exogenic source of sulfur implanted on Europa's surface and also to characterize the variability of Europa's plasma environment, for example.

The strongest Io plasma torus emission lines in the UV are at 68.0 nm due to $S^{++}$ and at 83.3 nm / 83.4 nm due to a combination of $O^+/O^{++}$ emission lines. The neutral sulfur and oxygen atoms that form the extended neutral cloud are also excited via electron impacts that cause emission in the UV. Under typical plasma and solar conditions, solar-resonant scattering is also thought to be responsible for ~1/3rd of the 130.4-nm neutral oxygen emission.

JUICE-UVS's neutral cloud and torus scans and stares are designed to constrain the strength of Jupiter's plasma environment and satellite plasma-interactions. This information supports the PEP (Barabash et al. 2026, this collection), RPWI (Wahlund et al. 2024) and J-MAG (Dougherty et al. 2026, this collection) instrument measurements.

Taking a nominal cylindrical radial centrifugal model of the Io torus plasma conditions using a combination of Cassini UVIS-derived warm torus properties (Nerney et al. 2017 & 2020) and Voyager PLS-derived cold torus and ribbon densities and temperatures of ions and electrons (Dougherty et al. 2017) defines our nominal Io Plasma Torus model between $\rho_c$ = 5-10 $R_J$ (Nerney et al. 2025). We use the JRM 33 (Connerney et al. 2021) and Connerney et al. (2020) current sheet model to define magnetic field lines using the community magnetic field code (Wilson et al. 2023) which we then use diffusive equilibrium (static pressure balance) to find the distribution of Ions along each field line (Nerney et al. 2025). Figure 18 shows the electron density in the meridional aligned system III longitude plane for our nominal model. Taking this and similar distributions of ion densities and electron temperatures, we find the volume emission rates over the line of sight using the CHIANTI atomic database version 10.1. These are then integrated over the line of sight to find in Fig. 8 the emission as viewed from an arbitrary unaligned longitude for the $S^{++}$ 68.0 nm emission in Rayleighs, showing the complete spatial structure. Then assuming the sources are filling the slit's field of view, we get the spectra as a function of wavelength in Fig. 19a in Rayleighs/nm and similarly in Fig. 19b in counts/pixel for a 30-minute observation for a case where the torus fills the 7.5° slit.

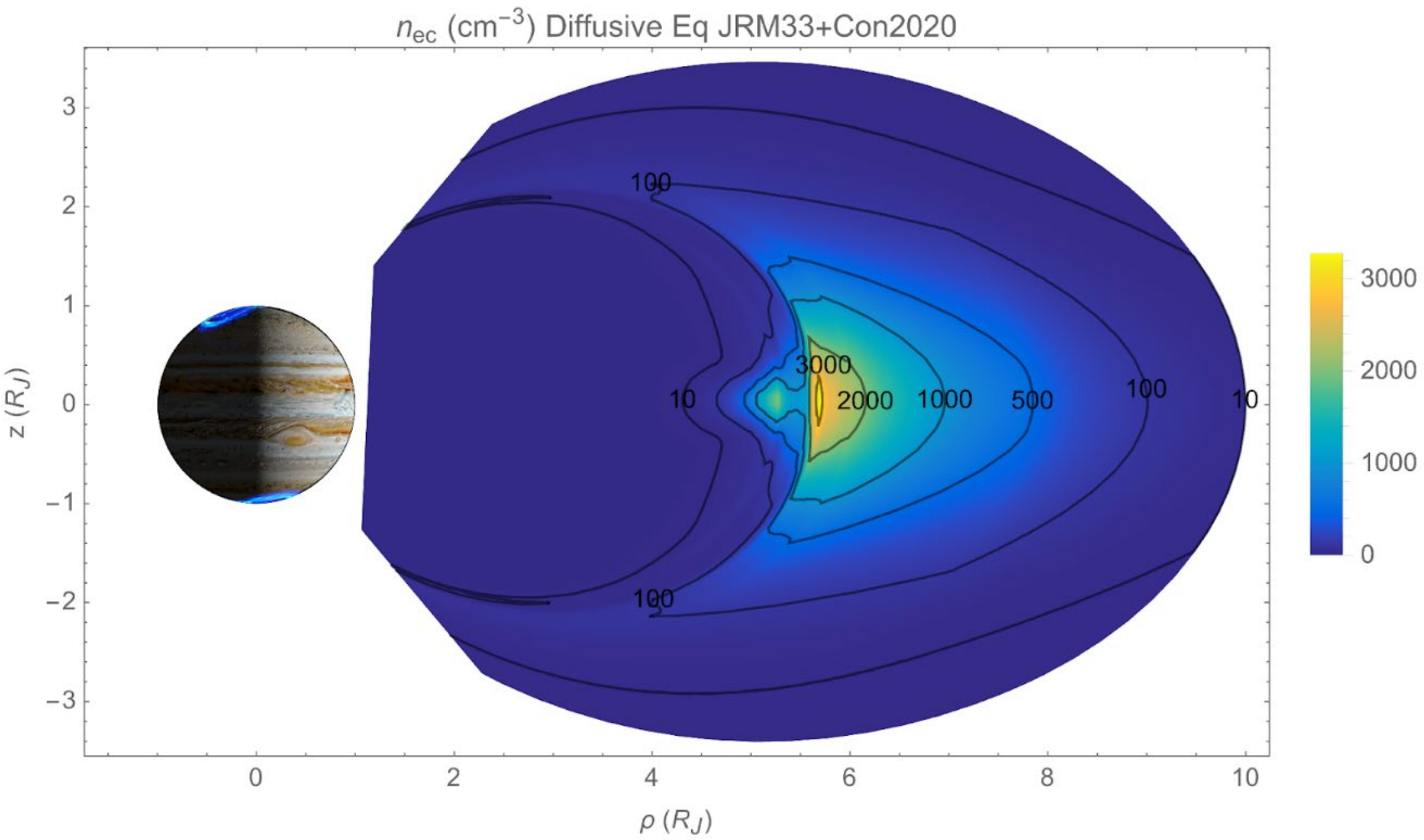

**Fig. 18** Nominal Io electron density profile in the aligned meridional aligned plane using pressure balance to find the distribution of plasma along each magnetic field line. The image of Jupiter's aurora is from Juno UVS superimposed on an image of Jupiter (Bonfond 2020/ULiège/SwRI/NASA)

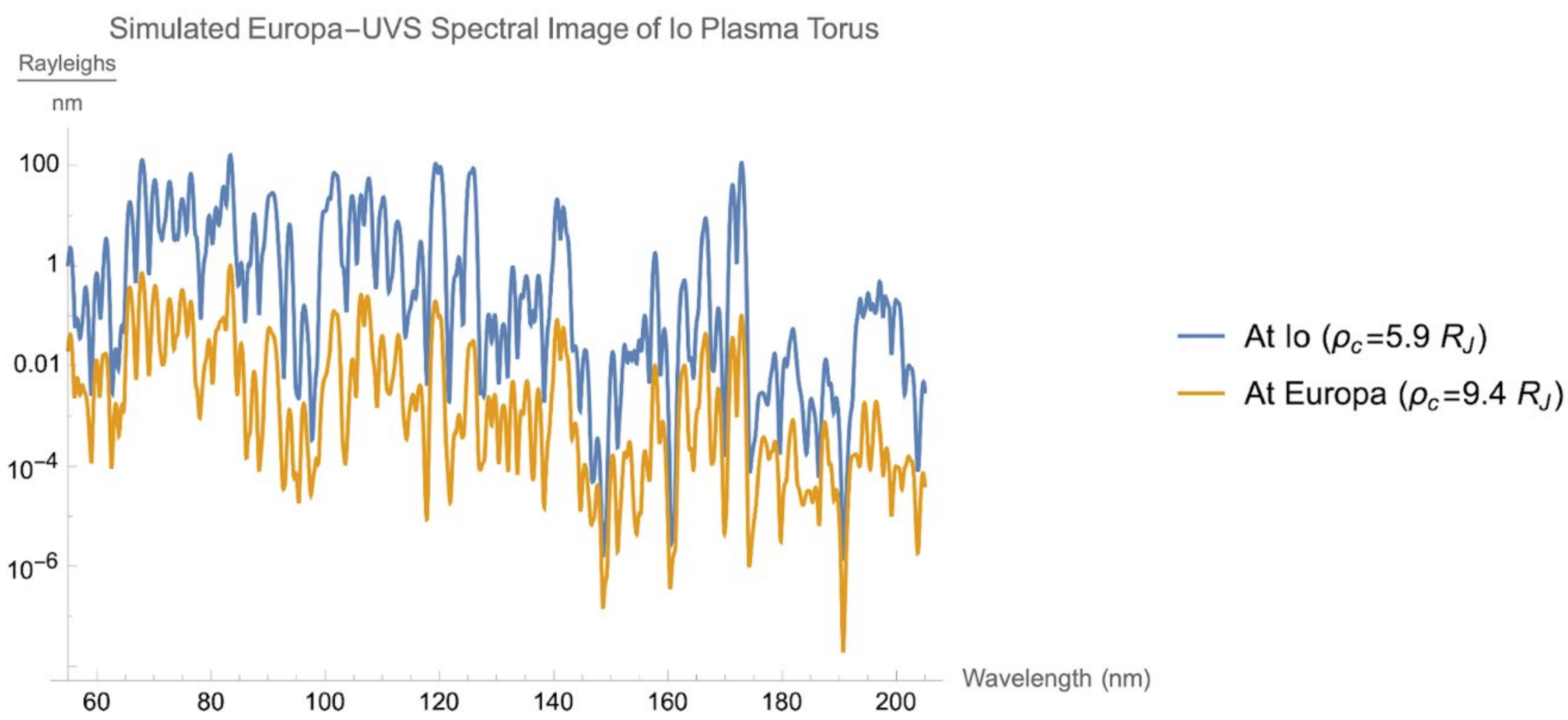

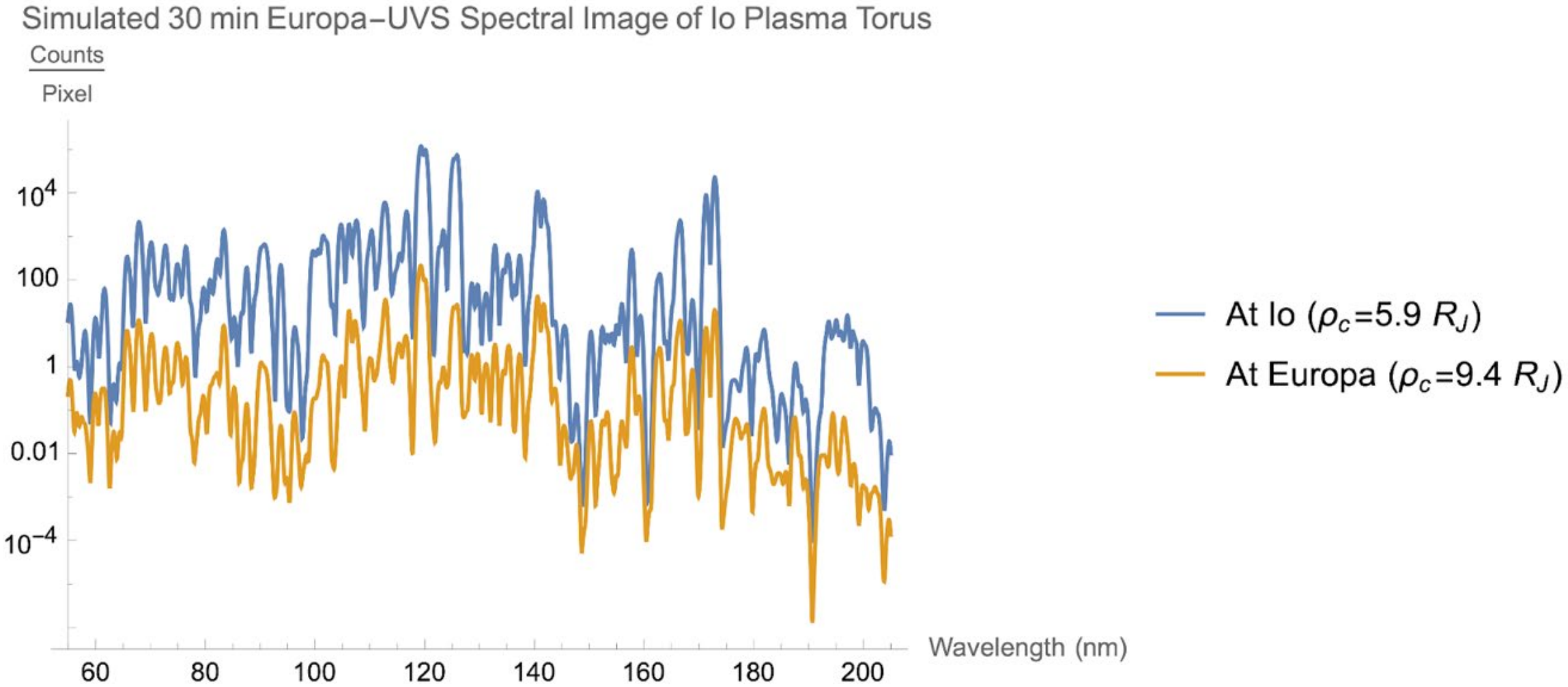

**Fig. 19** (a, top) Nominal Io Plasma Torus spectra at Io and Europa for ion species in Rayleighs/nm. Emission is proportional to the product of ion and electron density, so at Europa, emission is significantly reduced, and longer integration times are required. (b, bottom) Same spectrum as above with instrument response in total counts for a 30 minute observation integrating signal in the whole 7.5° slit. Emission is proportional to the product of ion and electron density, so at Europa, emission is significantly reduced, and longer integration times are required to get higher SNR due to Poisson counting statistics

The neutral cloud and torus scans are nominally scheduled during any Working Group 3 segment, which are distributed through each JUICE orbit, outside of the perijove and satellite flyby periods. A nominal observation comprises a ~90 minute raster scan covering the entire Io torus as minimum. In the segments closest to apojove, scans may extend out to Europa's orbit to target the Europa neutral cloud, whose relatively low brightnesses have most recently been constrained by Roth et al. (2023b). Based on Io and Europa neutral cloud models (Smith et al. 2019, 2022), the CHIANTI database (Dere et al. 1997) for electron impact rates expected in the torus (Nerney et al. 2017, 2020), and g-factors for solar resonant scattering the total expected ~0.2 and ~0.04 Rayleigh line brightness for the 130.4 nm and 135.6 nm oxygen features, respectively, are detectable as far as Europa with slit-integrated spectrum peaks of order ~220 counts/pixel and ~40 counts/pixel, respectively, despite the relative faintness of this feature.

In segments where scans are not possible, for example due to conflicting pointing requirements of other instruments, stares may be performed instead of scans. In this case, the UVS slit would perform a long integration of a small region of the torus/neutral cloud to better constrain fainter emission lines and to search for short-term variations in the intensity of brighter emissions. Ideally, the slit is aligned parallel or perpendicular to the local centrifugal equator to gather the most information, with improved deconvolution of line-of-sight density projection effects (i.e., 'peeling the onion' by fitting outer radial layers then moving inwards) or improved determination of effective ion-scale heights and ion parallel temperatures (parallel to magnetic field line) respectively. Matching models of the torus and neutral cloud emissions to the observations allows us to determine ion, electron, and neutral atom densities, and electron temperatures, and how they vary over the line of sight.

### 3.7.9 Radiation Observations

Nearly all of the UVS science activities naturally also collect information about Jupiter's radiation environment through the observed impact on the detector and the noise measurements within the data (Fig. 20). There are also designated Radiation Observation activities that occur pre- and post-flyby, nominally outside of the primary subphase time range when UVS scanning observations occur. During Radiation Observations, the JUICE-UVS aperture door is closed, similar to a "dark" calibration, in order to measure the high-energy electron and gamma-ray counts that penetrate through JUICE-UVS while within 20 $R_J$ of the planet. Each Radiation Observation lasts approximately 1 hour. These observations measure the radiation environment near the Galilean satellites in addition to providing a calibration of the instrument's background noise, complementing JUICE's other radiation environment studies (Hajdas et al. 2025).

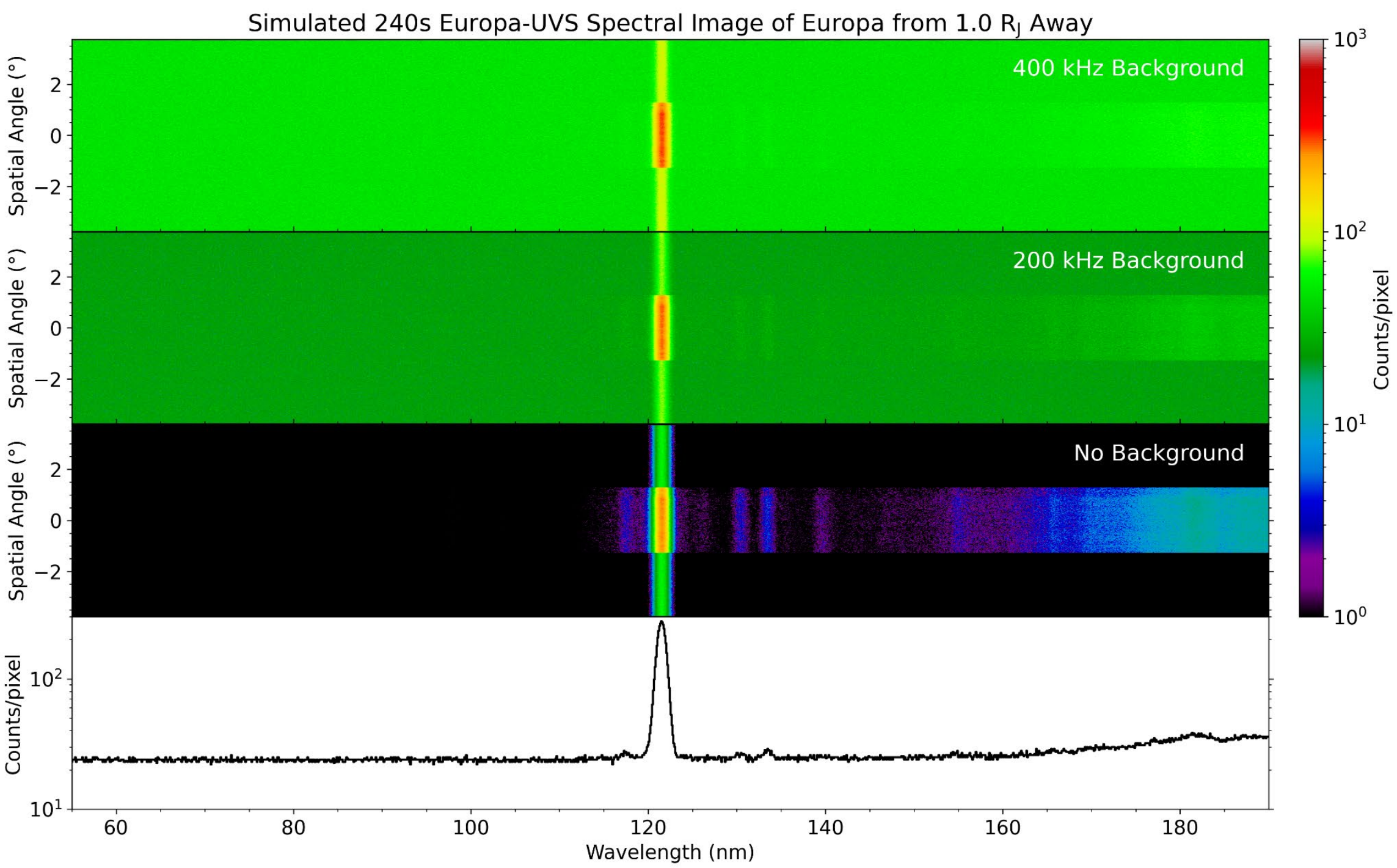

**Fig. 20** Proof-of-Concept spectral image with radiation noise shown for different cases. The bottom panel shows total counts integrated across the disk of Europa for an example case with 200,000 counts per second of dark background signals from the radiation environment; key features remain detectable

### 3.7.10 Ride-along Plans and Opportunistic Observations

The sets of standard UVS observation types just described are complemented by additional opportunities to ride-along with other instrument pointing activities, and such coordinated slew planning is currently under way with JANUS and MAJIS instruments especially. Sky slews, e.g., during large slews of the HGA towards Earth have proven invaluable during cruise for obtaining stellar spectra, and we anticipate several opportunities to obtain more Io plasma torus, Jupiter, and other data as unique geometries present themselves during tour phase. While signal levels for rings and small satellite UVS measurements, including possible Kalichore targeted observations, are expected to be low and of limited quality, UVS will operate and ride-along with other instrument measurements. Observations of interstellar comet 3I/ATLAS provide another good example of opportunistic events pursued by the operations team and successfully achieved and in preparations for publications.

## 4. Calibration

### 4.1. Ground-Calibration

Following assembly, alignments, and environmental testing, and prior to delivery to ESA for integration onto the JUICE spacecraft, JUICE-UVS was subject to a series of tests to characterize its radiometric performance. These tests included (1) dark-count measurements, (2) spatial point-spread-function (PSF) and wavelength calibration, (3) stray-light rejection, and (4) instrument sensitivity expressed as effective area. These measurements were performed in the Ultraviolet Radiometric Calibration Facility (UV-RCF) at SwRI, as initially reported in Davis et al. (2020).

#### 4.1.1 Ultraviolet Radiometric Calibration Facility (UV-RCF)

The UV-RCF system (Davis et al. 2014) is comprised of a differentially pumped Hollow Cathode Lamp (HCL) that uses arc discharge to excite various gases (typically He, $H_2$, Ne, Ar, $CO_2$, and mixtures thereof) to generate a broad spectral output overlain with a plethora of narrow emission lines over the JUICE-UVS 50–204-nm bandpass (Fig. 21). These emission lines are critical for JUICE-UVS wavelength calibration. Light from the HCL is fed to a dispersive monochromator (Acton VM-502), which can be operated in different grating modes to relay the light (broadband or at desired wavelengths) to a 4-inch OAP mirror. The OAP steers a collimated beam down the beamline towards the high-vacuum chamber that hosts JUICE-UVS on a 4-axis tip/tilt motorized stage. This stage allows for accurate on- and off-axis alignments of the JUICE-UVS apertures (airglow port + solar port) with respect to the incident beam essential for determining spatial resolution/PSF characterization.

The photon beam can be intercepted by a NIST-calibrated photodiode (Optodiode AUX-100G) mounted on a retractable linear stage to measure the absolute flux and intensity profile of the beam while the monochromator selects the target wavelength. Knowledge of the absolute flux as a function of wavelength is essential to measure the effective area. The optomechanical layout of the UV-RCF can be found in Davis et al. (2014). The entire beamline is evacuated with oil-free turbomolecular and non-evaporating getter (NEG) pumps to ~$10^{-7}$ Torr. The main chamber hosting JUICE-UVS is cryopumped to <$10^{-8}$ Torr during calibration runs.

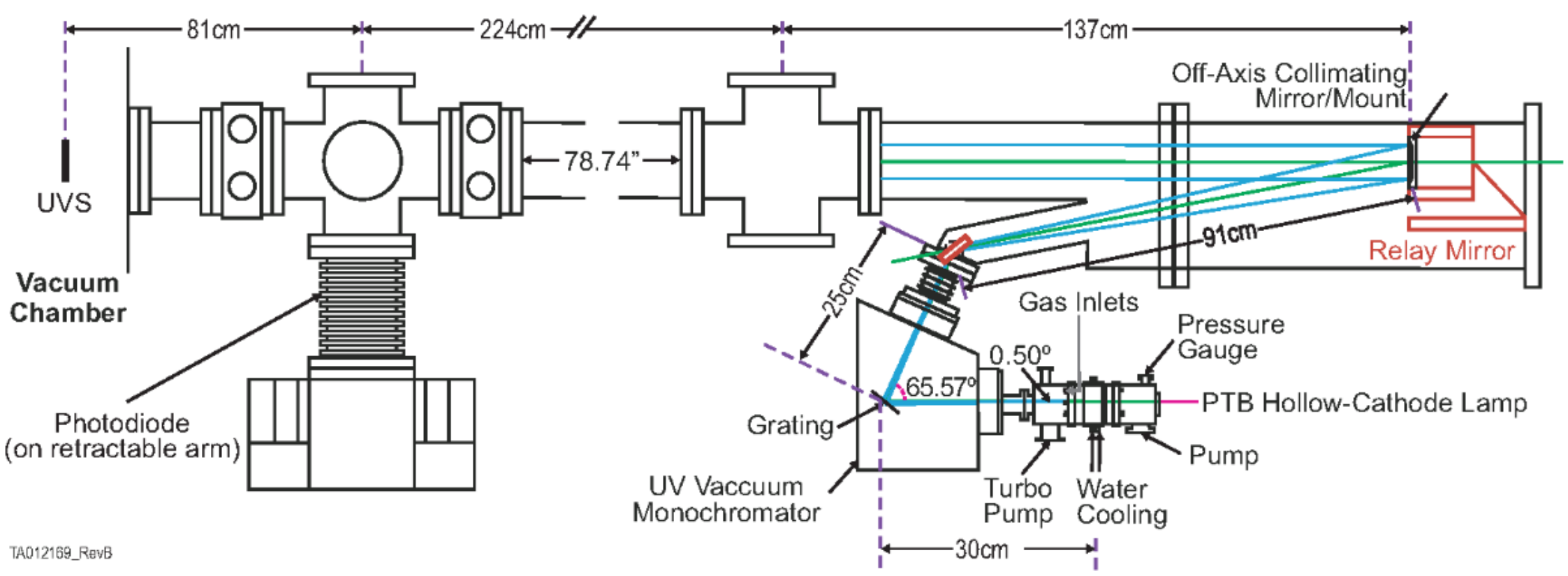

**Fig. 21** The SwRI Ultraviolet Radiometric Calibration Facility (UV-RCF). The hollow cathode lamp at bottom right emits light (much like a neon sign) that is collimated with a mirror for passage along the long tube depicted at top. UVS is located at far left and receives light from the collimated beam, which is calibrated using a retractable photodiode at bottom left

### 4.1.2 Wavelength scale

The JUICE-UVS wavelength calibration is derived from measurements of the emissions from multiple calibration gases. Images of emission-line features from these gases (Fig. 22) served to determine the spatial and spectral resolutions and the wavelength calibration. Twenty-four different lines produced by the three calibration gases form the basis of the wavelength solution. Variations in position across the slit (spectral direction) were corrected by comparing the relative position of Lyman-α and specific aluminum emission-line features present in all images (Fig. 23; ground-cal solution at y=1600). The cubic fit determines that the pixel scale varies from 0.095 nm to 0.11 nm across the detector. The cubic fit reported in Davis et al. (2020) is $\lambda = -5.46851\text{E-}8\ x^3 + 2.60886\text{E-}4\ x^2 + 0.664419\ x - 158.911$ and for easier use has a linear approximation of ~ $0.1064x – 35.29$ at y=1600, varying slightly over spatial position (Fig. 23).

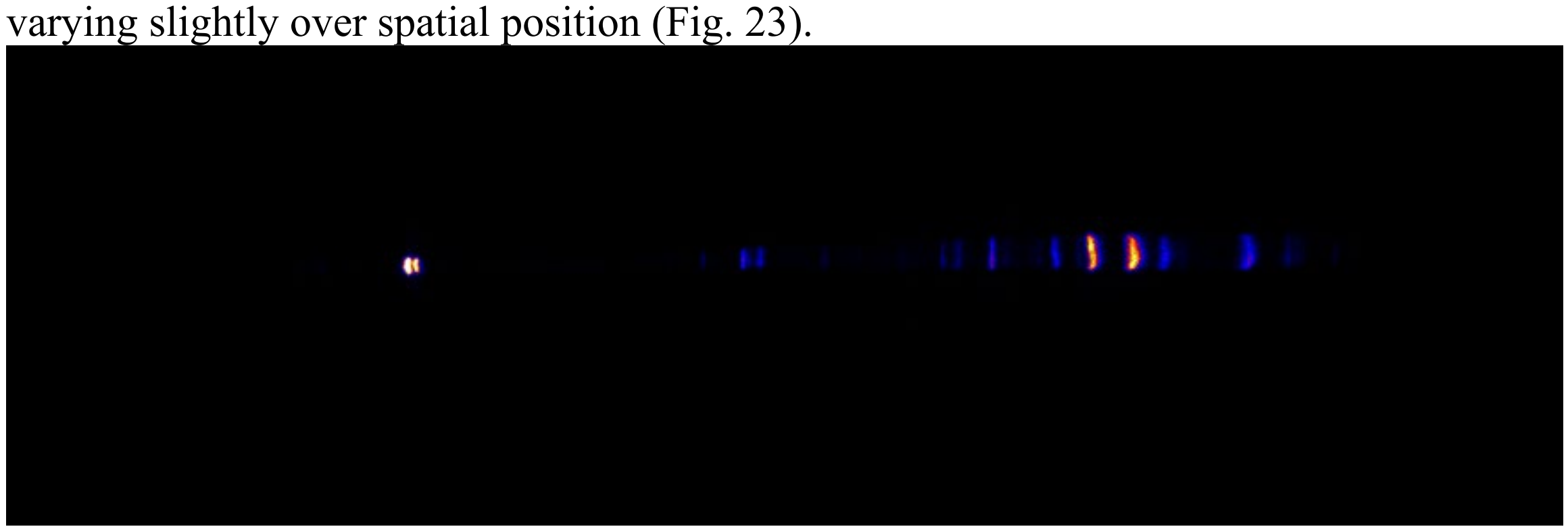

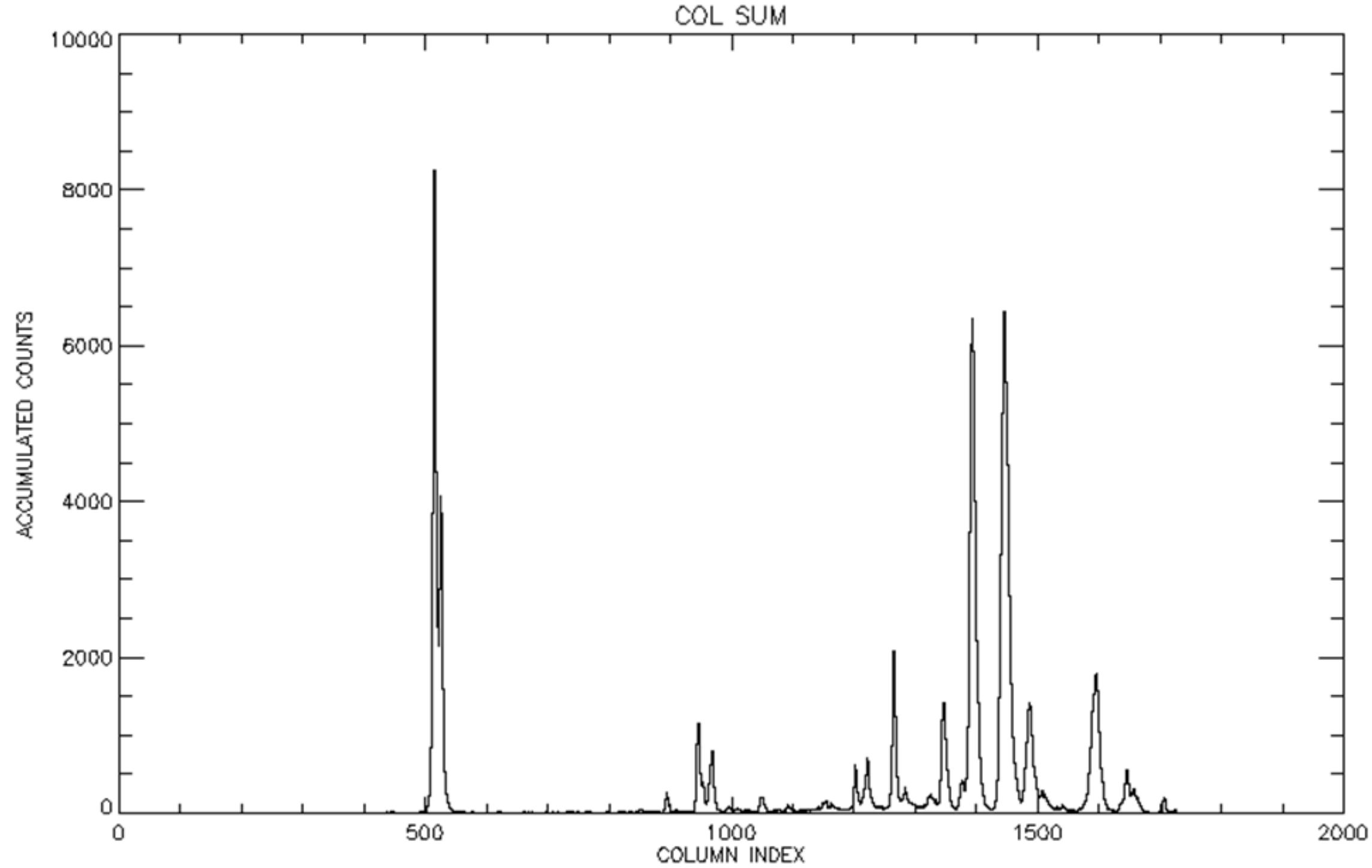

**Fig. 22** JUICE-UVS ground-based wavelength calibration (Davis et al. 2020). An example image (top) and spectral plot (bottom) of a Ne spectrum using an input collimated point source to JUICE-UVS at slit center

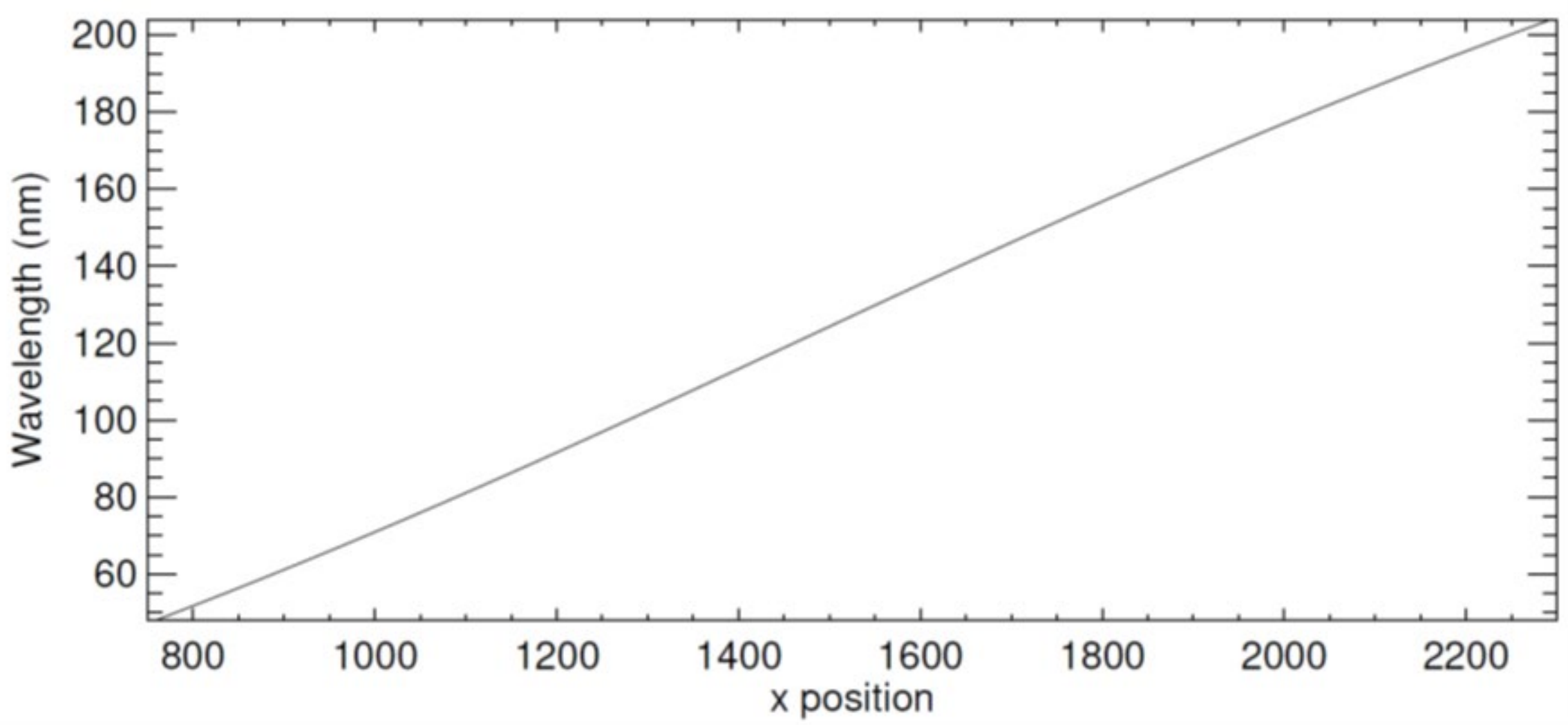

**Fig. 23** A mapping of the wavelength to detector x position is shown for an example y=1600 position near the boresight, with the thinner (and darker) straight line illustrating the deviations from a linear spectral plate scale near the left and right detector edges (bottom)

The total wavelength bandpass across the detector active area is computed with the measured plate scale and offset values. The active area covered spectral pixels 777 through 2263 in the ground-cal (but see Section 3.5 for the in-flight values 780-2222), corresponding to a total wavelength bandpass of 50–204 nm. This bandpass exceeds the instrument specification for a minimal bandpass of 55–180 nm with margin.

### 4.1.3 Effective Area

The effective area quantifies the effective collecting area of the instrument given the combined efficiencies of the optical and electronic elements. It translates the conversion efficiency of an incident flux of photons at the front end of the instrument into a detectable electronic signal at the back end of the detector. The effective area of the Airglow Port (AP) and Solar Port (SP) are given by:

$$E_{AP\ area} = AS \times R_{OAP} \times R_{Grating} \times G_{eff} \times QE \quad (1)$$

$$E_{SP\ area} = AS \times R_{SP} \times R_{OAP} \times R_{Grating} \times G_{eff} \times QE \quad (2)$$

where AS is the aperture size (16 $cm^2$ in the AP, 0.049 $cm^2$ in the SP); $R_{OAP}$ is the reflectance of the $MgF_2$-coated Al OAP mirror; $R_{Grating}$ is the grating reflectance; $G_{eff}$ is the grating efficiency; and QE is the detector quantum efficiency. The solar port has an additional bounce on a bare Au mirror and is characterized by a reflectance $R_{SP}$. The lab-measured effective areas for both the airglow port (Fig. 24) and the solar port (Fig. 25) are shown relative to defined requirements as a function of wavelength. Fig. 26 displays these quantities, as well as the resulting effective areas for the airglow and solar ports based on equations 1&2. The OAP mirror reflectance measurements were provided by Goddard Space Flight Center from 115–400 nm and are shown in dashed blue on Fig. 26a. At wavelengths shorter than 115 nm, $R_{OAP}$ is taken as the theoretical reflectance for a 25-nm-thick, $MgF_2$-coated Al mirror from IMD modeling (Windt 1998). The same coating is used on the grating, so $R_{Grating}$ is matched to $R_{OAP}$. The grating equation is used for $G_{eff}$ in Fig. 26b and set to the designed blaze angle of the optic. The as-delivered detector QE spectrum is shown in Fig. 26c. We compare the theoretical effective area with the measured one in Fig. 26d and e. We provide a fit of the ground-based measurements assuming a skewed gaussian function.

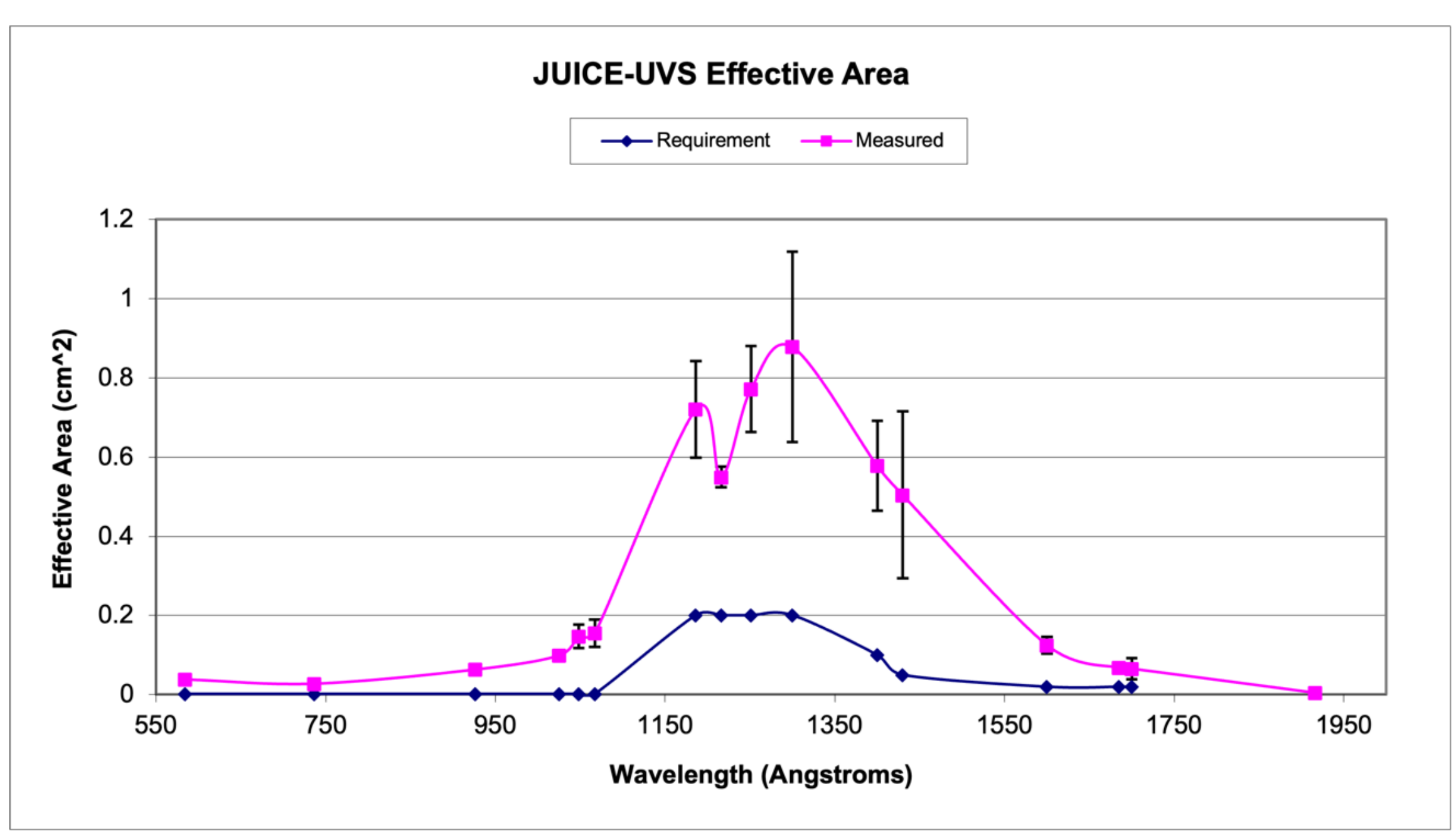

**Fig. 24** The ground-measured effective area (magenta squares) and goal (blue diamonds) for the airglow port mode plotted versus wavelength

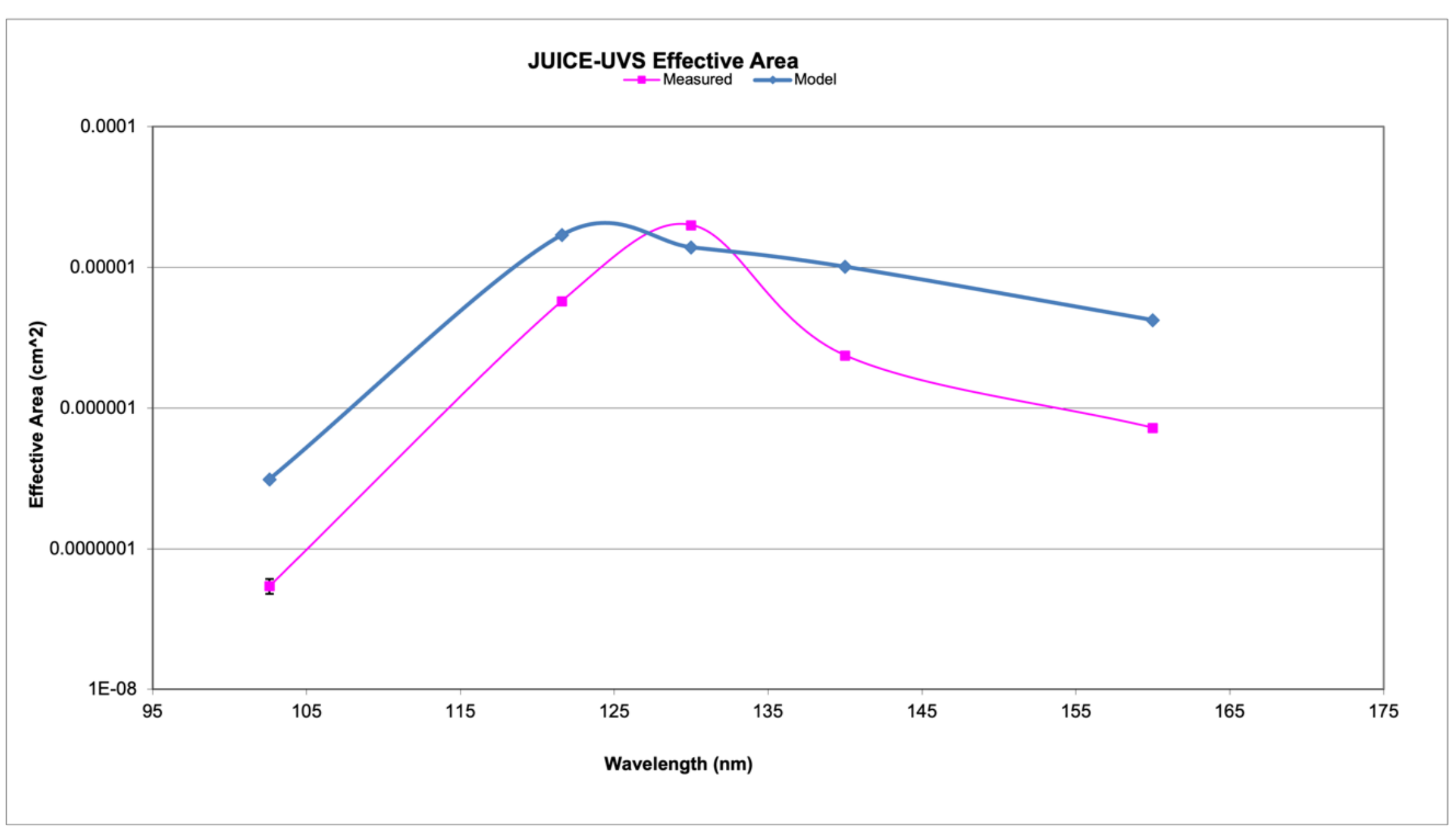

**Fig. 25** The ground-measured effective area for the solar port as a function of wavelength

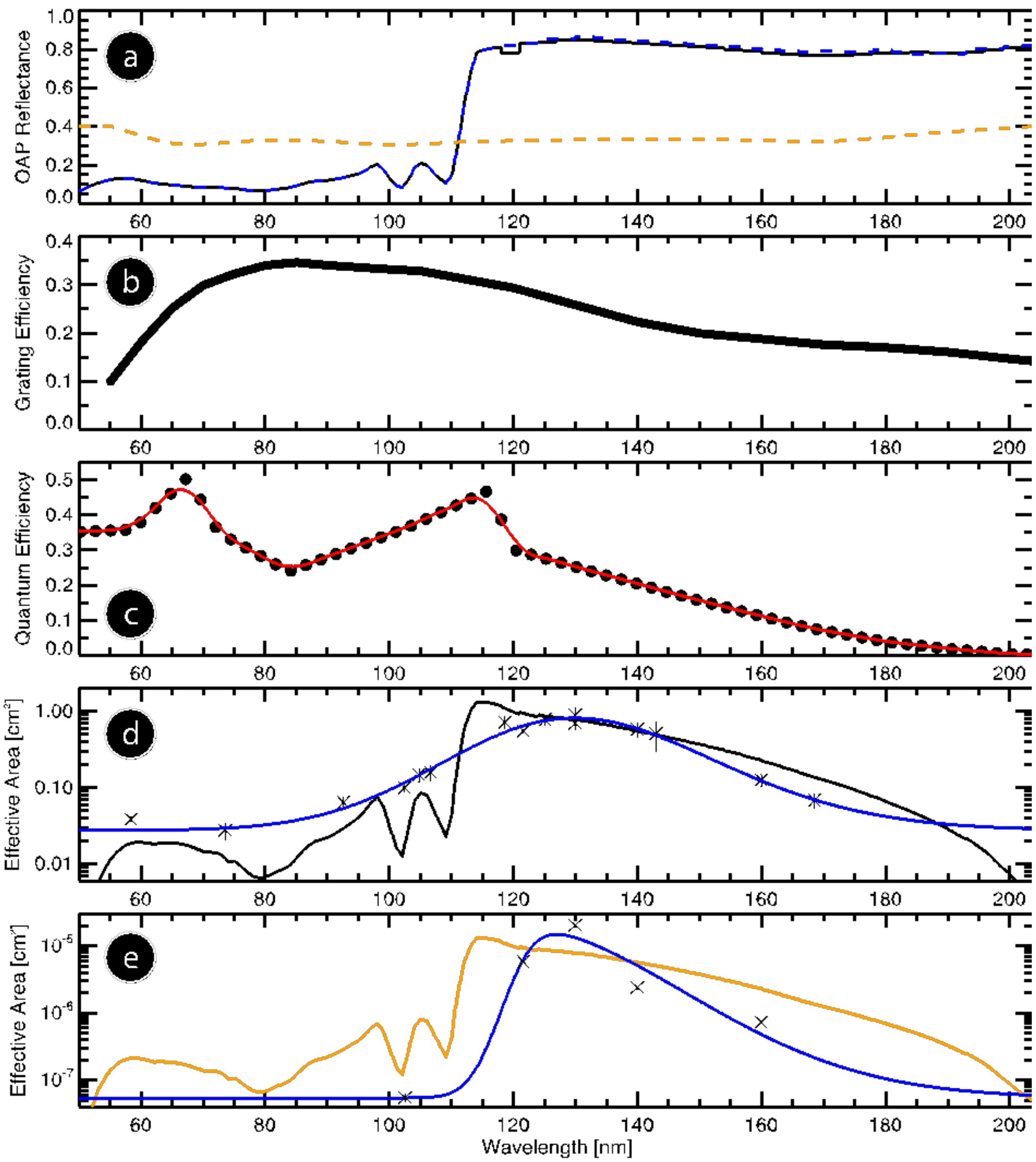

**Fig. 26** Final best estimate of pre-launch effective area performance based on an optical performance model of more detailed spectral variations in effective area scaled to match the measured points in both Fig. 24 (airglow port) and Fig. 25 (solar port). Panel a: OAP mirror reflectance (solid black), grating reflectance (dashed blue), and solar port mirror reflectance (orange). Panel b: grating efficiency. Panel c: quantum efficiency. Panel d: predicted AP effective area (black), compared with the measured one (black plusses), and a fit considering a skewed gaussian function. Panel e: predicted SP effective area (orange), compared with the measured one (black plusses), and a fit considering a skewed gaussian function. See Davis et al. (2025b) for initial assessments of post-launch effective area, which are in general agreement for the low count-rate regime

#### 4.1.4 Resolutions

Emission line spectra obtained at different spatial locations were mapped as shown in Fig. 27. These calibrations were used to assign the spatial and spectral resolutions as a function of detector location shown in Fig. 28-30.

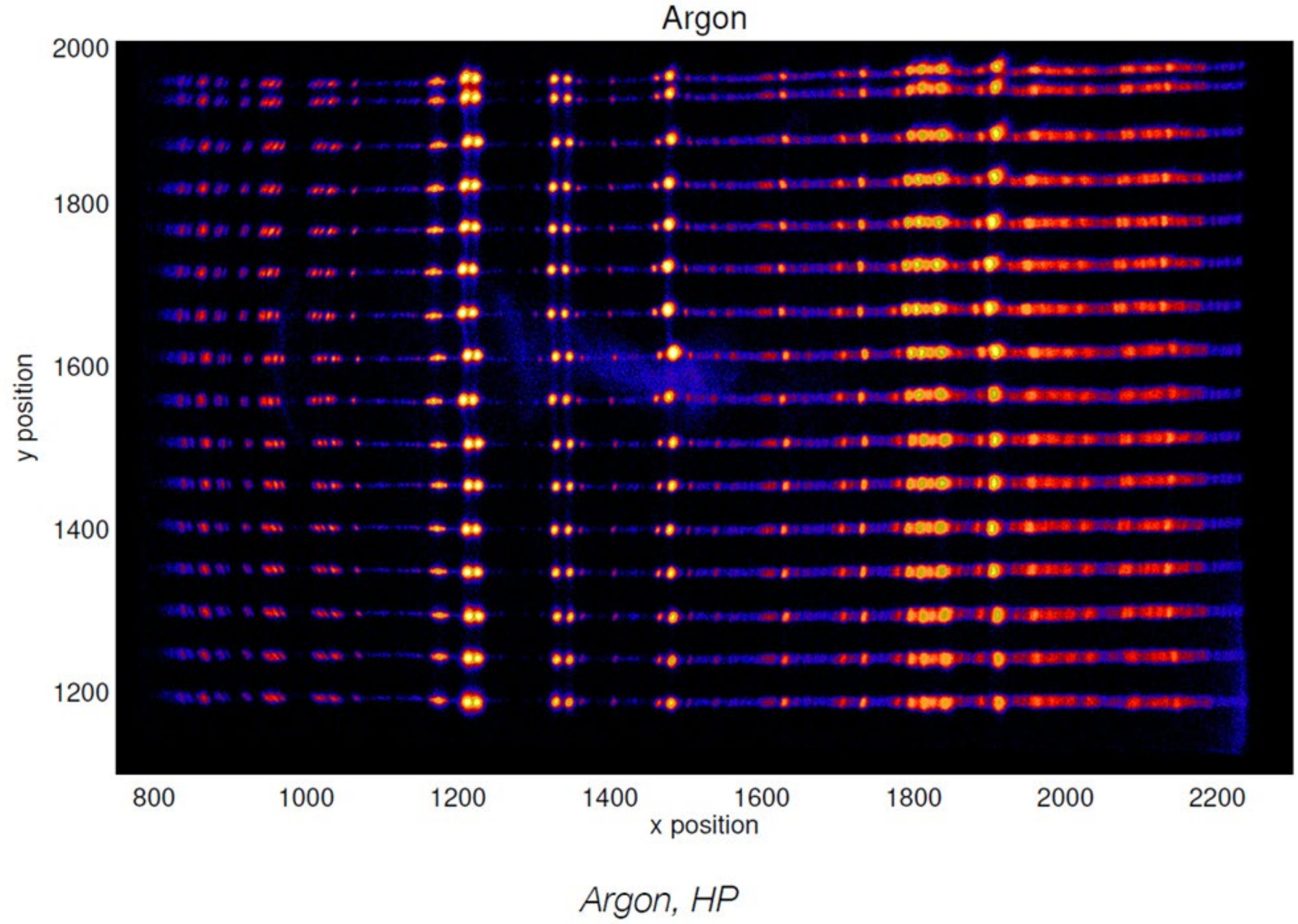

**Fig. 27** A composite of Ar emission measured at ~0.5° offset angles in y-position through the JUICE-UVS HP aperture. The derived spectral and spatial resolutions (Figs. 19 to 21) are based on these types of calibration data

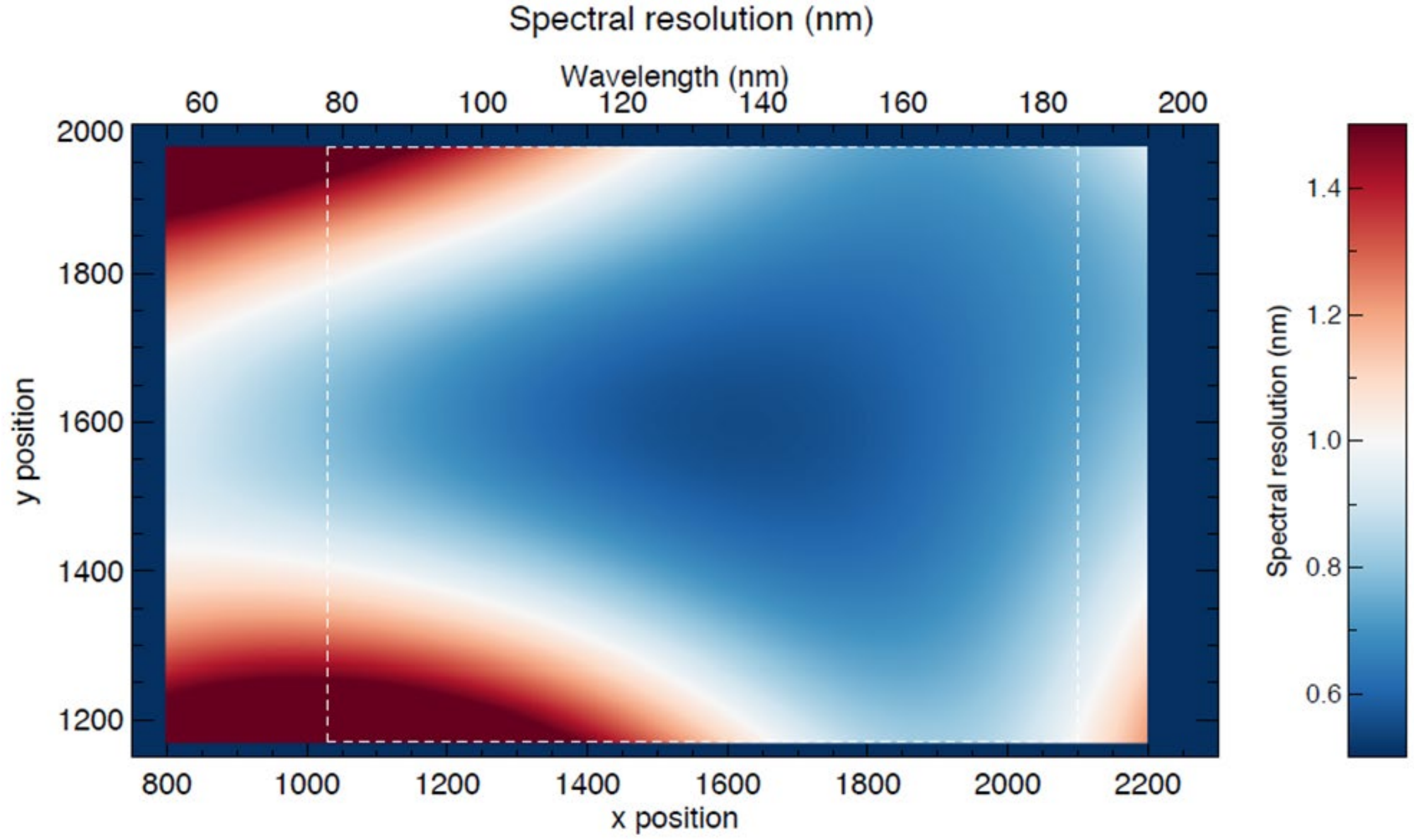

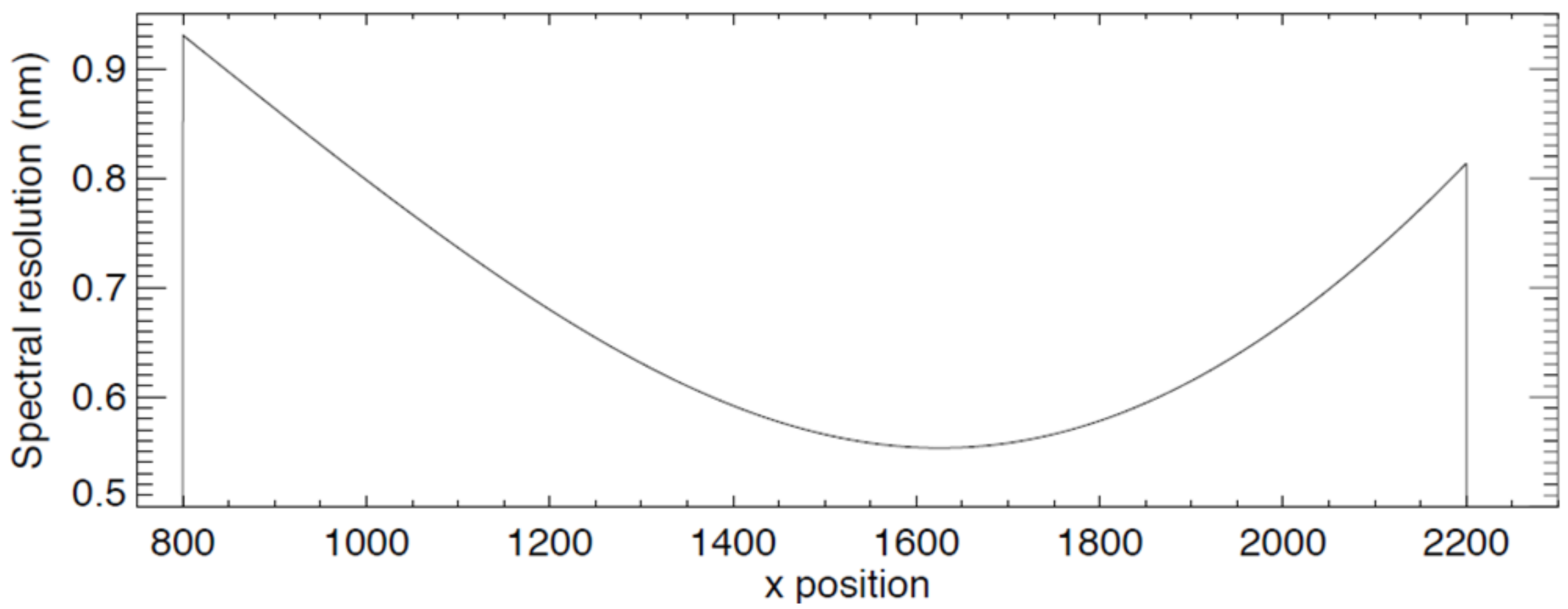

**Fig. 28** (top) A colorized plot of spectral resolution versus x and y position on the detector. Wavelength is labeled on the upper axis, and relative off-axis angle on the right axis. (bottom) All on-axis positions have better than 1–nm spectral resolution

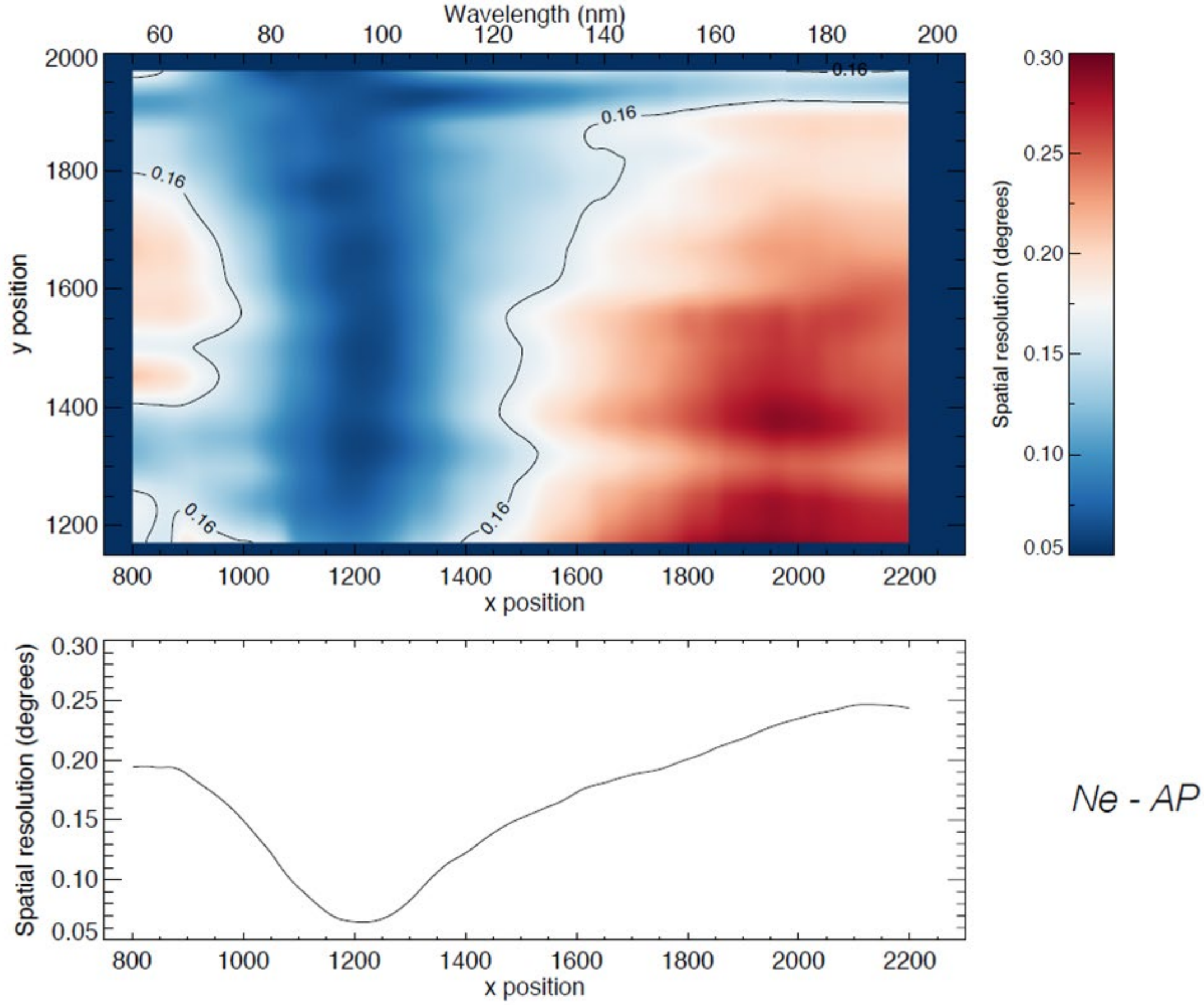

**Fig. 29** (top) The spatial PSF full-width half maximum (FWHM) as a function of wavelength and relative off-axis angle measured through the AP. (bottom) The spatial PSF FWHM near the center of the slit

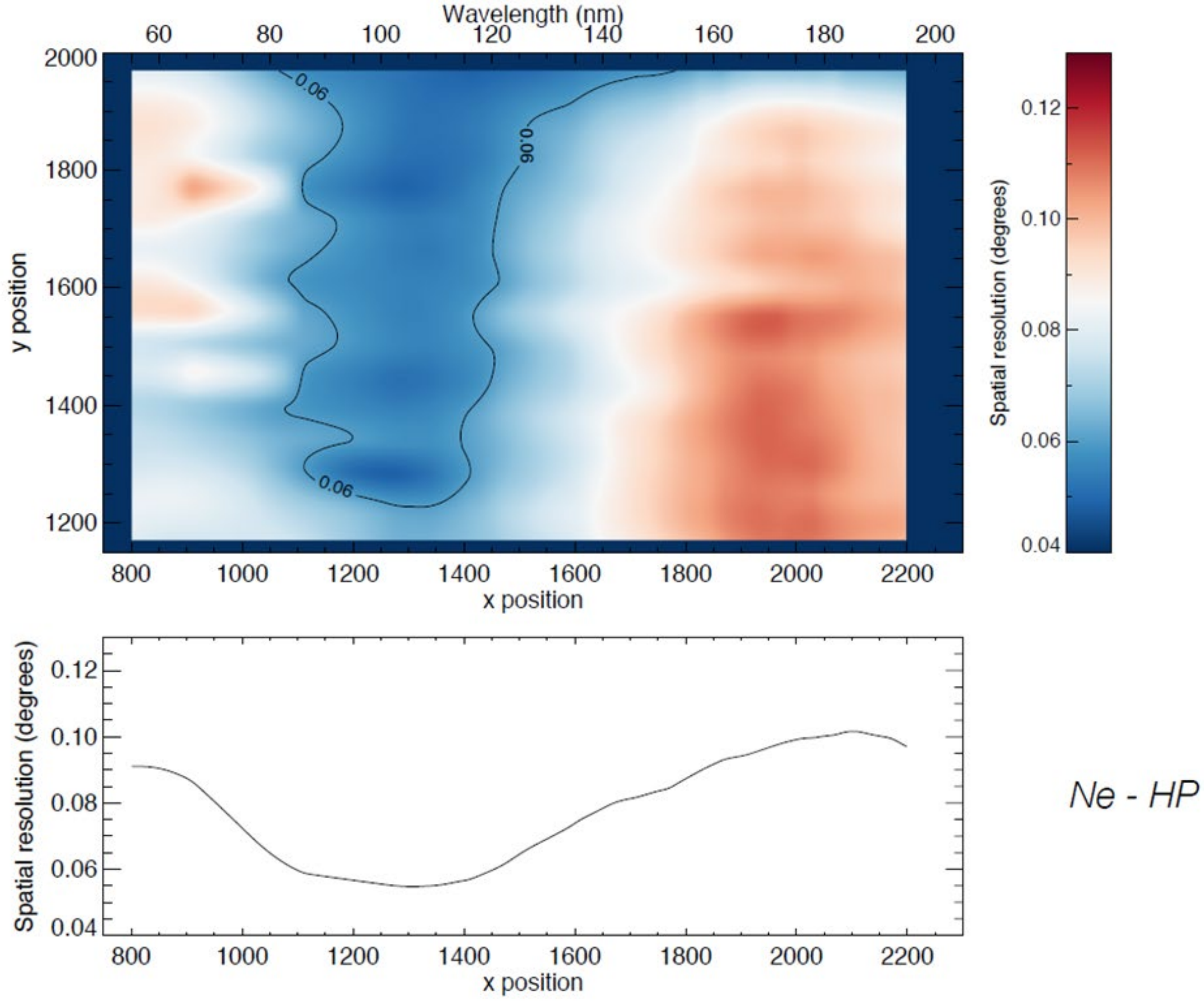

**Fig. 30** (top) The spatial PSF FWHM as a function of wavelength and relative off-axis angle measured through the HP. (bottom) HP spatial resolution versus x position near the center of the slit

### 4.1.5 Darks and Detector Linearity

A series of long dark observations were taken over multiple days and times throughout the ground-calibration campaign for a total of ~44 hours of dark acquisitions. The detector was operated at -4550V (the nominal HV level for room temperature at delivery) for these dark measurements. All chamber windows are covered with foil to minimize any stray light in the calibration chamber. Chamber pressures were well below $8 \times 10^{-6}$ Torr for all dark measurements.

Figure 31 shows the combined 44-hour dark image. The "warm" spots at both upper left and bottom right corners of the active area were also seen in a standalone detector dark test dataset, prior to instrument assembly. The overall dark rate of 24.5 counts/s measured pre-flight is lower than that measured during standalone detector testing due to a lower ambient pressure within the detector in the calibration chamber, versus the ground support equipment (GSE) pump station with a KF-16 port pumping on the otherwise sealed detector housing. The maximum accumulated counts in a single pixel were seven counts; the average pixel had no more than one count accumulated over the 44-hour exposure.

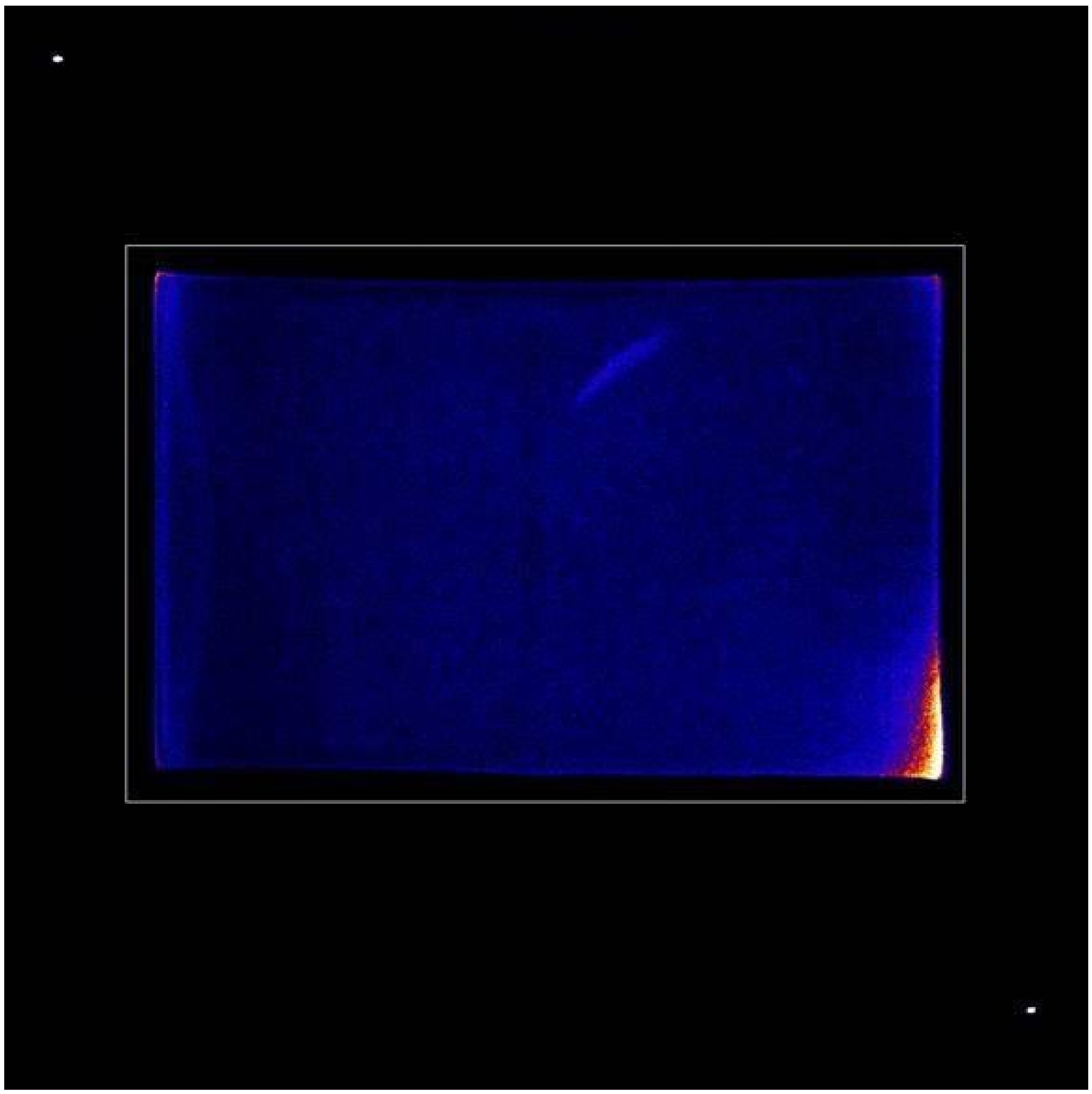

**Fig. 31** Image histogram of the dark pixel list exposures for 44 hours of accumulated exposure. The color bar represents total counts in a given pixel over the accumulated exposure

#### 4.1.6 Detector Dead-times and Gain Relationships

The measured count rates by JUICE-UVS, both analog and digital, are subject to dead-time effects, which decreases the measured signal at large measured count rates. The characteristic dead time for the digital count rate is higher due to the additional processing required by the electronics. Pre-launch measurement of the characteristic dead times allow us to correct the count rates using:

$$D_{corr} = \frac{D}{1 - D\tau_D} \quad (3),$$

where D and $D_{corr}$ are the digital and corrected digital count rate, respectively, and $\tau_D$ is the digital count rate. As Davis et al. (2021) report, $\tau_D$= 825 ns for JUICE-UVS. This results in a 10% loss of recorded events for the digital count rate at an input rate of 110 kHz.

**Global count-rate limits:** Based on the analog (raw) and digital (event) dead-time values, the reported count rates (analog) and captured science data (event) are lower than the true rates. With an analog count rate assuming a dead time of 150 ns ($A = R/(1+R\cdot\tau_{analog})$), the maximum count-rate levels off at 6.67 MHz. With a digital count rate assuming a dead time of 825 ns ($D = R/(1+R\cdot\tau_{digital})$), maximum count-rate levels off at 1.21 MHz.

**Detector "gain-sag":** Previous UVS/Alices experienced a gain degradation as a function of photon fluence when operating at high voltage in which the detector as a result of charge extraction (e.g., Grava et al. 2018 for LRO-LAMP). The ALD coating for MCP plates is designed to limit this degradation effect (McPhate et al. 2019). Owing to scheduling delays during the MCP fabrication process this ALD coating was applied to only one of the three MCP plates, making its effectiveness for JUICE-UVS more in question than for Europa-UVS and the subject of ongoing and future calibration efforts.

**Local "gain-droop":** An additional local gain degradation occurs when an especially bright point source or emission line feature draws charge from a microchannel plate pore faster than the charge can be replenished per its local RC time constant, resulting in fewer photoelectrons collected and therefore a lower measured local count rate (Ertley et al. 2017). The correction for this effect awaits analysis of in-flight calibration data.

### 4.1.7 Stray light

JUICE-UVS's stray-light rejection was characterized along both spectral and spatial axes at the UV-RCF. The test procedure involved illuminating the airglow entrance aperture with a collimated beam reflected by the OAP mirror at various off-axis angles. We performed the scattered-light analysis at Lyman-α because it is the brightest UV emission line in both the hollow cathode lamp spectrum and the solar spectrum.

The motorized stage supporting JUICE-UVS enabled us to swivel the AP view between -12° to +12° in the spectral plane (perpendicular to the slit length, while keeping the spatial axis fixed at the slit center). Similarly, the spectrograph was also tilted in the spatial plane from -11° to +11° (parallel to the slit length, while maintaining the spectral axis at slit center).

The acquired images by the MCP detector were analyzed to determine the point-source-transmittance (PST) as a function of angular offset from the JUICE-UVS airglow boresight axis. PST is the ratio of stray-light irradiance $E_{FP}$ on a MCP detector at the JUICE-UVS focal plane to source irradiance at the instrument aperture $E_{input}$. This expression (eqn. 4) can be rewritten in terms of the measured off-/on-axes count-rate ratio ($R_{off}/R_{on}$), the Lyman-α effective area ($A_{\text{eff-Ly}\alpha}$), the focal plane active area ($A_{FP}$), quantum efficiency ($QE$) of the detector at Lyman-α, and the geometric ratio of the beam size $A_b$ to the size of the entrance aperture $A_g$.

$$PST = \left(\frac{A_{eff-Ly\alpha}}{QE\ A_{FP}}\right)\left(\frac{A_b}{A_g}\right)\left(\frac{R_{off}}{R_{on}}\right) \tag{4}$$

The Lyman-α effective area and QE values used to compute the *PST* were measured during instrument radiometric and detector vendor tests ($A_{eff\text{-}Ly\text{-}\alpha}$ = 0.7 cm$^2$; $QE$= 0.4; $A_b/A_g$ = 0.25; $A_{FP}$ = 7.4 cm$^2$). With these input values, we estimate a *PST* of 0.05 at the on-axis position, aligned to the airglow boresight. The requirement of $PST < 10^{-5}$ at off-axis angles >7° was met in both axes (Fig. 32).

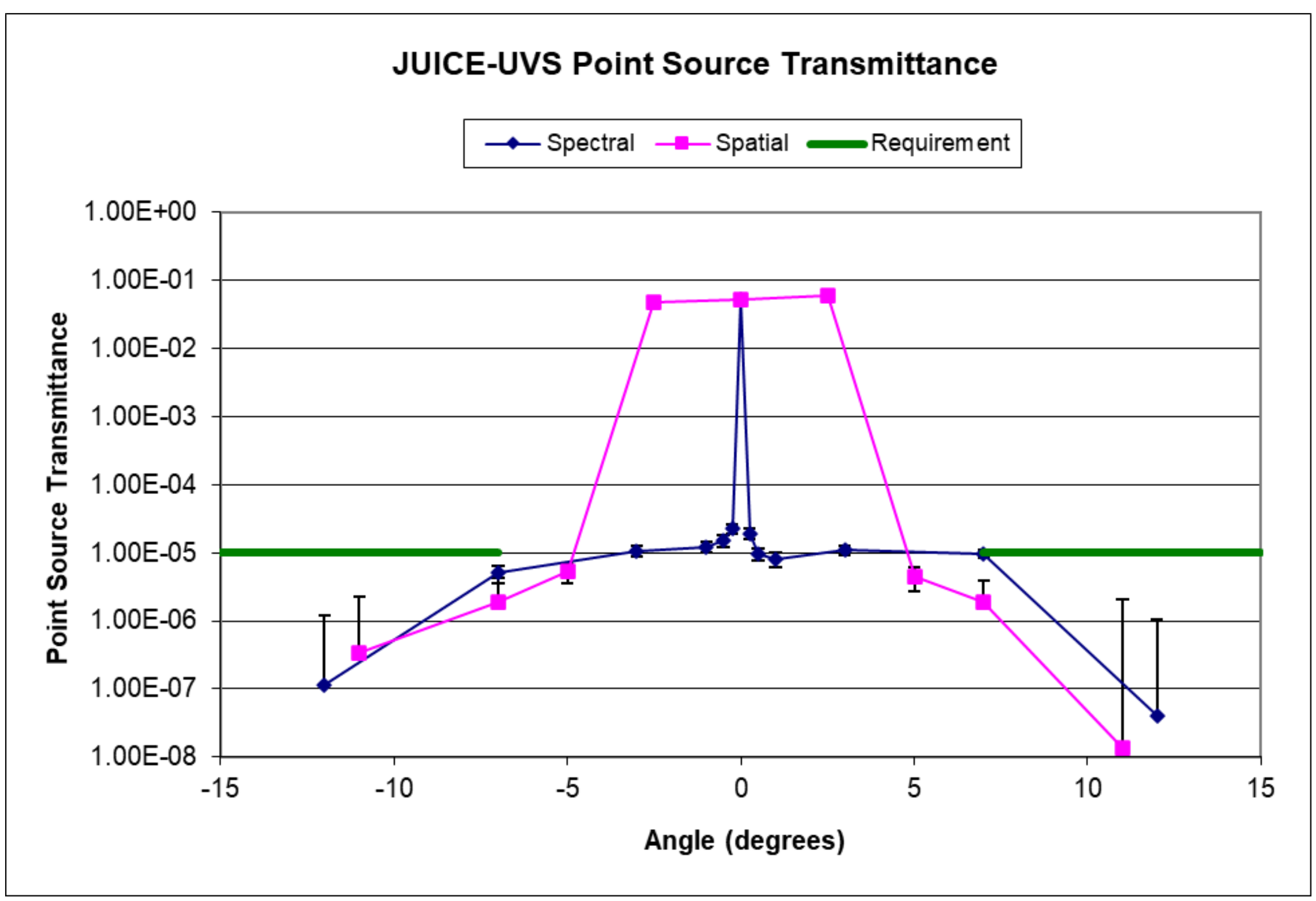

**Fig. 32** The measured stray-light rejection as a function of the input off-axis angle with respect to the boresight in both axes. Both the spectral data values (blue squares) and the spatial values (magenta squares) are shown along with the specification requirement (green line), which was achieved

### 4.2. In-Flight Calibrations and Plans

Two months after launch, in June 2023, JUICE-UVS performed a series of commissioning activities. Routine calibrations are performed regularly during the cruise phase, with more planned prior to Jupiter orbit insertion (JOI) and repetitively during the science tour phase. All of the performance properties described for the ground-calibration are reanalyzed and in almost all cases determined more precisely with the in-flight data calibrations.

#### 4.2.1. Commissioning Results

The initial commissioning of JUICE-UVS involved powering the LVPS to turn on the instrument, performing electrical functional tests, actuating the launch latches to release and open the aperture doors, followed by opening of the detector door with its one-time actuated wax pellet mechanism, ramping up the HVPS for the first time, and characterizing the detector dark signals. Once these activities were completed, JUICE-UVS was ready to perform its “first-light” observation. Since the spacecraft pointing during inner cruise is constrained thermally, the “first-light” spectrum was obtained opportunistically with sky Lyman-α emissions filling the slit. UVS then took advantage of its boresight pointing alignment with the spacecraft z-axis, riding along as the spacecraft performed a 360° roll around its high-gain antenna (which points in the spacecraft -x direction). During this roll, JUICE-UVS observed a 7.5° swath across the sky, including UV-bright stars and the Large Magellanic Cloud (Fig. 33). The data collected during the roll observation were used to better constrain the instrument pointing and to measure the effective area at wavelengths >100 nm.

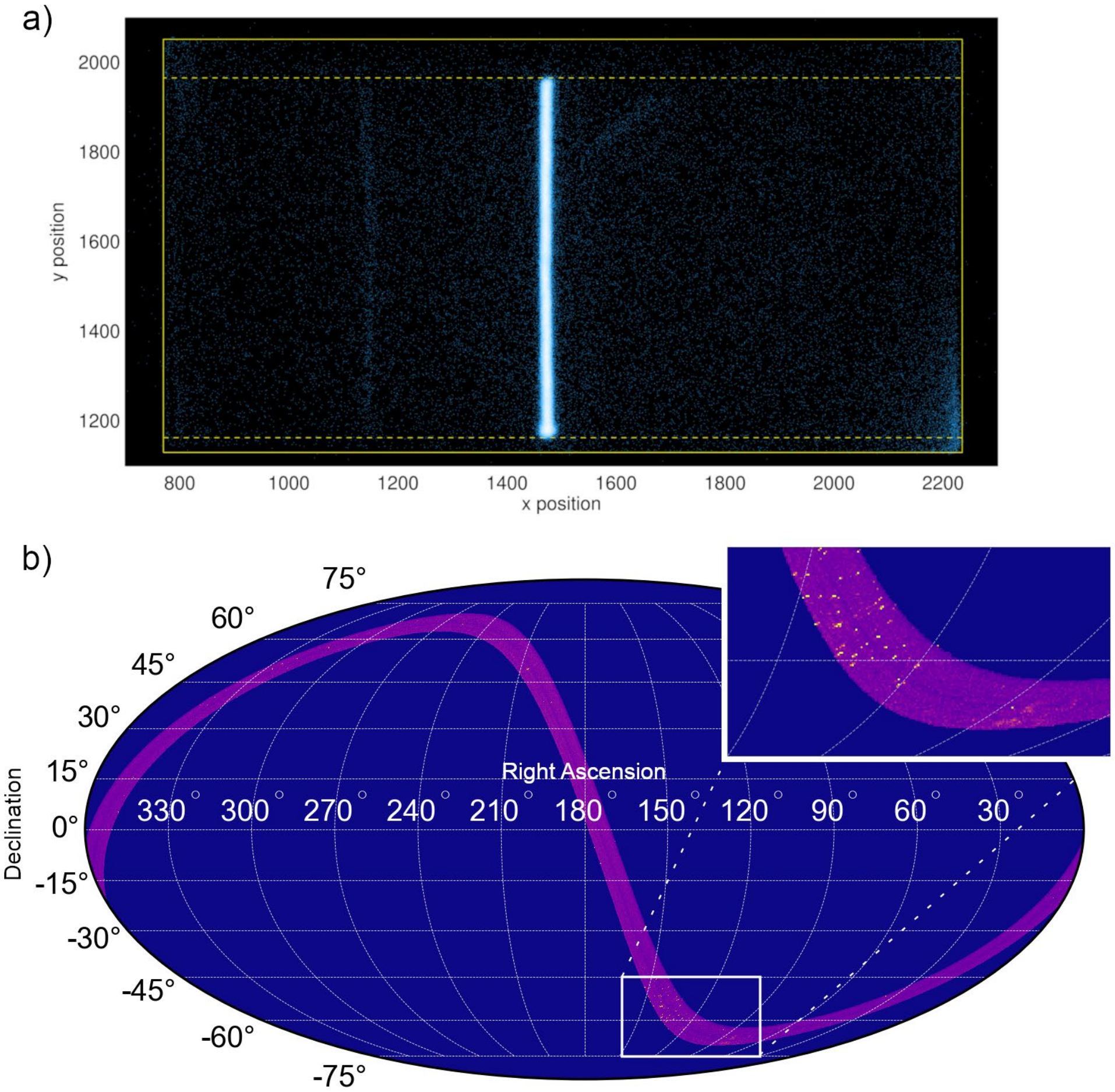

**Fig. 33** JUICE-UVS observations performed during near Earth commissioning: a) First light image of interplanetary Lyman-α emissions filling the UVS slit; b) UVS sky roll image with inset showing a region containing a concentration of UV-bright stars and the Large Magellanic Cloud

### 4.2.2. Cruise

**Payload Checkout Windows.** Since its successful commissioning in 2023, JUICE-UVS has continued to perform calibration and instrument monitoring activities during regular payload checkout windows (PCWs). There have been four payload checkout opportunities to date: PCW1 in January 2024, PCW2 in July 2024, PCW3 in April 2025, and PCW4 in February 2026. During each checkout window, UVS performs three standard activities: 1) a ~23-hour decontamination with heaters activated to remove any condensed material from the optics; 2) an engineering check to confirm instrument health and functionality; 3) a long dark exposure to monitor the detector background rate. Additional activities are then planned depending on the available data volume and calibration priorities. For example, previous checkout windows have included sky roll observations similar to that performed during commissioning using both the AP and HP ports, allowing a direct comparison of the instrument sensitivity in these two observing modes as the FOV passed over the same stars. Payload checkout periods also provide opportunities to test the on-board histogram binning functionality and to monitor the instrument response to temperature variations via measurements of the positions of the STIM pixels.

Future periods of the cruise phase with JUICE-Sun distance >1.4 AU include five opportunities for pointed checkout activities. The first of these pointing campaigns will be PCW4 in February 2026. During the first pointed checkout, JUICE-UVS will perform observations of two stars chosen to provide good SNR through the AP and HP ports, respectively. The AP observations will include: 1) an across-slit raster scan, such that the star crosses the slit multiple times, at 0.5° intervals along its 7.5° length; 2) a series of star stares at different high voltage settings (including one HP observation of the same star for a direct comparison); 3) an along-slit raster scan at 0.05° intervals across the slit. The across-slit raster and star stares will then be repeated through the HP aperture using a brighter star. These pointed observations will be used to accurately determine the position and shape of the UVS FOV on the sky and to optimize the high voltage settings for both modes. Finally, stare observations of a region of the sky without UV-bright stars will be performed in both AP and HP modes to track sensitivity to Lyman-α (121.6 nm) and He I (58.4 nm) emissions. Subsequent pointing campaigns will include similar sky and star stares to monitor any changes in instrument sensitivity, and smaller raster scans to track possible thermal shifts of the instrument relative to the spacecraft.

The JUICE-UVS solar port door cannot be opened until the JUICE-Sun distance is permanently >2 AU, in order to prevent accidental damage to the detector due to the high solar flux. Once the spacecraft reaches this point in its cruise, the SP calibration (pointing, resolution, throughput, and angular extent) will be performed in the next pointing campaign window.

**Gravity Assists.** In August 2024, JUICE performed a Lunar Earth Gravity Assist (LEGA), with a lunar closest approach altitude of 752 km on August 19th and an Earth closest approach altitude of 6836 km on August 20th. JUICE-UVS observed both bodies, detecting sunlight reflected by the lunar surface and UV airglow emissions at the Earth. These filled slit observations provide complementary information to the point source star observations performed during payload checkout periods. They allow the first measurement of the filled slit spectral resolution at wavelengths other than Lyman-α, and of the instrument effective area at wavelengths <91.2 nm where stellar flux is low due to absorption by the interstellar medium. The Earth airglow observations included emission lines of H, He, O, N, and $N_2$ at wavelengths ranging from 58.4 nm to 194.7 nm, contributing to an improved wavelength calibration (Fig. 34a). The final UVS observation during the LEGA period was another ride-along with a spacecraft roll, during which the Earth's extended hydrogen geocorona was imaged (Fig. 34b).

A Venus flyby in August 2025 was successful, but given the high temperatures the payload activities were restricted and no UVS observations were obtained. Two more Earth flybys in September 2026 and January 2029 JUICE are planned for cruise before heading to Jupiter. Further Earth observations will be useful both for calibration checks and for the development of data processing pipelines, since the observations of planetary targets performed during gravity assists are similar to observations UVS will perform of Jupiter and its moons during the science phase of the mission.

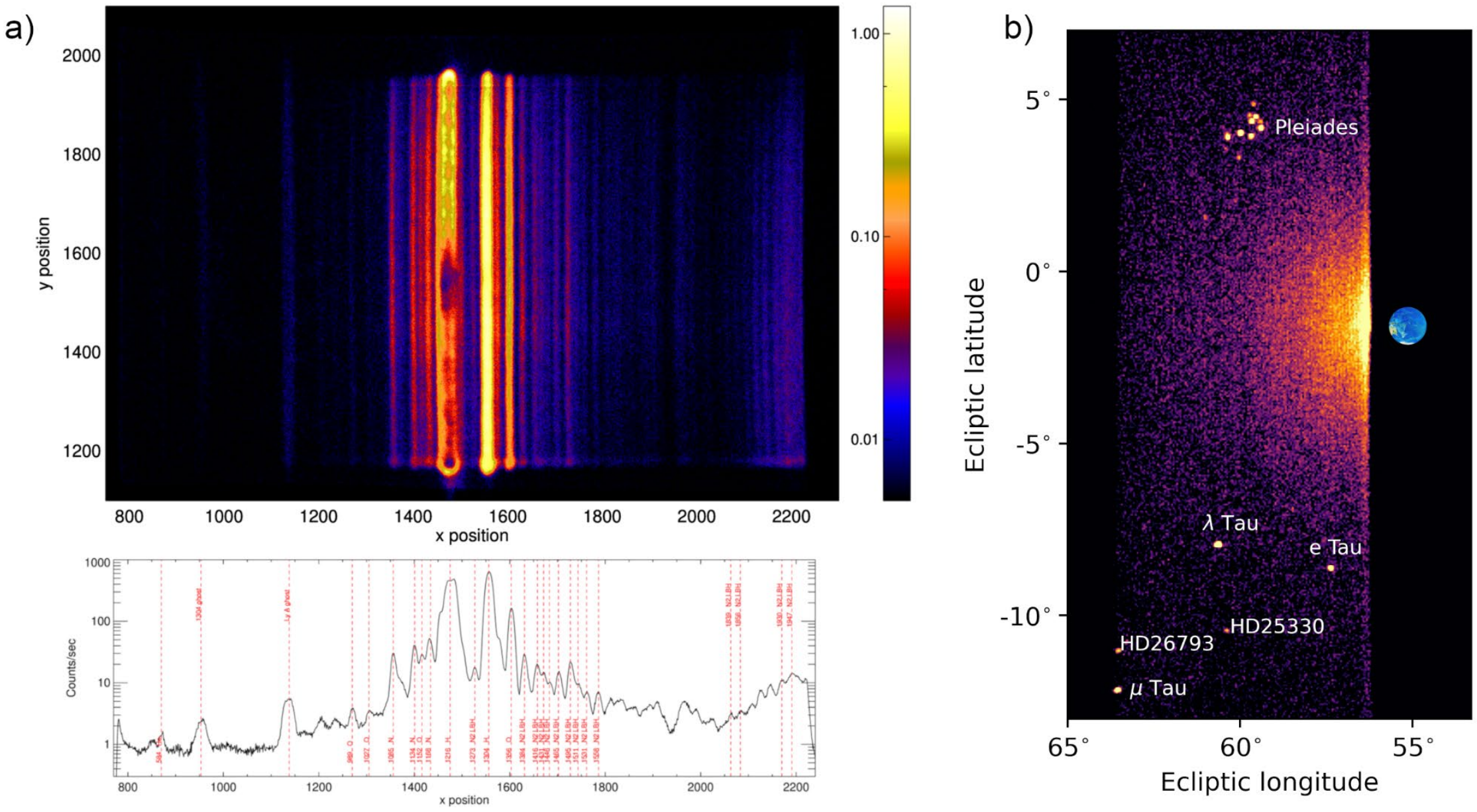

**Fig. 34.** Selected JUICE-UVS observations from the LEGA period: a) Histogram image showing Earth airglow emissions filling the UVS slit. The lower panel is a spectrum produced by summing the image over the y (spatial) dimension. b) Part of the Earth's hydrogen geocorona along with nearby UV-bright stars (the Earth was outside of the UVS FOV and is represented by an animation scaled to the correct size and position, and rotated to the correct sub-observer longitude and latitude)

### 4.2.3. Pre-JOI

Pre-JOI campaigns observe the EUV/FUV emission from the Io plasma torus simultaneously with the FUV emissions from Jupiter's aurora to improve the calibration at EUV wavelengths, which cannot be calibrated with stars due to interstellar absorption <91.2 nm. These campaigns begin 6 months before JOI, and in addition to their importance for calibration they have scientific value, extending studies of the torus variability performed during the Jupiter tour (Section 3.7.8).

### 4.2.4. Jupiter Tour

Throughout the tour phase of the mission, most of the in-flight calibrations that occurred during cruise are repeated; some occur repeatedly with each orbit of Jupiter, while others are conducted on a yearly basis. Starting at 6 months prior to JOI several campaigns of Jupiter pointed Io Torus observations will consist of consecutive acquisitions for week long periods, tailored to available downlink. These observations will serve as additional EUV calibration.

During each orbit, a Dark Observation near apojove is planned. This observation is similar to the Dark Observations conducted during cruise, but shorter, at ~1 hour. A Radiation Observation (Section 3.7.9) is conducted once or more per flyby during encounters to assess radiation noise backgrounds.

The sensitivity of the detector is tracked, in part, by observing the overall brightness of the stars used for the stellar occultation observations (Section 3.7.3). The frequency of stellar and solar occultation event observations is expected to be sufficiently high during Jupiter tour and during Ganymede Elliptical Orbit (GEO) to allow these observations to provide sufficient calibration. UVS may also conduct shorter, 10-minute star stare nods in both AP and HP mode for routine calibrations near apoapsis as needed. Sun stares through the SP are conducted in pixel list mode approximately yearly, and in histogram mode quarterly to help calibrate spectral albedo datasets, provided the time and resources are available to the spacecraft.

#### 4.2.5. Ganymede Orbit

During Ganymede orbit, the radiation environment background signals will be indistinguishable from "normal" detector dark counts, so the sum will be measured for subtraction from Ganymede observations with a frequency of once per week and a duration of ~10-30 min. Once in low-altitude circular orbit phases, inertially pointed stellar observations are largely precluded by nadir pointing and yaw-steering requirements for power management, and SP solar occultations are enabled only after the initial non-eclipse period ends.

## 5. Instrument Operations Planning

The JUICE project science planning process is described in Witasse et al. 2026 and Altobelli et al. 2026, with the JUICE spacecraft system design detailed in Sarri et al. 2026. Since JUICE is a complex mission with multiple science targets, the first step of strategic science planning involves segmentation of the trajectory by the Working Groups to define which periods are allocated to each target, and to prioritize specific science goals. For example, WG4 segments focused on Jupiter science may be further segmented to prioritize studies of Jupiter's atmospheric chemistry, zonal wind mapping, auroral morphology studies, etc. Once the science priority of each segment is defined, the instrument teams design the optimum timeline of observations to address that priority, including pointing requirements for each instrument. UVS maintains a database of standard observations for planning purposes (Table 5). These observations, however, assume a default set of LUTs-defining parameters such as the spatial, spectral, and sampling time resolution for the observations that the JUICE-UVS team can choose to update. For example, the spatial resolution and time sampling may be varied based on target size and flyby speed to optimize the required data volume. An online planning tool developed for JUICE-UVS ingests JUICE's observation timeline for each orbit, placing those activities on an interactive timeline for edits to these preset LUT and related parameters as needed. Along with basic geometric information such as altitude and phase angle, the JUICE-UVS online tool provides a sequence of images to show the overall perspective of the spacecraft as it views its current target. It also calculates in real-time the anticipated data volume produced by the sequence of activities.

Requested edits to the flyby sequence investigated within the webtool can include adjusting the timing and length of the observations. Each defined activity can also be directly edited to change the LUTs to optimize the detector parameters for each observation. Editing the activity itself is done in a separate, connected webtool that can implement interactive decisions for the spectral and spatial dimensions. This observation-card designing tool can then be used to estimate the expected SNR across the detector, accounting for background radiation and other parameters.

As described in Section 3.7.3, the stellar occultation prioritization scheme we plan is semi-automated, with available stellar occultation events given an initial adjectival priority ranking automatically. The JUICE-UVS team may increase or decrease the relative ranking of schedulable events based on these example criteria for icy moon occultations as first reported in Velez et al. (2024): 1) Fills global location and time coverage needs; 2) Passes through candidate plume locations (for Europa) or other regions of particular interest; 3) Previously observed with JUICE-UVS; 4) Has IUE or other known spectra, not only models; 5) Low flux variability and/or high accuracy of parameters such as astrometry; 6) Emphasizes extreme-UV (< 115 nm) capability and sensitivity to additional species; 7) Pair of events, two cords at once (lines connecting ingress and egress for astrometric calculations), or proximity to another UV-bright star; 8) Times when background radiation levels near the spacecraft are low; 9) Follows the JUICE trajectory line of sight for best in situ (PEP + RPWI + J-MAG) comparisons; 10) Lines up with radar (RIME) ground-track for best astrometry correlation; 11) Coincides with images (JANUS + MAJIS) of high phase limb plume search locations; 12) Relatively less auroral brightness, impacting cases with borderline SNR; 13) Near-simultaneity with Europa Clipper encounters; 14) Near-simultaneity with Europa-UVS occultation observations; and 15) Avoids scheduling conflicts with other key observations or mission events. Stellar occultations of Jupiter will be ranked using a similar set of criteria with some minor differences. For example, occultations of Jupiter's auroral region are prioritized to constrain the unique chemistry at high latitudes, despite the additional complexity of data analysis due to the presence of the bright and variable auroral emissions.

Once all activities for the period of interest have been investigated by the JUICE-UVS team and edited as needed, the tool can be used to download a csv file containing a list of all planned observations, including start and end times and the LUT definition used. The format of this file is designed to be compatible with the software used to generate the command load for JUICE-UVS, facilitating a smooth transition between the science planning and instrument operations.

## 6. Data Processing and Products

### 6.1. Instrument Science Operations Center (iSOC)

The JUICE-UVS instrument Science Operations Center (iSOC) produces data products (raw and calibrated) that are delivered to ESA's Planetary Science Archive (PSA). PSA then sends the data to NASA's Planetary Data System (PDS) for a mirrored distribution to the public and archiving. The JUICE-UVS iSOC pipeline consists of two sequential elements, named "Lima" and "Mike". First, "Lima" converts the telemetry data files into raw data products, and then "Mike" calibrates and spatially locates the data contained in the raw data products, resulting in the calibrated data products. "Lima" is also referred to as the "tm2raw" process and "Mike" the "raw2cal" process.

A JUICE-UVS iSOC pipeline "executive" program periodically checks for newly delivered data files. The periodicity of this check can be set based on operational needs, e.g., as the cadence of orbits increases. Once the received files are cataloged, the "executive" program initiates the JUICE-UVS SOC data processing pipeline. The first element of the pipeline ("Lima") converts the raw data files into raw data products (PDS4 Raw). These products represent the lowest processing level of the JUICE-UVS data and are delivered to the PSA. No calibrations are applied to the science data at this stage. Raw telemetry values are converted to engineering units where applicable; however, both raw and converted values are included in the raw data product. Multiple versions of the output raw products are made available if software bugs affecting the output data are uncovered and corrected. In the event of an error whose correction alters released data, the data are reprocessed by the revised software and then delivered to the PSA. The second element of the pipeline ("Mike") calibrates and spatially locates the data contained in the raw data products, resulting in the calibrated data products (PDS4 Calibrated). The calibrated files are similarly delivered to the PSA.

### 6.2. Lower-level items and formats

The JUICE-UVS PDS4 raw data product (created from instrument telemetry packets) contains data from histogram and pixel list science packets, as well as count-rate, housekeeping, event, and memory-dump packets. One file is produced per instrument turn-on. The raw data products are stored using the Flexible Image Transport System (FITS) file format, useful for multi-dimensional array table based formats (i.e., one "extension" per table within the same file). A JUICE-UVS PDS4 raw data file includes multiple image extensions for science data and binary table extensions for other raw packet information. Retherford et al. (2024) lists the FITS file structure expected for raw data products, which is similar for both Europa-UVS and JUICE-UVS. See the software interface specification document in the Planetary Data System for the detailed definitions for each calibrated data product extension, currently under development.

The calibrated data product uses a similar FITS file structure. One FITS file is produced for each acquisition recorded in the raw product FITS file. Relevant geometry information is also included in the calibrated data products, including several coordinate frames such as planetographic latitude and longitude, sky right-ascension and declination, moon-nearest-point altitudes, and so on. Calibrated product FITS extensions include:

**Count/second Histogram Image**: This image summarizes the counts/second data for the acquisition for both histogram and pixellist modes.

**Spectral vs Time Image**: This image summarizes the spectral data over time for the acquisition for both histogram and pixellist modes.

**Acquisition List**: This table summarizes metadata about the acquisition. Each acquisition may use distinct configuration items such as LUTs and masking definitions, which are listed here.

**Wavelength Map**: This map defines the central wavelength for each pixel defined by the acquisition's lookup table (LUT).

**Delta Wavelength Map**: This map defines the delta wavelength from the right side of each LUT pixel to the left side.

**Effective Y Map**: This map defines the effective Y position of each LUT pixel.

**Delta Effective Y Map**: This map defines the delta effective Y position of each LUT pixel, from the top edge of the pixel to the bottom edge.

**Effective Area**: This map defines the effective area of each LUT pixel of the acquisition.

**Frame List**: This dataset contains a list of the generated frames. The packets of each acquisition are organized into one or more transfer frames for downlink. This extension summarizes information at the frame level.

**Engineering Count-Rate Packet Data**: This dataset contains deadtime corrected countrate data. The counts have been converted to frequency based on the sample period.

**Count-Rate Band Definitions**: The count rates are integrated over bands (ranges of pixel columns). For each band definition this extension lists the time the definition becomes active along with the high and low column numbers that define the band.

**Engineering Housekeeping Data**: This dataset contains the complete housekeeping dataset in calibrated engineering units, where applicable.

**Stim Observation Data**: This table summarizes information on the positions of the stim pixels during the acquisition. Data are collected into time bins for each row, nominally 30 second bins.

**Acquisition Data**: For pixel list acquisitions the final FITS file extension is a table of information calculated for each detected photon. Histogram acquisitions are calibrated from counts to counts/s/nm/sr/cm$^2$. For each pixellist acquisition a table is created to list the recorded photons. For each histogram acquisition two image extensions are created, one for the array-totaled histogram image and one for the pulse height distribution during the integration time interval. These include the primary science information for the data user.

### 6.3. Higher-level items

Each of the science activities described in Section 3.7 and Table 5 contributes to higher-level data products. Aurora and airglow spectral images and spectrograms are produced from the dedicated stares and scans for both Jupiter and the Galilean satellites (Fig. 4). From these calibrated data products, two types of maps are produced: (1) UV auroral-brightness maps at a few wavelengths of interest; and (2) surface-brightness maps at a few wavelengths of interest. These maps can be made from both dayside and nightside observations of the primary targets for UVS (Ganymede, Callisto, Europa, Jupiter, and Io).

The observed morphology and brightness of the UV emissions in the auroral-brightness maps, and how they change with time, are connected to the properties of the magnetospheric environment and can be used to constrain how the satellites interact with the local plasma environment (e.g., Roth et al. 2016, Greathouse et al. 2022). Atmospheric abundances are determined from the auroral-brightness maps, with knowledge of the satellite's position in Jupiter System III coordinates and/or in situ electron density and temperature measurements by other JUICE or Europa Clipper instruments. Changes in the relative abundances of the atmospheric constituents can be used to trace how the tenuous atmospheres responds to daily temperature variations and should reflect sporadic events including atmospheric variations.

Jupiter auroral-brightness maps provide similar atmospheric abundance constraints, and benefit from similar products currently produced for Juno-UVS. These Juno products are often reported with magnetic field locations and other relevant fields and particles data mapped on the images (e.g., Giles et al. 2025, Hue et al. 2021a) and we anticipate similar usage of UVS Jupiter datasets, albeit at lower spatial resolutions (Fig. 7).

Surface-brightness maps enable assessment of properties across each satellite's surface, including grain size and composition. Near-global surface-ratio maps of UV off-band (155-190 nm) to on-band (130-155 nm) albedos, similar to those produced by LAMP (e.g., Byron et al. 2021, Magaña et al. 2022), can be used to map relative surface maturity. Maps of the spatial distribution of Lyman-α (121.6 nm) reflectance, which currently appears to be roughly inverted in brightness from visible maps for Europa (McGrath et al. 2009, Roth et al. 2014a) and Ganymede (Alday et al. 2017), can be compared with global and regional images from JANUS and MAJIS to further explore surface composition, maturity, and other regolith properties in identified geologic regions (Stephan et al. 2021). Comparative maps near the 165-nm water ice spectral edge (e.g., on-band, edge, and off-band as defined in Hendrix et al. 2019) can also be used to identify fresh deposits of water that would stand out against the rest of the irradiated, UV-darkened surface (e.g., Becker et al. 2018, Raut et al. 2023). Spectral-albedo image cubes at 2–10-nm resolution and spectral-slope map products are also envisioned. As with LAMP observations, nightside surface-reflectance maps could be made using ambient starlight, interplanetary medium skyglow, and Jupitershine, which could be most useful if there are permanently shadowed regions found near Europa's poles.

To constrain surface composition using the surface reflectance spectra, continuing laboratory studies of the far-UV reflectance of water ice, mixtures, and irradiated samples are planned. The low albedo of Europa's surface reported by Becker et al. (2018) using HST observations presents an interesting problem since water ice-rich surfaces are expected to be bright longward of the water ice absorption edge at 165 nm (e.g., Hendrix et al. 2010). Could the presence of trace constituents like $CO_2$, $SO_2$ and their irradiated byproducts serve to darken the surfaces of Juptier's icy satellites? Planned experiments will characterize the FUV darkening of Ganymede-ice analogs processed by energetic irradiation. Similar studies previously constrained the single scattering albedo and scattering anisotropy in light reflected by the canonical Apollo 10084 soil and lunar simulants, as demonstrated in Raut et al. (2018b) and Gimar et al. (2022). These results were valuable in the interpretation of LRO-LAMP datasets.

Raw stellar occultation profiles of signal vs. time are converted to transmission profiles as a function of altitude above the surface for another type of higher level product (Figs. 5&6). Each transmission profile provides a high-resolution probe of the relative amount of absorption by the slice of the atmosphere that occulted the star. Compositional abundances can then be determined from these transmission profiles, and the overall structure of the atmosphere can be explored by assessing these transmission profiles at different altitudes above the surface. Because the stellar occultations are expected to occur over the duration of the mission and through different regions of the atmospheres of five primary objects, comparisons between the profiles serve to explore local variability and long-term changes in the composition and structure of each atmosphere.

Jupiter transit observations result in a global view of the Galilean moons while backlit by Jupiter. From these raw observations, atmospheric-absorption profiles at Lyman-α (121.6 nm) are derived. The absorption profiles from this silhouette are then used to observe the extent of the satellite atmosphere above its limb to characterize its structure and search for current activity such as plumes. Lyman-α is the brightest wavelength reflected from Jupiter and therefore the wavelength with the highest SNR, better enabling the detection of absorptions by the satellite's atmosphere and plumes (if present).

The Io torus and neutral cloud stares each result in a number of spectral images of the torus and neutral clouds that may exist near both Io's (Fig. 19) and Europa's orbit. Emissions detected within the acquired spectral images are used to produce a higher-level data product of compositional abundances and the overall extent of the torus and/or neutral clouds. These data are reflective of the rate of loss of material from Io and the icy satellites into the local space environment, either via sputtering of the surface, ionization of the atmosphere, or possibly directly from ongoing activity such as plumes.

## 6.4. Quick-look and Collaborative Items

To facilitate quality checks and provide information for providing tactical changes to observation plans several data products are generated quickly upon receipt of the data at the iSOC. These "quick-look products" are immediately useful for collaborative science purposes, with the caveat that more accurate or detailed higher-level products more suitable for science publications will eventually be available. The lower-level count-rate product list, raw count-rate packet data, and its corresponding count-rate band definition (see Section 6.2) are a primary tool for validating the performance of JUICE-UVS during acquisitions at glance. These three higher level quick-look products will be of more interest to a general scientist: 1) preliminary full-disk maps and pushbroom observations, 2) preliminary surface brightness maps, and 3) transmission profiles with wavelength for candidate plume search locations. Lastly, a map of surface composition for features of interest will prove useful for collaborative overlays with the other remote sensing datasets, and are iteratively improved throughout the mission.

### 6.5. Housekeeping/Engineering

JUICE-UVS's housekeeping data stream is transmitted to two destinations on the spacecraft: one to the spacecraft computer directly and one to the Bulk Data System (BDS) recorder. This telemetry is sent to both destinations at a nominal 1 Hz rate. JUICE-UVS tracks the number of detected photon events in one analog counter and multiple digital counters at configurable rates and this is also sent with the housekeeping telemetry (Section 3.5). The digital counters are distinguished by different onboard processing levels and five parallel counters for configurable spectral bands. These count rates are transmitted to the ground in JUICE-UVS count-rate science packets, then incorporated into archival data products. The BDS-recorded housekeeping is kept intact as a whole packet, ensuring sampling continuity throughout all fields.

### 6.6. Archiving

Data products produced by the JUICE-UVS iSOC are archived in ESA's Planetary Science Archive (PSA) and mirrored in NASA's PDS for long-term access. Data are archived according to the PDS4 standard and the JUICE Science Data Management Plan.

The JUICE Mission Operations System (MOS) delivers spacecraft telemetry data packets to the JUICE-UVS iSOC. The iSOC is then responsible for processing the telemetry into PDS4 raw, calibrated, and derived archive data products. In addition to the FITS files containing the data products, the iSOC also produces extensible markup language (XML) label files in accordance with the PDS4 standard, along with accompanying documentation and ancillary information files. These are then assembled into archive bundles and submitted to PDS for long-term archival storage.

Peer reviews of sample data products are organized throughout the life of the mission. JUICE-UVS raw data products are peer reviewed before launch, with another peer review scheduled late in the cruise phase for calibrated data products.

Archiving activities are organized through periodic meetings of JUICE's Data and Archive Working Group (DAWG). The DAWG consists of representatives from each investigation team, along with project and PDS personnel.

## 7. Summary

JUICE's ultraviolet instrument, JUICE-UVS, is designed to investigate key questions informing Jupiter's icy moons, Jupiter, and the Io system. JUICE-UVS is ready to explore the atmospheres, plasma interactions, and surfaces of the Galilean satellites; to determine the dynamics, chemistry, and vertical structure of Jupiter's upper atmosphere, from equator to pole; and to investigate the Jupiter-Io connection by quantifying energy and mass flow in the Io atmosphere, neutral clouds, and torus. The JUICE-UVS design is similar to the Europa-UVS instrument; they are planned to join each other during their tours at Jupiter starting in 2031 and 2030, respectively. Both UVS instruments have broader wavelength coverage, lower radiation noise susceptibility at Jupiter, improved dynamic range in count-rate and gain stability with fluence, and up to ten times better spatial resolution at certain wavelengths than predecessor UVS/Alice spectrograph concepts. With its planned sequences of nadir-pointed and off-nadir pointed observation activity types, the data sets expected from JUICE-UVS are sure to change our view of ice covered ocean worlds habitability and numerous processes and phenomena in the Jupiter system, an archetype of giant planet systems throughout our Universe.

## Acknowledgements

This work was supported by NASA through the JUICE Project, managed by the project management program office at Marshall Space Flight Center under contract NNM13AA38C. JUICE is a mission under ESA leadership with contributions from its Member States, NASA, JAXA and the Israel Space Agency. It is the first Large-class mission in ESA's Cosmic Vision programme. VH acknowledges support from the French government under the France 2030 investment plan, as part of the Initiative d'Excellence d'Aix-Marseille Université – A*MIDEX AMX-22-CPJ-04. We thank the ESA national funding agencies and supporting institutions whose continuous support of our UVS coinvestigators and the JUICE mission as a whole have made this ambitious programme possible. The authors thank the Belgian Federal Science Policy Office (BELSPO) for the provision of financial support in the framework of the PRODEX Programme of the European Space Agency (ESA) under contract number PEA 4000115321.

**Conflict of Interest:** This work was financially supported by NASA through the JUICE project.

## Authors and Affiliations

K. D. Retherford
kretherford@swri.edu

**Retherford, K. D.[1,2], P. M. Molyneux[1], T. K. Greathouse[1], M. W. Davis[1], G. R. Gladstone[1], S. C. Persyn[1], M. H. Versteeg[1], M. F. Araujo[1], F. Bagenal[3], T. M. Becker[1,2], A. Beth[4], R. K. Black[1], B. Bonfond[5], A. Bouabdellah[6], S. Brooks[7], E. J. Bunce[8], A. Carapelle[9], S. Cortinas[1], G. J. Dirks[1], B. Esquivel[1], J. Eterno[1], P. D. Feldman[+], S. Ferrell[1], M. Ferris[1], L. N. Fletcher[8], M. A. Freeman[1], M. Galand[4], R. S. Giles[1], D. Grodent[5], V. Hue[10], E. Johnson[1], J. A. Kammer[1], C. Kempe[1], L. Lamy[10], A. Martin[11], M. A. McGrath[12], E. G. Nerney[3], C. Nuñez[1], Z. Olsen[1], N. Pelletier[1+], B. Perez[1], K. B. Persson[1], E. Quémerais[6], R. Raffanti[11], U. Raut[1,2], R. Rickerson[1], L. Roth[13], O. H. W. Siegmund[11], J. R. Spencer[14], A. J. Steffl[14], B. J. Trantham[1], J. V. Vallerga[11], T. J. Veach[1,2], M. A. Velez[1], M. Villanueva[1], B. C. Walther[1] and the JUICE-UVS Team**

1. Southwest Research Institute, San Antonio, TX, USA
2. University of Texas at San Antonio, San Antonio, Texas, USA
3. LASP, University of Colorado at Boulder, Boulder, CO, USA
4. Imperial College London, London, UK
5. Université de Liège, Liège, Belgium
6. LATMOS, Guyancourt, France
7. Jet Propulsion Laboratory, California Institute of Technology, Pasadena, CA, USA
8. University of Leicester, School of Physics and Astronomy, Leicester, UK
9. Centre Spatial de Liège (CSL), Space sciences, Technologies and Astrophysics Research (STAR) Institute, Université de Liège, Liège Science Park, Belgium
10. Aix Marseille Université, CNRS, LAM (Laboratoire d'Astrophysique de Marseille), Marseille, France
11. Sensor Sciences LLC, Pleasant Hill, CA, USA
12. SETI Institute, Mountain View, CA, USA
13. KTH Royal Institute of Technology, Stockholm, Sweden
14. Southwest Research Institute, Boulder, CO, USA

ORCID's

Fletcher: 0000-0001-5834-9588